\documentclass[%
  showpacs,preprintnumbers,amssymb,%
  aps,reprint,%
  superscriptaddress,showkeys,%
  amsmath,floatfix%
]{revtex4-2}

\usepackage{graphicx}
\usepackage{dcolumn}
\usepackage{bm}
\usepackage{amsmath}
\usepackage{amssymb}
\usepackage{comment}
\usepackage{amsfonts}
\usepackage{hyperref}
\usepackage{array}
\usepackage{booktabs}
\usepackage{microtype}
\usepackage{multirow}
\usepackage[english]{babel}
\usepackage{subcaption}
\usepackage[normalem]{ulem}
\newcommand{\color}[1]{}

\def \l{\lambda}
\def \L{\Lambda}

\def \O{\Omega}

\def \ben{\begin{equation}\begin{aligned}}
\def \een{\end{aligned}\end{equation}}

\def \La{\mathcal{L}}

\def \n{\\}

\begin{document}

\title{Multi-Probe Cosmological Constraints on Clustering DBI Dark Energy Scenario}

\author{Samit Ganguly}
\email{samitgphy07@gmail.com}
\affiliation{Department of Physics, University of Calcutta, 92, A.P.C. Road, Kolkata-700009, India}
\affiliation{Department of Physics, Haldia Government College, Haldia, Purba Medinipur 721657, India}

\author{Surhud More}
\email{surhud@iucaa.in }
\affiliation{Inter-University Centre for Astronomy and Astrophysics (IUCAA), Pune, India}

\author{Goutam Manna$^a$}
\email{goutammanna.pkc@gmail.com \\$^a$Corresponding author}
\affiliation{Department of Physics, Prabhat Kumar College, Contai, Purba Medinipur 721404, India} 
\affiliation{Institute of Astronomy, Space and Earth Science, Kolkata 700054, India}

\author{Kazuharu Bamba}
\email{bamba@sss.fukushima-u.ac.jp}
\affiliation{Faculty of Symbiotic Systems Science, Fukushima University, Fukushima 960-1296, Japan}

\author{Eduardo Guendelman}
\email{guendel@bgu.ac.il}
\affiliation{Department of Physics, Ben-Gurion University of the Negev, Beer-Sheva, Israel}
\affiliation{Frankfurt Institute for Advanced Studies (FIAS), Ruth-Moufang-Strasse 1, 60438 Frankfurt am Main, Germany.}
\affiliation{Bahamas Advanced Study Institute and Conferences, 4A Ocean Heights, Hill View Circle, Stella Maris, Long Island, The Bahamas.}

\author{Abhijit Bhattacharyya}
\email{abhattacharyyacu@gmail.com}
\affiliation{Department of Physics, University of Calcutta, 92, A.P.C. Road, Kolkata-700009, India}

\date{\today}

\begin{abstract}
We present a multi-probe observational analysis of a clustering tachyonic Dirac-Born-Infeld~(DBI) dark energy model. Its noncanonical kinetic term gives a subluminal sound speed, so dark energy perturbations can grow on subhorizon scales. Working in a multifluid framework that tracks both the background expansion and linear perturbations, we constrain the model with nested sampling, using Planck 2018 compressed CMB distance priors, DESI DR2 BAO, cosmic chronometers, $f\sigma_8$ growth data, and three Type Ia supernova compilations: Pantheon$+$, DES-SN5YR, and Union3. The standard cosmological parameters agree with $\Lambda$CDM within $2\sigma$, with DBI preferring a slightly lower $H_0$ and higher $\Omega_{m0}$, with the largest shift being $\simeq1.8\sigma$ in $H_0$ for Pantheon$+$. The Bayesian evidence favours DBI in every compilation, moderately for Pantheon$+$ ($\Delta\ln\mathcal{Z}\simeq3.4$) and inconclusively for DES-SN5YR and Union3 ($\Delta\ln\mathcal{Z}\simeq0.6$-$0.7$). The field thaws naturally. Dark energy perturbations stay several orders of magnitude below matter perturbations, yet the noncanonical sound speed still alters the Hubble friction term, raising $S_8$ by only $\approx0.5\sigma$ relative to $\Lambda$CDM. Four independent diagnostics, namely the CPL parametrization, $Om(z)$, the statefinder hierarchy, and the growth index, all point to genuinely evolving dark energy rather than a cosmological constant. The thermal history is left intact, with negligible early dark energy. However, one limitation remains. Because $w_\phi=\dot\phi^2-1\geq-1$, the phantom regime is out of reach, which is in line with the no-go theorem for a single minimally coupled scalar field, so the model cannot follow the DESI DR2 preference for an effective equation of state crossing $w=-1$.
\end{abstract}

\keywords{%
$k$-essence geometry, dark energy, dark matter, late-time acceleration, observational data analysis}

\pacs{04.50.Kd, 98.80.-k, 98.80.Bp, 98.80.Es} 

\maketitle

\section{Introduction}

The late-time accelerated expansion of the Universe, established through Type~Ia supernovae, baryon acoustic oscillations (BAO), and the cosmic microwave background (CMB)~\cite{Riess, Perlmutter, Eisenstein, EisensteinHu1998, HuSugiyama1996, Planck2018}, remains one of the most compelling unsolved problems in modern cosmology. Within the standard $\Lambda$CDM framework, this phenomenon is attributed to a cosmological constant, but the associated fine-tuning and coincidence problems motivate dynamical alternatives in which an evolving field describes the dark sector. Among the many possibilities in the dynamical dark-energy sector, a particularly well-studied class of candidates is provided by noncanonical scalar fields ($k$-essence)~\cite{Picon, Picon1}, where the kinetic structure can lead to rich background dynamics and provide a distinctive signature in the growth of cosmic structures.

Among $k$-essence theories, the Dirac-Born-Infeld (DBI) model~\cite{Born, Dirac} offers a theoretically motivated framework, originally arising from the effective description of D-branes in string theory~\cite{Tong, Alishahiha}. In particular, the tachyonic DBI scalar field, arising from the effective action of a non-BPS D-brane, provides a natural dark-energy candidate with a bounded kinetic term and a well-defined pressure. In cosmology, DBI fields can behave as dark energy or unify dark matter and dark energy, depending on the choice of Lagrangian. An important phenomenological feature of DBI-type Lagrangians is the emergence of a nontrivial effective sound speed~($c_s^2$)~\cite{Mukhanov}, which typically depends on time. Since the sound speed controls the propagation of perturbations, it determines whether the dark component remains smooth on subhorizon scales or clusters with matter.

The clustering properties of DBI dark energy are therefore central to confronting the model with large-scale-structure data. If the effective sound speed is close to unity, pressure support suppresses perturbations, and the field affects structure formation mainly through the background expansion history. In contrast, a small sound speed allows the DBI component to develop inhomogeneities, potentially modifying the matter power spectrum, the growth rate $f\sigma_8$, and gravitational-lensing observables~\cite{Hu, Dore}. These effects can introduce degeneracies with standard cosmological parameters and offer new avenues to discriminate between DBI scenarios and $\Lambda$CDM. Moreover, the modified growth history of DBI dark energy may offer a path toward relieving the $S_8$ tension, which has emerged as a persistent discrepancy between CMB-inferred and weak-lensing measurements of the matter fluctuation amplitude.

We note that the Dark Energy Spectroscopic Instrument's Data Release 2 (DESI-DR2)~\cite{DESI, DESI2} in combination with data from the CMB and Type Ia supernovae (regardless of the chosen compilation) show evidence for an evolving dark energy component. The constant dark energy model is at odds with the observations at 2.8-4.2 $\sigma$ depending upon the combination of data sets used. Non-parametric reconstructions also show a preference for an effective evolving dark energy equation of state~\cite{Lodha}, where $w>-1$ at the current epoch, but shows a phantom-like behaviour with $w<-1$ at redshifts beyond $0.5$. It is interesting to understand if this behaviour is a side-effect of the adopted parameterization or a requirement for explaining the data~\cite{Mishra}.

The goal of this study is to investigate whether a dynamical dark-energy component with a time-dependent equation of state and a nontrivial sound speed can address phenomenological shortcomings of $\Lambda$CDM - in particular the $S_8$ tension between CMB and large-scale-structure observations, and the $H_0$ discrepancy between early- and late-Universe probes. A varying equation of state offers a degree of freedom absent in the case of a rigid cosmological constant, thereby allowing modifications to the late-time expansion history, while dark-energy clustering can leave additional signatures on the growth of structure beyond purely background effects. Since background and perturbation observables are strongly degenerate and can be obscured when relying on a single data set, we perform a consistent multiprobe analysis that combines geometric measurements (CMB distance priors, BAO, SNe~Ia, and cosmic chronometers) with growth data ($f\sigma_8$). This enables us to confront the model simultaneously with both expansion history and structure formation observations.

In this work, we investigate a clustering tachyonic DBI dark-energy scenario in a spatially homogeneous and isotropic Friedmann--Lema\^{i}tre--Robertson--Walker spacetime, focusing on the interplay between the background expansion history and linear perturbations induced by the underlying $k$-essence scalar field. We derive the background and perturbation equations, identify the physical conditions, chiefly controlled by the time-dependent effective sound speed, under which the DBI component can cluster on subhorizon scales, and quantify the resulting imprint on late-time structure formation through the growth observable $f\sigma_8$. We constrain the model using a broad suite of current observations, including cosmic chronometers, Type~Ia supernovae (Pantheon$+$ excluding SHOES calibrations, DES-SN5YR, and Union3), large-scale-structure information, BAO measurements from DESI~DR2, compressed CMB distance priors, and $f\sigma_8$ data. We then perform a comparative parameter inference study to assess how the constraints shift when different supernova and large-scale-structure compilations are adopted. Finally, we obtain posterior distributions via \textsc{dynesty} nested sampling.

Note that throughout this work, we use the metric signature $(-,+,+,+)$. Also, we adopt a mixed-unit convention. In all theoretical expressions, including the action, the Friedmann equations, the scalar field equation of motion, and the linear perturbation equations, we set $c = 1$ implicitly. Whereas the factor of $c$ is restored explicitly in all of observational and distance-measure formulas. Since parameter estimation proceeds entirely through the observational sector, where $c$ is consistently retained, this convention introduces no ambiguity in the likelihood and does not affect any inferred parameter values. We also set the reduced Planck mass $M_{\rm Pl} = 1$ in our analysis.

The paper is organized as follows. In Sec.~\ref{S2} we introduce the tachyonic DBI ($k$-essence) framework, present the action and Lagrangian derive the background evolution and scalar-field equation of motion, and formulate the linear-perturbation equations in a multifluid description, together with the distance- and BAO-related quantities and the autonomous dynamical system used for numerical evolution. In Sec.~\ref{S3} we summarize the observational inputs employed in our joint likelihood analysis. In Sec.~\ref{S4} we describe our inference pipeline that enables efficient exploration of the full parameter space across multiple dataset combinations. In Sec.~\ref{S5} we present and discuss the resulting constraints, the implied background and perturbation dynamics, and the associated statistical consistency measures between the different supernova compilations. Finally, we conclude in Sec.~\ref{S6}. 

\section{DBI-type Lagrangian and formulation of the associated equations}
\label{S2}

In this section, we present the theoretical framework of the tachyonic DBI dark energy (k-essence) model. We begin by specifying the action and subsequently derive the background cosmological equations and relevant perturbative quantities governing the dynamics of the model.

\subsection{Action and Lagrangian}

We consider a spatially flat universe described by the action \cite{Copeland}
\ben
S = \int d^4x \sqrt{-g} 
\bm{\Big[}\frac{1}{2} \mathcal{R} + \mathcal{L}_\phi(\phi,X) \bm{\Big]} + S_{\mathrm m} + S_{\mathrm r}.
\label{1}
\een
Here, $M_{\mathrm{Pl}}$ denotes the reduced Planck mass, which we set to unity, and $\mathcal{R}$ is the Ricci scalar. The quantity $\mathcal{L}_\phi$ represents the tachyonic DBI Lagrangian density, while $S_{\mathrm m}$ and $S_{\mathrm r}$ correspond to the matter (including both baryonic and dark matter) and radiation sectors, respectively. The action in Eq.~\eqref{1} describes a non-interacting cosmological framework in which dark energy, modeled by the scalar field, evolves independently of the matter and radiation components~\cite{Copeland, Tsujikawa2013DE}. In particular, there are no direct non-gravitational couplings or energy exchange between these sectors (in contrast to interacting/coupled dark-energy scenarios; see, e.g.,~\cite{Amendola2000CoupledDE}). Their interaction arises solely through gravity, where the matter and radiation sectors couple minimally to the metric $g_{\mu\nu}$, while the scalar field influences the spacetime geometry via its noncanonical kinetic structure encoded in $\mathcal{L}_\phi$.

The tachyonic DBI Lagrangian, we are particularly interested in is given by~~\cite{Scherrer1, Sen, Padmanabhan, gm1, gm2, gm3, Ganguly, Ganguly1, Panda, Panda1}
\ben
\mathcal{L}_\phi\equiv \La_{\phi}(\phi,X) = - V(\phi) \sqrt{1 - 2X},
\label{2}
\een
where
\ben
X = -\frac{1}{2} g^{\mu\nu} \partial_\mu \phi \partial_\nu \phi,
\label{3}
\een
is the canonical kinetic term, $V(\phi)$ is the scalar field potential.

In this work, we consider a spatially homogeneous scalar field $\phi \equiv \phi(t)$ \cite{Vikman, Das, Ganguly1, Ganguly}, consistent with the isotropic and homogeneous FLRW background. As a result, spatial derivatives vanish, $\partial_i \phi = 0\, (i=1,2,3)$, and the kinetic term simplifies to
$X = -\tfrac{1}{2} g^{\mu\nu} \partial_\mu \phi \partial_\nu \phi = \tfrac{1}{2} \dot{\phi}^2.$ This setup isolates the background expansion dynamics, while inhomogeneities come separately at the level of linear perturbations.\\

A distinctive feature of the DBI Lagrangian is its non-canonical square root kinetic structure, $\sqrt{1 - 2 X}$,
which fundamentally modifies the dynamical behavior of the scalar field. Unlike canonical quintessence models, where the kinetic term is quadratic in derivatives, the DBI form resembles the relativistic action of a particle and naturally introduces a Lorentz-like factor defined as
\ben
\gamma = \frac{1}{\sqrt{1 - \dot{\phi}^2}}
\label{4}
\een
The requirement that the action remains real imposes the condition
$1 - \dot{\phi}^2 > 0,$ which leads to the constraint $\dot{\phi}^2 < 1$. This condition effectively bounds the field velocity, ensuring that it never exceeds the limiting value of unity. Furthermore, it is important to acknowledge that in our scenario, $\dot\phi^{2}\neq 0$, as assuming $\dot\phi^{2}$ to be zero makes the k-essence theory meaningless. Consequently, the scalar field dynamics exhibits a relativistic behavior, with $\gamma$ playing a role analogous to the Lorentz boost factor in special relativity.

This intrinsic speed limit is a defining characteristic of tachyonic DBI models and is directly responsible for the emergence of a non-trivial effective sound speed,
\ben
c_{s,\phi}^2 = \frac{1}{\gamma^2} = 1 - \dot{\phi}^2,
\label{5}
\een
which controls the propagation of perturbations~~\cite{Gibbons, Bamba, Copeland}.
As a result, the DBI structure not only modifies the background cosmological evolution but also determines the clustering properties of dark energy at the perturbative level. 

\subsection{Background Cosmological Equations}

We assume a spatially flat FLRW metric,
\ben
ds^2 = -dt^2 + a^2(t) d\vec{x}^2,
\label{6}
\een
where $a(t)$ is the scale factor.

For a homogeneous scalar field $\phi = \phi(t)$, the pressure ($\bar{p}_\phi$) of the scalar field is identified with the Lagrangian:
\ben
\bar{p}_\phi = \mathcal{L}_\phi.
\label{7}
\een
The corresponding energy density ($\bar{\rho}_\phi$) is obtained from
\ben
\bar{\rho}_\phi = 2X \mathcal{L}_{,X} - \mathcal{L}_\phi,
\label{8}
\een
which yields
\ben
\bar{\rho}_\phi =
\frac{V(\phi)}{\sqrt{1- 2X}}.
\label{9}
\een
Therefore, the dark energy equation of state~(EoS) parameter can be written as
\ben
w_\phi = \frac{\bar{p}_\phi}{\bar{\rho}_\phi} = - (1 - \dot{\phi}^2).
\label{10}
\een
The Friedmann equations are then given by
\ben
3 H^2 = \bar{\rho}_m + \bar{\rho}_\phi + \bar{\rho}_r,
\label{11}
\een
\ben
2\dot{H} = - 
\big(\bar{\rho}_m + \bar{\rho}_\phi + \bar{\rho}_r + \bar{p}_\phi + \bar{p}_r \big).
\label{12}
\een
Here, $H = \dot{a}/a$ represents the Hubble parameter. We take the matter Lagrangian, which correspond to dust-like, pressureless matter, implying $p_m = 0$. In this setting, $\bar{\rho}_m$ denotes the total matter energy density, while $\bar{\rho}_r$ and $\bar{p}_r$ refer to the radiation energy density and radiation pressure, respectively, corresponding to the action \eqref{1}. In this context, we note that the radiation energy density and pressure are related by $\bar{p}_{r} = \frac{\bar{\rho}_{r}}{3}$, so the radiation EoS can be written as $w_r = \frac{\bar{p_r}}{\bar{\rho_r}} = 1/3$.

The continuity equation for the $i^{th}~(i=m, \phi, r)$ component in flat FLRW background takes the form
\ben
\frac{d\bar{\rho}_{i}}{dt} = - 3H(\bar{\rho}_{i} + \bar{p}_{i})
\label{13}
\een.

\subsection{Scalar Field Equation of Motion}

Varying the action defined in Eq.~\eqref{1} with respect to $\phi$ yields the scalar-field equation of motion:
\ben
\ddot{\phi} + \frac{3H}{\gamma^2}\dot{\phi}
+ \frac{1}{\gamma^2} \frac{V_{,\phi}}{V} = 0.
\label{14}
\een
where $V_{,\phi}\equiv \frac{dV(\phi)}{d\phi}$. This equation \eqref{14} governs the background evolution of the tachyonic DBI field. 

Physically, this equation exhibits three distinct contributions. The first term $\ddot{\phi}$ represents the intrinsic acceleration of the scalar field. The second term, $\frac{3H}{\gamma^2}\dot{\phi}$, acts as a Hubble friction term arising from cosmic expansion. The factor $1/\gamma^2$ modifies the effective damping relative to a canonical scalar field, which essentially reflects the non standard DBI kinetic structure. The final term, $\frac{1}{\gamma^2}\frac{V_{,\phi}}{V}$, corresponds to the effective force generated by the scalar potential.

From a cosmological perspective, Eq.~\eqref{14} determines how the DBI field interpolates between different dynamical regimes. In the slow-roll limit ($\dot{\phi}^2 \ll 1$, $\gamma \approx 1$), the equation reduces to a damped Klein-Gordon form, allowing accelerated expansion. In contrast, when $\dot{\phi}^2 \to 1$, the field behaves effectively as pressureless matter, thereby enabling tracking or scaling behavior in the early Universe.

\subsection{Linear Perturbations: Fluid Description of Each Component}

To study structure formation and the impact of the tachyonic DBI field on observables, we consider linear scalar perturbations around the spatially flat FLRW background. In the Newtonian gauge, also known as the longitudinal gauge, the perturbed metric is written as~\cite{Bertschinger, Bardeen, Kodama, Dodelson, MFB, Weinberg, Bamba}
\ben
\mathrm{d}s^2 = -\big(1+2\Psi\big)\mathrm{d}t^2 + a^2(t)\big(1-2\Phi\big)\delta_{ij}\,\mathrm{d}x^i\mathrm{d}x^j,
\label{15}
\een
where $\Psi$ and $\Phi$ are the two scalar metric potentials often referred to as the Bardeen potentials which, in the conformal Newtonian gauge, coincide with the corresponding gauge invariant combinations of scalar perturbations~\cite{Bertschinger,Dodelson}. The potential $\Psi$ perturbs the time-time component of the metric, while $\Phi$ characterizes the scalar perturbation of the spatial curvature. In the absence of appreciable anisotropic stress as for pressureless matter and minimally coupled scalar fields at linear order, the Einstein equations imply $\Phi=\Psi$~\cite{Bertschinger,Dodelson}.

In this gauge setup, the scalar perturbations described by $\Phi$ and $\Psi$ fully encode the gravitational response of all matter components, which includes the tachyonic DBI scalar field, pressureless matter, and radiation. The perturbations of the scalar field enter the Einstein equations through their contributions to the perturbed energy-momentum tensor, which act as a source to the Bardeen potentials. Consequently, the metric potentials act as mediators of structure formation \cite{MFB, Kodama, Dodelson, Weinberg}. They determine the evolution of matter density perturbations via the Poisson and dynamical constraint equations, while simultaneously reflecting the clustering properties, sound speed, and EoS behavior of the DBI field. In this way, all modifications induced by the non-canonical scalar sector, whether at the level of background expansion or perturbation dynamics, are gravitationally encoded in $\Phi$ (or equivalently $\Psi$ in the absence of anisotropic stress), which directly governs observable quantities such as the growth rate $f\sigma_8$, and large scale structure formation.

It is worth emphasizing that, although the tachyonic DBI model belongs to the class of k-essence theories whose linear scalar perturbations propagate on an effective acoustic metric determined by the kinetic structure of the Lagrangian, this emergent metric is not conformally equivalent to the underlying gravitational metric $g_{\mu\nu}$~\cite{Mukhanov, Picon1}. In the present analysis, we adopt the unitary gauge, defined by $\delta\phi=0$, so that the scalar field fluctuation is absorbed into the scalar metric perturbations denoted by $\Phi$ and $\Psi$ \cite{Zumalacárregui, Cheung, Gleyzes, Roy}. For a single minimally coupled scalar field there is only one physical scalar degree of freedom at linear order, meaning different gauge choices simply distribute this mode between $\delta\phi$ and the metric variables~\cite{MFB, Kodama, Weinberg}. Consequently, instead of evolving $\delta\phi$ explicitly, we evolve the metric potentials sourced by the perturbed energy-momentum tensor, which fully captures the scalar perturbation dynamics in our single-field setup. We implement this using an effective fluid description of the DBI component, in which the non-canonical kinetic structure enters only through the background EoS parameter $w_\phi$ and the effective rest-frame sound speed $c_{s,\phi}^2$~\cite{Hu, Dore}.

Within this framework, the DBI component behaves as a perfect fluid at linear order, and its perturbative effects enter the Einstein and conservation equations only through $(\delta_\phi, \theta_\phi)$ defined later in Eq.~\eqref{25} together with $w_\phi$ and $c_{s,\phi}^2$~\cite{Hu, Dore}. Consequently, while the effective acoustic metric governs the internal propagation properties of the scalar fluctuations, the spacetime geometry itself remains described by perturbations of the FLRW metric sourced by the density and velocity perturbations of the different sectors discussed above. The use of the conformal Newtonian gauge is therefore fully consistent in our analysis~\cite{Bertschinger}.

In our analysis, we do not explicitly account for perturbations in the radiation component. This simplification is appropriate because we are mainly interested in the growth of cosmic structures after recombination and we do not essentially modify the key physics governing the early universe. Before recombination, the universe is radiation dominated. At very high redshift the energy density is dominated by relativistic species~($\rho_r\propto a^{-4}$), and density fluctuation in the coupled photon-baryon plasma undergo acoustic oscillation rather than the gravitational growth which is characteristic of pressureless matter~\cite{Dodelson}. As the expansion proceeds, the matter component ($\rho_m\propto a^{-3}$) eventually overtakes radiation at matter-radiation equality. After this the background expansion is well described by matter domination~\cite{Dodelson, Weinberg}.

At redshift $z\sim 1100$, corresponding to the epoch of recombination, the expansion is already deep into the matter-dominated era as matter-radiation equality occurs around $z_{\mathrm{eq}}\simeq 3400$, so that $\rho_m\gg\rho_r$ and the radiation component provides only a small correction to the background and to the late-time growth of matter perturbations~\cite{Dodelson,Weinberg,Planck2018}. Dark energy is even more strongly suppressed at these redshifts and is dynamically negligible compared to matter in the Friedmann equation, so it does not affect the evolution of perturbations near recombination.

Therefore, the subsequent evolution of density perturbations from $z=1100$ to $z=0$ is predominantly governed by matter and dark energy dynamics. Neglecting radiation perturbations introduces only a subleading correction to the growth history at late times and does not significantly affect the observational quantities considered in this work.

An effective fluid energy momentum tensor of each component $i\in\{m, r,\phi\}$ can be described by
\ben
T^{\mu}{}_{\nu(i)} = \big(\rho_i+p_i\big)u^{\mu}_{(i)}u_{\nu(i)} + p_i\,\delta^{\mu}{}_{\nu},
\label{16}
\een
with background quantities $\bar\rho_i(t)$ and $\bar p_i(t)$, and first-order perturbations
\ben
\rho_i(t,\vec{x}) = \bar\rho_i(t)\big(1+\delta_i(t,\vec{x})\big) \\
 p_i(t,\vec{x}) = \bar p_i(t) + \delta p_i(t,\vec{x}),
\label{17}
\een
where $\delta_i = \frac{\delta\rho_i}{\bar\rho_i}$ is the dimensionless density contrast of the $i$-th component, defined as the fractional perturbation of its energy density with respect to the homogeneous background $\bar\rho_i$~\cite{Bertschinger, Kodama, MFB, Dodelson}. The scalar part of the perturbed 4-velocity can be written in terms of a velocity potential $v_i$ (in Fourier space), and we define the velocity divergence ~\cite{Bertschinger, Kodama, MFB, Dodelson}
\ben
\theta_i \equiv \frac{k^2}{a}\,v_i
\label{18}
\een
where $k$ is the comoving wavenumber. We also define the EoS parameter $w_i\equiv \bar p_i/\bar\rho_i$ and the rest frame sound speed $c^2_{s,i}\equiv (\frac{\delta \bar{p}_i}{\delta\bar{\rho}_i})_{\mathrm{rf}}$.

In linear order, energy momentum conservation $\nabla_\mu T^{\mu}{}_{\nu(i)}=0$ yields the standard continuity and Euler equations for each fluid mode as \cite{Bertschinger, Kodama, MFB, Dodelson, Fahimi}:
\ben
\dot\delta_i = -(1+w_i)\big(\frac{\theta_i}{a}-3\dot\Phi\big)-3H\big(c^2_{s,i}-w_i\big)\delta_i,
\label{19}
\een
\ben
\dot\theta_i = -H\big(1-3c^2_{s,i}\big)\theta_i + \frac{k^2}{a}\big(\Psi+\frac{c^2_{s,i}}{1+w_i}\,\delta_i\big),
\label{20}
\een
where overdots indicate derivatives with respect to the cosmic time $t$, and the above equations are expressed in Fourier space by using $\nabla^2 = -k^2$.

The fluid equations must be complemented by the linearized Einstein equations, which connect the metric potentials $(\Phi,\Psi)$ to the perturbations in the matter/fields via the relation $\delta G^{\mu}_{\nu}=\delta T^{\mu}_{\nu}$. These are the Einstein equations at linear order in perturbation theory, which assert that the first order perturbation of the Einstein tensor ($\delta G^{\mu}_{\nu}$) is equal to the first order perturbation of the stress–energy tensor ($\delta T^{\mu}_{\nu}$). In the conformal Newtonian gauge and after transforming to Fourier space, the time-time component of this equation yields a generalized Poisson equation,
\ben
-k^2\Phi + 3a^2H\big(\dot\Phi + H\Psi\big)= \frac{a^2}{2}\,\sum_{i}\delta\rho_{i},
\label{21}
\een
while on subhorizon scales ($k\gg aH$) this reduces to the familiar form
\ben
k^2\Phi \simeq - \frac{a^2}{2}\,\sum_{i}\delta\rho_{i}.
\label{22}
\een
For completeness, we may write the time-space constraint as
\ben
\dot\Phi + H\Psi = \frac{a}{2}\sum_{i}(\bar{\rho}_{i}+\bar {p}_i)v_i,
\label{23}
\een
with $v$ the total velocity potential ~\cite{Bertschinger,Dodelson,Mukhanov}.

In the following analysis, we adopt the quasi-static approximation on subhorizon scales ($k \gg aH$), under which spatial gradients dominate over time derivatives of the metric perturbations. 
In this regime, the gravitational potential evolves slowly compared to the Hubble time, satisfying $|\dot{\Phi}| \ll H|\Phi|$. 
This behavior is exact during matter domination, where $\Phi$ remains constant, and remains a good approximation at late times for modes relevant to large-scale structure observations. 
Consequently, we neglect the $\dot{\Phi}$ term in the continuity equation, which introduces only subleading corrections to the growth of matter perturbations on the scales considered in this work.

\subsubsection{Matter (pressureless dust)}

For nonrelativistic matter, $w_{\mathrm{m}}=0$ and $c^2_{\mathrm{s,m}}=0$, so Eqs.~\eqref{19} and \eqref{20} reduce to
\ben
\dot\delta_{\mathrm{m}} = -\frac{\theta_{\mathrm{m}}}{a} ,\qquad
\dot\theta_{\mathrm{m}} = -H\theta_{\mathrm{m}} + \frac{k^2}{a}\Psi,
\label{24}
\een

\subsubsection{Tachyonic DBI field as an effective fluid}

The tachyonic DBI $k$-essence field can be reformulated as a perfect fluid at both the background and linear-perturbation levels. The background energy density and pressure, together with the dark-energy equation of state and the sound speed~($c_{s,\phi}^2$), are given in Eqs.~\eqref{5}, \eqref{7}, \eqref{9}, and \eqref{10}. Consequently, the perturbations of the scalar field can be modeled as a fluid described by $(\delta_\phi,\theta_\phi)$, supplemented by the time-dependent quantities $w_\phi(t)$ and $c^2_{s,\phi}(t)$. Within this fluid framework, we can employ Eqs.~\eqref{19} and \eqref{20} with $i=\phi$ to evolve the tachyon density contrast and velocity divergence in a self-consistent manner alongside matter. The perturbation equations for the scalar field then become

\ben
\dot{\delta}_{\phi}=-(1 + w_{\phi})\frac{\theta_{\phi}}{a}-3H(c_{s,\phi}^2 - w_{\phi})\delta_{\phi} , \n 
\dot\theta_{\phi} = -H\big(1-3c^2_{s,\phi}\big)\theta_{\phi} + \frac{k^2}{a}\big(\Psi+\frac{c^2_{s,\phi}}{1+w_{\phi}}\,\delta_{\phi}\big),
\label{25}
\een

\subsection{Comoving distance and BAO scales}

For a spatially flat FLRW background, the (line-of-sight) comoving distance to redshift $z$ is defined by~\cite{Weinberg,Dodelson}
\ben
\chi(z)=c\int_{0}^{z}\frac{\mathrm{d}z'}{H(z')}\,,
\label{26}
\een
This quantity converts a redshift interval into a comoving radial separation and serves as the basic building block for distance measures used in large-scale structure analyses. The transverse comoving distance is $D_M(z)=\chi(z)$ (flat case), and the angular diameter distance is $D_A(z)=\frac{D_M(z)}{1+z}$. The luminosity distance, relevant for standard candle observations, is~\cite{Weinberg, Dodelson}
\ben
D_L(z)=(1+z)\,D_M(z)=(1+z)^2\,D_A(z)\,,
\label{27}
\een
It is also convenient to define the Hubble distance $D_H(z)=\frac{1}{H(z)}$, which sets the radial conversion scale.
For an object of absolute magnitude $M$ and apparent or observed magnitude $m$, the distance modulus relates them to $D_L$ via 
\ben
\mu(z)\equiv m-M = 5\log_{10}\!\big(\frac{D_L(z)}{10\,\mathrm{pc}}\big) \\
=5\log_{10}\!\big(\frac{D_L(z)}{\mathrm{Mpc}}\big)+25,
\label{28}
\een
which is the quantity that Type-Ia supernova Hubble diagrams~\cite{Riess, Perlmutter} directly constrain.\\

BAO provides a ``standard ruler'' set by the comoving sound horizon in the photon-baryon plasma prior to recombination. The relevant physical scale for late-time clustering is the sound horizon at the baryon drag epoch~\cite{EisensteinHu1998, HuSugiyama1996, Planck2018}
\ben
r_{\mathrm d}\equiv r_{\mathrm s}(z_{\mathrm d})=\int_{z_d}^{\infty}\frac{c_s(z)}{H(z)}\,\mathrm{d}z,
\label{29}
\een
with $c_s(z)$ being defined as the baryon-photon sound speed and $z_d$ the drag redshift. In configuration space, this scale appears as a broad peak in the galaxy two-point correlation function at a comoving separation $\sim \mathcal{O}(10^2)\,\mathrm{Mpc}$, which reflects the maximum distance travelled by the primordial sound wave before baryons decouple from photons~\cite{Eisenstein, Planck2018}.

Observationally, BAO measurements constrain combinations of $D_M(z)$ and $D_H(z)$ relative to $r_d$. In isotropic analyses we often uses,
\ben
D_V(z)=\bm{\Big[}D_M^2(z)\,\frac{cz}{H(z)}\bm{\Big]}^{1/3},
\label{30}
\een
whereas anisotropic analyses separately probe the transverse and radial dilation through $D_M(z)/r_d$ and $D_H(z)/r_d$. These relations provide a direct geometrical test of the expansion history $H(z)$ and, consequently, of the underlying dark energy and modified gravity dynamics.

\subsection{Autonomous system of differential equations}

Although the differential equations are currently written with respect to cosmic time $t$, it is convenient to rewrite them in terms of the number of e-folds $N$, defined by $N\equiv \ln a$. Since $\frac{\mathrm dN}{\mathrm dt}=H$, time derivatives can be re-expressed for derivatives with respect to $N$ via $\frac{\mathrm d}{\mathrm dt}=H\,\frac{\mathrm d}{\mathrm dN}$. This choice of independent variable is particularly useful in cosmology because it factors out the background expansion and often turns the evolution equations into an autonomous dynamical system. To this end, we introduce the following dimensionless variables that capture the scalar field kinetic and potential contributions relative to the Hubble scale: $x\equiv \dot{\phi}$ and $y\equiv \frac{\sqrt{V(\phi)}}{\sqrt{3}H}$. We also define the potential slope and curvature parameters $\lambda\equiv -\frac{V'(\phi)}{V^{\frac{3}{2}}(\phi)}$ and $\Gamma\equiv \frac{V''(\phi)V(\phi)}{\big[V'(\phi)\big]^2}$, which determine how the potential drives the phase-space evolution.

Hence, the autonomous system of differential equations can be written as

\ben
\frac{dx}{dN} &= - (1-x^2)\bm{\Big(}3x - \sqrt{3}\,\l y\bm{\Big)}, \\
\frac{dy}{dN} &= -y\Big(\frac{\sqrt{3}}{2}\,\l x y - \frac{dB}{dN}\Big), \\
\frac{d\lambda}{dN} &= \sqrt{3} \l^2 x y \Big(\frac{3}{2} - \Gamma\Big), \\
\frac{dF}{dN} &= - \frac{e^{-N}}{E}, \\
\frac{d\delta_{\mathrm{m}}}{dN} &= - u_{\mathrm{m}}, \\
\frac{du_{\mathrm{m}}}{dN} &= - u_{\mathrm{m}} \Big(2 + \frac{dB}{dN}\Big) + Q\Phi, \\
\frac{d\delta_{\phi}}{dN} &= - u_{\phi}(1 + w_{\phi}) - 3(c_s^2 - w_{\phi})\delta_{\phi}, \\
\frac{du_{\phi}}{dN} &= -u_{\phi}\Big(2 - 3c_s^2 + \frac{dB}{dN}\Big) + Q\bm{\Big(}\Phi + \frac{c_s^2}{1 + w_{\phi}}\delta_{\phi}\bm{\Big)}, \\
\frac{dQ}{dN} &= -2\Big(1 + \frac{dB}{dN}\Big)Q.
\label{31}
\een
Here, $\O_{\mathrm{m}}=\frac{\bar{\rho}_{\mathrm{m}}}{3H_0^2}$, $\O_{\mathrm{r}}=\frac{\bar{\rho}_{\mathrm{r}}}{3H_0^2}$, and $Q=\Big(\dfrac{ck}{aH}\Big)^2$; moreover, $u_{\mathrm{m}}=\dfrac{\theta_{\mathrm{m}}}{aH}$, $u_{\phi}=\dfrac{\theta_{\phi}}{aH}$, and $c_{s,\phi}^2\equiv c_s^2$. The scalar field density parameter is written in the dimensionless form as $\O_{\phi}=\frac{y^2}{\sqrt{1-x^2}}$.

The quantities $\O_{\mathrm{m}}$ and $\O_{\mathrm{r}}$ follow from the continuity equation in Eq.~\eqref{13} for matter and radiation and can be expressed in terms of the corresponding present-day values ($\O_{\mathrm{m0}}$ and $\O_{\mathrm{r0}}$) as

\ben
\O_{\mathrm{m}}=\O_{\mathrm{m0}}e^{-3N}~ ; ~ \O_{\mathrm{r}}=\O_{\mathrm{r0}}e^{-4N}.
\label{32}
\een
Defining $B=\ln(H)$, we may write
\ben
\frac{dB}{dN} = -\frac{3}{2}\bm{\Big(}\frac{\O_m + \frac{4}{3}\O_r}{E^2} + \O_{\phi}(1 + w_{\phi})\bm{\Big)},
\label{33}
\een
with
\ben
E^2 = \frac{\O_m + \O_r}{1 - \O_{\phi}},
\label{34}
\een
obtained from Eq.~\eqref{11} using the definitions of $\O_m$, $\O_r$, and $\O_{\phi}$.\\

We also introduce the dimensionless comoving distance variable
\ben
F(N)\equiv H_0\bm{\Big[}\chi(N)-\chi(N_i)\bm{\Big]},
\label{35}
\een
where $\chi$ represents the comoving distance along the line of sight. Using the relation $1+z=e^{-N}$ and introducing $E(N)\equiv H(N)/H_0$, Eq.~\eqref{26} yields the form of $\frac{dF}{dN}$ given in Eq.~\eqref{32}. Once $F$ is obtained, the comoving distance can be expressed as $\chi(N)=\chi(N_i)+F(N)/H_0$. With this definition, the initial condition $F(N_i)=0$ follows automatically.

The autonomous system in Eq.~\eqref{31} is integrated by specifying initial data at $a=a_i$ (or, equivalently, $N=N_i$). Hereafter, the subscript $i$ denotes evaluation at these initial values. We set $x_i=0.2$ and $y_i=\sqrt{\O_{\phi,i}}$, ensuring that the background variables reproduce the assumed initial dark-energy abundance and satisfy the Friedmann constraint via Eq.~\eqref{34}. The choice $x_i=0.2$ is not finely tuned: for a broad range of sufficiently small initial values, the dynamics is driven into the same early-time ``frozen'' regime by Hubble damping, so the resulting late-time observables are nearly insensitive to the precise value of $x_i$. In particular, at $z_i\approx1100$ the friction term $3H\dot\phi/\gamma^2$ in Eq.~\eqref{14} typically dominates over the potential gradient and rapidly damps the field velocity, $\dot\phi\to0$, on a timescale much shorter than a Hubble time. Consequently, different (small) initial conditions converge to $w_\phi\approx-1$ well before dark energy becomes dynamically relevant, indicating an attractor-like behavior of evolution in this $k$-essence scenario. As $H$ decreases at late times, the field subsequently thaws, but the inferred cosmological constraints remain effectively unchanged throughout this attractor regime.\\

For the perturbations, we impose adiabatic initial conditions. For the dominant matter component, we take $(\delta_m)_i=a_i$ and $(u_m)_i=-a_i$, and we initially set the scalar-field fluctuations to zero, $\delta_{\phi i}=u_{\phi i}=0$. This choice selects the growing mode and ensures that any subsequent dark-energy clustering arises dynamically from the coupled evolution rather than being introduced by hand. Finally, for each Fourier mode we define $Q_i=\big(ck/(a_i H_i)\big)^2$, where $H_i=H_0 E_i$ is obtained from Eqs.~\eqref{32} and \eqref{34}. In the numerical implementation, we adopt a representative linear scale $k=0.05\,h\,\mathrm{Mpc}^{-1}$, comfortably within the quasistatic range $0.01\,h\le k\le 0.1\,h\,\mathrm{Mpc}^{-1}$ relevant for large-scale-structure analyses.\\

The growth observable used in our large-scale structure likelihood is $f\sigma_8(z)$, which required the redshift evolution of the rms matter fluctuation $\sigma_8(z)$ on sphere of radius $8\,h^{-1}\,\mathrm{Mpc}$. In our numerical analysis this quantity is obtained directly from the gauge invarient linear matter density contrast $\delta_{\mathrm m}(z)$ evolved using the perturbation system in Eq.~\ref{31}. We define the linear growth factor
\ben
D(z)\equiv\frac{\delta_{\mathrm m}(z)}{\delta_{\mathrm m}(0)}
\label{35a}
\een
and normalize the amplitude of fluctuation by treating the present-day value $\sigma_{8,0}$ as a free parameter in the inference. The redshift dependent normalization then follows as,
\ben
\sigma_8(z)=\sigma_{8,0}\,D(z)= \sigma_{8,0}\,\frac{\delta_{\mathrm m}(z)}{\delta_{\mathrm m}(0)}
\label{35b}
\een
The corresponding linear growth rate is
\ben
f(z)\equiv\frac{d\ln\delta_{\mathrm m}}{dN}= \frac{1}{\delta_{\mathrm m}}\frac{d\delta_{\mathrm m}}{dN}=-\frac{u_{\mathrm m}}{\delta_{\mathrm m}}
\label{35c}
\een
the definition $d\delta_{\mathrm m}/dN=-u_{\mathrm m}$ comes from Eq.~\ref{31}. Combining these expressions (Eqs.~\ref{35b} and  \ref{35c}) yields the theoretical prediction for $f\sigma_8(z)$ used in Sec.~\ref{S3}.

\section{Observational datasets}
\label{S3}
In this section, we describe the observational data sets used to constrain both the model parameters and the standard cosmological parameters. For each observational probe, we compare the corresponding theoretical predictions with the measured quantities and construct an associated likelihood function. Under the assumption that the data sets are statistically independent, the total likelihood is obtained by summing the individual contributions, and the resulting joint constraints are then employed to infer the posterior distributions of the parameters.

We now summarize the main probes considered: background expansion history measurements, including $H(z)$ determinations and distance modulus data from Type~Ia supernovae; geometric information from BAO; and early-universe constraints encapsulated in CMB-derived distance priors. In addition, we incorporate large-scale structure observables, encapsulated by $f\sigma_8$, to constrain the growth of matter perturbations.

\subsection{Hubble Data}
The Hubble parameter measurements provide direct constraints on the background expansion rate at different redshifts. We use observational determinations of $H(z)$ obtained from cosmic chronometers which measures differential ages of passively evolving galaxies~\cite{Zhang, Simon, Moresco1, Moresco2, Ratsimbazafy, Stern, Borghi, Moresco3}. The corresponding contribution to the total likelihood is constructed from
\ben
\chi^2_{\mathrm{H}} = \sum_{i}\frac{\bm{\Big[}H_{\mathrm{th}}(z_i)-H_{\mathrm{obs}}(z_i)\bm{\Big]}^2}{\sigma_{\mathrm{H},i}^2},
\label{36}
\een
where $H_{\mathrm{obs}}(z_i)$, $H_{\mathrm{th}}$, and $\sigma_{\mathrm{H},i}$ denote the measured values, theoretical predictions, and uncertainties, respectively. The data set used in this work is provided in Appendix~\ref{A1}.

\subsection{DES SN5YR Data}
The Dark Energy Survey Supernova Program has released a five-year sample of photometrically classified Type~Ia supernovae (DES-SN5YR), together with binned/unbinned Hubble-diagram products and a full statistical$+$systematic covariance matrix~\cite{DESSN5YRCosmo,DESSN5YRDR}. In this work, we use the updated, recalibrated DES-SN5YR Hubble-diagram release (DES-Dovekie)~\cite{DESDovekie}.

Since DES-SN5YR is provided on a relative distance scale, we model the theoretical distance-modulus vector as
\ben
\boldsymbol{\mu}_{\mathrm{th}}(z)=\bm{\Big[}5\log_{10}D_L(z)\bm{\Big]}\mathbf{1}+\mu_0\,\mathbf{1},
\label{38}
\een

\subsubsection{Analytic marginalization}
The DES-SN5YR contribution to the joint likelihood is
\ben
\chi^2_{\mathrm{DES}}(\mu_0)=\Delta\boldsymbol{\mu}^{\,T}\,\mathbf{C}_{\mathrm{DES}}^{-1}\,\Delta\boldsymbol{\mu},
\label{39}
\een
where $\Delta\boldsymbol{\mu}=\boldsymbol{\mu}_{\mathrm{th}}-\boldsymbol{\mu}_{\mathrm{obs}}$ and $\mathbf{C}_{\mathrm{DES}}$ is the covariance matrix supplied with the DES-SN5YR release. To reduce the sampling dimensionality, we analytically profile over the nuisance parameter $\mu_0$. Defining
\ben
\mathbf{y}_{\mathrm{DES}}\equiv \bm{\Big[}5\log_{10}D_L(z)\bm{\Big]}\mathbf{1}+ \boldsymbol{\mu}_0-\boldsymbol{\mu}_{\mathrm{obs}},
\label{40}
\een
the statistic becomes a quadratic function of $\mu_0$,
\ben
\chi^2_{\mathrm{DES}}(\mu_0)=A_{\mathrm{DES}}+2\mu_0 B_{\mathrm{DES}}+\mu_0^2 C_{\mathrm{DES}},
\label{41}
\een
with
\ben
&& A_{\mathrm{DES}}\equiv\mathbf{y}_{\mathrm{DES}}^T\mathbf{C}_{\mathrm{DES}}^{-1}\mathbf{y}_{\mathrm{DES}},\quad
B_{\mathrm{DES}}\equiv\mathbf{y}_{\mathrm{DES}}^T\mathbf{C}_{\mathrm{DES}}^{-1}\mathbf{1}, \n  &&
C_{\mathrm{DES}}\equiv\mathbf{1}^T\mathbf{C}_{\mathrm{DES}}^{-1}\mathbf{1}.
\label{42}
\een
Minimizing with respect to $\mu_0$ gives
\ben
\hat\mu_0=-\frac{B_{\mathrm{DES}}}{C_{\mathrm{DES}}},
\label{43}
\een
and the statistic entering the joint likelihood after marginalization is
\ben
\chi^2_{\mathrm{DES}}=A_{\mathrm{DES}}-\frac{B_{\mathrm{DES}}^2}{C_{\mathrm{DES}}}.
\label{44}
\een
Here, we use $\mathrm{DES}$ in the subscript to indicate DES-SN5YR.

\subsection{Pantheon$+$ Data}
Type~Ia supernovae (SNe~Ia) act as standardized candles and constrain the luminosity distance through the distance modulus $\mu(z)$ defined in Eq.~\eqref{28}. In this work, we adopt the Pantheon$+$ supernova compilation~\cite{Brout}; consequently, the SN likelihood constrains only relative distances and provides no direct information on $H_0$.

Given the observed distance modulus vector $\boldsymbol{\mu}_{\mathrm{obs}}$ and its full covariance matrix $\mathbf{C}$, which includes both statistical ($\mathbf{D}_{\mathrm{stat}}$) and systematic ($\mathbf{C}_{\mathrm{sys}}$) contributions, the SN contribution to the joint likelihood is modeled as a multivariate Gaussian,
\ben
\chi^2_{\mathrm{SN}}(\mu_0)=\Delta\boldsymbol{\mu}^{\,T}\,\mathbf{C}^{-1}\,\Delta\boldsymbol{\mu},
\label{37}
\een
where $\Delta\boldsymbol{\mu}=\boldsymbol{\mu}_{\mathrm{th}}-\boldsymbol{\mu}_{\mathrm{obs}}$ and $\mathbf{C}=\mathbf{D}_{\mathrm{stat}}+\mathbf{C}_{\mathrm{sys}}$. The theoretical distance moduli are given by Eq.~\eqref{28}.

Since we do not impose the SH0ES absolute calibration, the global zero-point of the SN Hubble diagram is not fixed: adding a constant offset $\mu_0$ to the theoretical moduli $\boldsymbol{\mu}_{\mathrm{th}}$ is exactly degenerate with shifting the SN absolute magnitude $M_B$. We therefore introduce $\mu_0$ as a nuisance parameter and analytically marginalize over it with a flat prior (as for DES-SN5YR), obtaining the usual compressed form
\ben
\chi^2_{\mathrm{Pantheon+}}=A_{\rm Pantheon+}-\frac{B_{\rm Pantheon+}^2}{C_{\rm Pantheon+}},
\label{37a}
\een
where $A_{\rm Pantheon+}=\Delta\boldsymbol{\mu}^{T}\mathbf{C}^{-1}\Delta\boldsymbol{\mu}$, $B_{\rm Pantheon+}=\mathbf{1}^{T}\mathbf{C}^{-1}\Delta\boldsymbol{\mu}$, and $C_{\rm Pantheon+}=\mathbf{1}^{T}\mathbf{C}^{-1}\mathbf{1}$ (cf. Eq.~\eqref{44}). Thus, Pantheon$+$ constrains only relative distances; information on $H_0$ comes from the CMB distance priors together with BAO and CC.

\subsection{Union3 Data}
The Union3 compilation provides a homogenized set of 2087 cosmologically useful SNe~Ia standardized on a common distance scale using SALT3 within the UNITY1.5 framework~\cite{Union3}. For cosmological inference, the public release provides a compressed Hubble diagram consisting of 22 binned distance-modulus measurements and an associated covariance matrix.

Union3 is treated in the same way as DES-SN5YR. It is a relative distance indicator, so the theoretical prediction and likelihood are given by Eqs.~\eqref{38} and \eqref{39}, respectively, with $\mathbf{C}_{\mathrm{DES}}\rightarrow\mathbf{C}_{\mathrm{U3}}$. The analytic profiling over $\mu_0$ follows Eqs.~\eqref{40}--\eqref{44}, with the replacements ${\mathrm{DES}}\rightarrow{\mathrm{U3}}$ (and $\mathbf{C}_{\mathrm{DES}}\rightarrow\mathbf{C}_{\mathrm{U3}}$).

The resulting statistic used in the joint likelihood evaluation after marginalization is
\ben
\chi^2_{\mathrm{Union3}}=A_{\mathrm{U3}}-\frac{B_{\mathrm{U3}}^2}{C_{\mathrm{U3}}}.
\label{45}
\een

\subsection{Compressed CMB data}
We incorporate early-universe information through a set of CMB ``distance priors,'' which provide an accurate compression of the full CMB likelihood for late-time background analyses. The basic idea is that, for a broad class of late-time extensions, which includes dynamical dark energy models in which the pre-recombination physics is unchanged, the dominant CMB information relevant to background expansion is carried by a small set of geometrical combinations that control the mapping between physical length scales at $z\simeq z_*$ and observed angular scales. In particular, the locations of the acoustic peaks are primarily set by the ratio of the angular-diameter distance to last scattering, $D_A(z_*)$, to the sound-horizon scale, $r_s(z_*)$, while the overall matter content enters through the ``shift'' of the peak pattern encapsulated by $R$. The baryon density $\omega_{\mathrm{b}}$ is included because it fixes the pre-recombination sound speed and hence the calibration of $r_s$.

The distance before construction, therefore, compresses the high-dimensional CMB power spectrum likelihood into the few parameters most directly tied to late-time geometry, while marginalizing over the detailed perturbation shape information, e.g., peak heights and damping tail, that is either less relevant for our purpose or would require a full Boltzmann treatment~\cite{Chen,Planck2018,Zhai}. This compression is well suited to multi-probe analyses in which CMB data is primarily used as an early time anchor, as it reproduces the late time distance constraints of the full Planck likelihood with good accuracy at a fraction of the computational cost when the early universe sector is standard. Since in our framework, the tachyonic scalar field is assumed to be homogeneous and subdominant at the early time, it does not contribute significantly to the evolution of metric perturbations prior to recombination \cite{Roy, Weinberg, Dodelson}. Therefore, the acoustic peak structure and CMB anisotropy power spectra remain unchanged to leading order.

In this work, we therefore adopt the Planck~2018 distance priors and their covariance as an efficient substitute for the full CMB likelihood.

Following Planck~2018, we use the data vector~\cite{Chen, Zhai}
\ben
\mathbf{v}_{\mathrm{CMB}}\equiv (R,\,\ell_A,\,\omega_{\mathrm{b}}),\qquad \omega_{\mathrm{b}}\equiv\Omega_b h^2,
\label{46}
\een
where the shift parameter $R$ and the acoustic angular scale $\ell_A$ are defined at photon decoupling ($z_*$) as
\ben
&& R\equiv \sqrt{\Omega_{\mathrm{m0}}}\,H_0\,(1+z_*)\,D_A(z_*),\n  &&
\ell_A\equiv \pi\,\frac{(1+z_*)\,D_A(z_*)}{r_s(z_*)}.
\label{47}
\een
Here $D_A$ is the angular-diameter distance and $r_s(z_*)$ is the comoving sound horizon at decoupling, defined as~\cite{EisensteinHu1998}
\ben
&& r_s(z)\equiv \int_z^{\infty}\frac{c_s(z')}{H(z')}\,\mathrm dz',\, c_s(z)=\frac{c}{\sqrt{3\bm{\big[}1+R_b(z)\bm{\big]}}},\n  && R_b(z)=\frac{3\rho_b(z)}{4\rho_\gamma(z)}.
\label{48}
\een
We evaluate the redshift of photon decoupling $z_*$ using the standard fitting formula~\cite{HuSugiyama1996, Planck2018}
\ben
z_* = 1048\big(1+0.00124\,\omega_{\mathrm{b}}^{-0.738})\bm{\big[}1+g_1\,\omega_m^{\,g_2}\bm{\big]},
\label{49}
\een
where $h\equiv H_0/(100\,\mathrm{km\,s^{-1}\,Mpc^{-1}})$, $\omega_{\mathrm{b}}\equiv\Omega_b h^2$, $\omega_m\equiv\Omega_{\mathrm{m0}}h^2$, and
\ben
g_1=\frac{0.0738\,\omega_{\mathrm{b}}^{-0.238}}{1+39.5\,\omega_{\mathrm{b}}^{0.763}},\qquad g_2=\frac{0.560}{1+21.1\,\omega_{\mathrm{b}}^{1.81}}.
\label{50}
\een
In our likelihood, we adopt the Planck~2018 TT, TE, EE+lowE distance prior means~\cite{Chen}
\ben
&&\mathbf{v}_{\mathrm{CMB}}^{\mathrm{obs}}=(R,\,\ell_A,\,\omega_{\mathrm{b}})\n &&=(1.750235,\,301.4707,\,0.02235976),~
\label{51}
\een
with the corresponding inverse covariance matrix~\cite{Chen}
\ben
&&\mathbf{C}_{\mathrm{CMB}}^{-1}\n &&=\begin{pmatrix}
94392.3971 & -1360.4913 & 1664517.2916\\
-1360.4913 & 161.4349 & 3671.6180\\
1664517.2916 & 3671.6180 & 79719182.5162
\end{pmatrix},\n
\label{52}
\een
The contribution to the total likelihood is then
\ben
\chi^2_{\mathrm{CMB}}=\Delta\mathbf{v}_{\mathrm{CMB}}^{\,T}\,\mathbf{C}_{\mathrm{CMB}}^{-1}\,\Delta\mathbf{v}_{\mathrm{CMB}},
\label{53}
\een
where $\Delta\mathbf{v}_{\mathrm{CMB}}=\mathbf{v}_{\mathrm{CMB}}^{\mathrm{th}}-\mathbf{v}_{\mathrm{CMB}}^{\mathrm{obs}}$.

\subsection{DESI DR2 BAO Data}
BAO provides a robust standard ruler set by the sound horizon at the drag epoch, $r_d$, and therefore yields geometric constraints on the expansion history through distance measurements at multiple redshifts. In this analysis, we use the BAO measurements from the DESI Data Release~2 (DR2), including the galaxy and quasar samples as well as the high-redshift Ly$\alpha$ forest results~\cite{DESI,DESI2}.

For our cosmological model, the key theoretical quantities are the transverse comoving distance $D_M(z)$ and the Hubble distance $D_H(z)=c/H(z)$, along with the volume-averaged distance $D_V(z)$ defined in Eq.~\eqref{30}. These are typically expressed in the dimensionless forms $D_M(z)/r_d$, $D_H(z)/r_d$, and $D_V(z)/r_d$. The DESI DR2 BAO measurements employed in this work (data points and quoted $1\sigma$ uncertainties) are summarized in Table~\ref{TableIV} in Appendix~\ref{A2}, from which we construct the observed BAO data vector $\mathbf{d}_{\mathrm{obs}}$.

We formulate the BAO likelihood assuming Gaussian, uncorrelated errors for the entries listed in \emph{Appendix \ref{A2},~Table~\ref{TableIV}},
\ben
\chi^2_{\mathrm{BAO}}=\sum_i\frac{\big[d^{\mathrm{th}}_i-d^{\mathrm{obs}}_i\big]^2}{\sigma_i^2},
\label{54}
\een
where $d^{\mathrm{obs}}_i$ and $\sigma_i$ are the measured BAO quantities and their quoted $1\sigma$ uncertainties, and $d^{\mathrm{th}}_i$ denotes the corresponding theoretical predictions (either $D_V/r_d$ or the anisotropic observables $D_M/r_d$ and $D_H/r_d$ at the effective redshifts).

\subsection{f$\sigma_8$ Data}

Redshift space distortions (RSD) provide measurements of the combination $f\sigma_8(z)$, where $f\equiv (d\ln\delta_{\mathrm{m}}/d\ln a$) is the linear growth rate and $\sigma_8(z)$ is the rms matter fluctuation in spheres of radius $8\,h^{-1}\,\mathrm{Mpc}$. This observable is particularly sensitive to the clustering properties of dark energy through its impact on the growth of matter perturbations. High-redshift RSD measurements, such as FastSound at $z\sim1.4$~\cite{Okumura2016}, extend these growth tests into the regime where dark energy begins to dominate.

In this work, we use a compilation of $63$ $f\sigma_8$ data points spanning low to intermediate redshifts~\cite{Nesseris, Kazantzidis}. For a given cosmological model we compute the theoretical prediction $f\sigma_8^{\mathrm{th}}(z_i)$ and construct the likelihood assuming Gaussian errors,
\ben
\chi^2_{f\sigma_8}=\sum_{i=1}^{63}\frac{\bm{\Big[}f\sigma_8^{\mathrm{th}}(z_i)-f\sigma_8^{\mathrm{obs}}(z_i)\bm{\Big]}^2}{\sigma_{i}^2},
\label{55}
\een
where $f\sigma_8^{\mathrm{obs}}(z_i)$ and $\sigma_i$ are the observed values and their $1\sigma$ uncertainties.

\section{Methodology of Data Analysis}
\label{S4}

The goal of this work is to place robust constraints on extended dark energy scenarios and to quantify current cosmological tensions within a unified joint likelihood framework. To assess the impact of Type~Ia supernova (SN~Ia) calibration systematics, we perform three parallel inferences in which a fixed ``Base'' compilation of non-SN probes is combined with one of three modern SN~Ia samples.

All non-SN datasets and likelihood terms (CMB distance priors, BAO, CC, and RSD) are implemented as described in the preceding subsections with the addition of three SN~Ia alternatives treated in separate runs:
(i) Pantheon$+$ (excluding the SH0ES calibration)~\cite{Brout};
(ii) DES-SN5YR, implemented with the full inverse covariance matrix and marginalising over a global distance modulus (absolute magnitude) offset $\mu_0$ to absorb global light curve calibration/SALT3 pipeline offsets \cite{DESSN5YRCosmo};
(iii) Union3, treated as a relative distance indicator and marginalising over a fully free offset parameter $\mu_0$. 

For a robust statistical analysis, it is essential to treat each dataset with care, especially when the associated observables are extracted under explicit assumptions about the underlying cosmological background, as is the case for the $f\sigma_8$ measurements employed here, which are derived assuming a fiducial $\Lambda$CDM cosmology. In this context, particular attention must be paid to the Alcock–Paczynski (AP) correction~\cite{AlcockPaczynski}, which has to be incorporated into the theoretical prediction for $f\sigma_8$ before any meaningful comparison with the data can be made.

RSD constraints are derived in practice by translating galaxies’ observed redshifts and angular coordinates into comoving distances while adopting a particular fiducial cosmological model. If the actual cosmology, or the cosmological model under consideration, differs from this fiducial choice, the resulting transformation from observables to comoving separations becomes distorted. This geometric mismatch induces a spurious anisotropy in the clustering pattern, referred to as the AP effect, which is purely geometric in origin and distinct from the physical RSD signal itself.\\

To guarantee that the theoretical model is evaluated in the same effective coordinate system as that used in the survey’s data reduction pipeline, we implement a rescaling of the theoretical $f\sigma_8$ prediction at every point in parameter space probed by the likelihood. Concretely, we compute \cite{Kazantzidis} (\textit{Eq.(1.7)}):
\ben
\bm{\Big[}f\sigma_8(z)\bm{\Big]}_{\mathrm{pred}}=\bm{\Big[}f\sigma_8(z)\bm{\Big]}_{\mathrm{th}}\,\bm{\Big(}\frac{H_{\mathrm{th}}(z)\,D_{A,\mathrm{th}}(z)}{H_{\mathrm{fid}}(z)\,D_{A,\mathrm{fid}}(z)}\bm{\Big)},
\label{57}
\een
where $H_{\mathrm{fid}}(z)$ and $D_{A,\mathrm{fid}}(z)$ denote, respectively, the Hubble parameter and the angular diameter distance evaluated in the fiducial $\Lambda$CDM cosmology specified by the survey, while $H_{\mathrm{th}}(z)$ and $D_{A,\mathrm{th}}(z)$ are computed in the trial cosmology under consideration.

This rescaling effectively translates the theoretical growth rate prediction into the same “fiducial-geometry frame” used in the original RSD analysis. In doing so, it systematically incorporates the dependence on both $H(z)$ and $D_A(z)$ into the RSD likelihood evaluation and ensures that any geometric distortions are consistently accounted for. As a result, we avoid spurious shifts in the inferred cosmological parameters that would otherwise arise from a naive comparison of $f\sigma_8$ values obtained under different background assumptions, thereby preventing biased constraints due to mismatched cosmological geometries.

Take note of that, a complete treatment of the AP effect requires recomputing the full anisotropic power spectrum multipoles under independent transverse and line-of-sight dilations $\alpha_\perp$ and $\alpha_\parallel$~\cite{Ballinger}, as implemented in Boltzmann-based pipelines such as \textsc{Cobaya}~\cite{Torrado}. Since our inference pipeline evolves the autonomous ODE system of Eq.~\eqref{31} directly without computing the full matter power spectrum $P(k,z)$, we adopt the standard first-order rescaling of Eq.~\eqref{57}, following the practice established in~\cite{Macaulay} and used in compiled $f\sigma_8$ analyses~\cite{Nesseris,Kazantzidis}. A posterior validation of this approximation using the inferred posterior mean parameters is presented in \emph{Appendix~\ref{A3}}.

\subsection{eBOSS LRG Spatial Clustering Comparison}

\begin{figure}[htbp]
    \centering
    \includegraphics[width=\columnwidth]{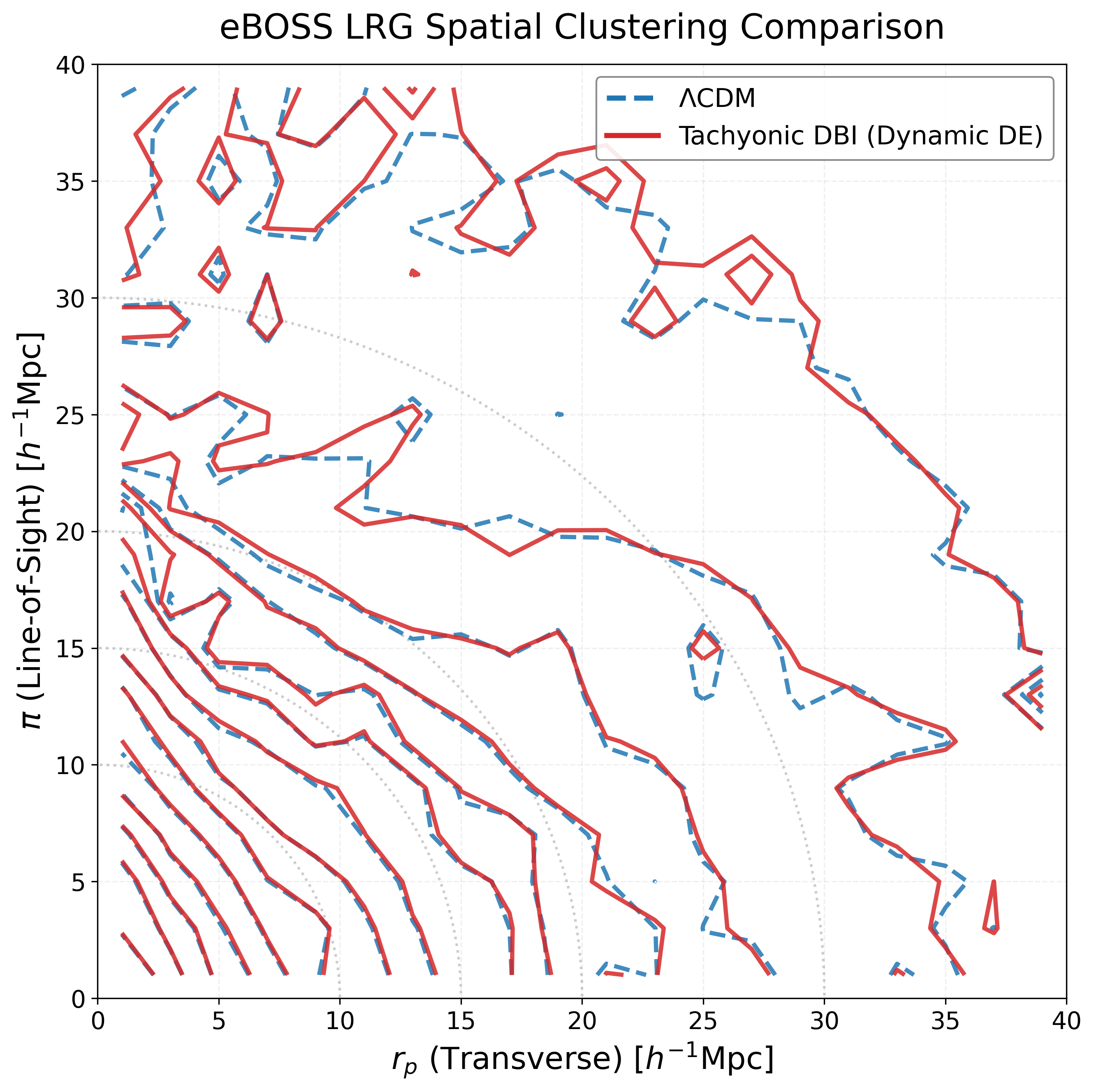}
    \caption{Direct contour comparison of the 2D correlation function $\xi(r_p, \pi)$ based on eBOSS LRG clustering data. The red solid contours represent the tachyonic DBI model, while the blue dashed contours represent the baseline $\Lambda$CDM model. Notably, this visualization maps the intrinsic RSD effects prior to the application of Alcock-Paczynski (AP) geometric corrections.}
    \label{Fig1A}
\end{figure}

Fig.~\ref{Fig1A} overlays the constant $\xi$ contours for the tachyonic DBI model (red, solid) and $\Lambda$CDM (blue, dashed). Because this specific visualization isolates the clustering data without applying the AP geometric scaling, it allows us to directly examine the dynamical clustering differences between the two models. We observe the following physical distinctions. First, at very small separations ($r_p, \pi < 5 \, h^{-1}\mathrm{Mpc}$), the contours remain highly degenerate. The nonlinear physics dominating this regime, specifically the virialized Fingers-of-God effect, effectively masks the subtle linear perturbation differences introduced by the tachyonic dark energy. Secondly, as we examine intermediate scales ($10 \, h^{-1}\mathrm{Mpc} < r_p, \pi < 30 \, h^{-1}\mathrm{Mpc}$), some differences in contour are visible. The red solid contours of the tachyonic model exhibit slightly different curvature and spatial reach compared to the $\Lambda$CDM model.

These contour shifts are not statistically significant regarding the growth of structure. It shows that within the current observational precession the tachyonic field's noncanonical sound speed ($c_s^2 < 1$) is not making any significant contribution to subtly modifies the Hubble friction, and therefore the evolution of the density contrast $\delta_{\mathrm{m}}$ and the linear growth rate $f$ remain mostly unaltered.\\

Although Fig.~\ref{Fig1A} isolates the pure RSD clustering signatures, conducting robust Bayesian parameter inference requires bridging these theoretical predictions with actual observational surveys. In practice, translating observed angular coordinates and redshifts into comoving separations relies on assuming a fiducial cosmology. If the true cosmology or our tachyonic model possesses a different Hubble parameter $H(z)$ or angular diameter distance $D_A(z)$ than the survey's fiducial $\Lambda$CDM model, it induces a purely geometric distortion known as the Alcock Paczynski (AP) effect, which we have already discussed above.

Therefore, although the contours visualized above demonstrate insignificant \textit{intrinsic} structural differences, our full statistical pipeline explicitly applies the AP rescaling detailed in Eq.~\eqref{57}. This crucial step ensures that the geometric volume distortions driven by the tachyonic model's distinct background expansion are rigorously disentangled from the physical RSD signal during likelihood evaluation. The separate 2D correlation heatmap plots for the eBOSS LRG sample, corresponding to both the tachyonic DBI model and the baseline $\Lambda$CDM model, are presented in Appendix~\ref{A3}.

\subsection{Bayesian Cosmological Inference}

We construct a fully Bayesian inference framework to constrain the parameters of a dynamical dark energy cosmological model using multiple late-time and early-time cosmological observational datasets. The framework combines numerical integration of cosmological evolution equations with nested sampling to evaluate posterior distributions, Bayesian evidence, and model selection statistics.

The analysis incorporates measurements of Type Ia supernova luminosity distances from different SN compilations along with cosmic chronometer measurements of the Hubble expansion rate $H(z)$, growth-rate measurements $f\sigma_8(z)$, BAO distance measurements, and compressed CMB likelihood constraints.

While dealing with the Pantheon$+$ supernova sample, we apply a redshift cut $z \ge 0.01$ in order to suppress low-redshift peculiar velocity systematics. The full covariance matrix provided with the each SN dataset is used in the likelihood evaluation.

The combined dataset provides a comprehensive probe of both the background expansion history and the growth of cosmological structure.

\subsection{Cosmological Model}

We consider the dynamical dark-energy model of Eq.~\eqref{1}, where the late-time acceleration is sourced by a scalar field. The parameter space explored in the inference is
\ben
\Theta =
\{\Omega_{\mathrm{m0}},\, H_0,\, \omega_{\mathrm{b}},\, \sigma_{8,0},\, \lambda_{\mathrm{ini}},\, u\equiv1/\Gamma\}.
\label{58}
\een
Here, $\Omega_{\mathrm{m0}}$ denotes the present day matter density parameter, $H_0$ the Hubble constant, $\omega_{\mathrm{b}}$ the physical baryon density, and $\sigma_{8,0}$ the present amplitude of matter fluctuations on $8\,h^{-1}\mathrm{Mpc}$ scales. The parameters $\lambda_{\mathrm{ini}}$ and $\Gamma$ characterise the scalar field dynamics, corresponding to the initial slope and the tracking parameter, respectively. Here $\Gamma$ is sampled through $u\equiv1/\Gamma$, which remains finite in the cosmological-constant limit $\Gamma\to\infty$. The tachyonic model therefore has six free parameters, two more than $\Lambda$CDM. The initial dark energy density parameter $\Omega_{\mathrm{de,ini}}$ is not a free parameter: for each point in $\Theta$ it is fixed by a shooting method that enforces spatial flatness today, $\Omega_{\mathrm{de}}(z=0)=1-\Omega_{\mathrm{m0}}-\Omega_{\mathrm{r0}}$. We solve this condition for $\log_{10}\Omega_{\mathrm{de,ini}}$ with a secant iteration (five iterations, tolerance $10^{-4}$ in the logit of $\Omega_{\mathrm{de}}(0)$), each step integrating the full background system, and report $\log(\Omega_{\mathrm{de,ini}})$ as a derived parameter.

The background and linear perturbation equations are written as an autonomous system (Eq.~\eqref{31}) and evolved using the e-folding time variable $N \equiv \ln a$. We solve the coupled system for the scalar-field variables together with matter and dark-energy perturbations, yielding consistent predictions for both the expansion history and the linear growth of structure.

The numerical evolution is initialized at $z_{\mathrm{ini}}\simeq 1100$ (matter domination), with initial conditions chosen such that the dark-energy density fraction is negligible at early times. We solve the ODE system with the \texttt{diffrax} integrator~\cite{Kidger} using the adaptive fifth-order Runge-Kutta scheme Tsit5 and a PID-controlled step-size controller, which ensures stable integration across the matter, and dark-energy dominated epochs.

\subsection{Cosmological Observables}

From the numerical solutions, we compute the observables entering the joint likelihood. The luminosity distance follows from the comoving distance as $D_L(z)=(1+z)D_C(z)$, giving the Type~Ia supernova distance modulus $\mu(z)$. The growth observable is obtained from the evolution of the matter density contrast $\delta_{\mathrm{m}}$~\cite{Linder,SongPercival},
\ben
f\sigma_8(z)=\frac{d\ln\delta_{\mathrm{m}}}{d\ln a}\,\sigma_{8,0}\,\frac{\delta_{\mathrm{m}}(z)}{\delta_{\mathrm{m}}(0)},
\label{59}
\een
For BAO, we compute the transverse comoving distance $D_M(z)=D_C(z)$, the Hubble distance $D_H(z)=c/H(z)$, and the volume-averaged distance defined in Eq.~\ref{30}.

Finally, the compressed CMB distance-prior likelihood is evaluated using the shift parameter $R$ and acoustic scale $\ell$ defined in Eq.~\eqref{51}.

\subsection{Likelihood Construction}

Assuming Gaussian observational uncertainties, the likelihood function is constructed from the total chi-square
\ben
\chi^2_{\mathrm{tot}} =
\chi^2_{\mathrm{H(z)}}
+
\chi^2_{\mathrm{SN}}
+
\chi^2_{f\sigma_8}
+
\chi^2_{\mathrm{CMB}}
+
\chi^2_{\mathrm{BAO}}.
\label{61}
\een
The log-likelihood used for Bayesian inference is therefore
\ben
\ln \mathcal{L} = -\frac{1}{2}\chi^2_{\mathrm{tot}} ,
\label{62}
\een
Solutions that produce numerically unstable cosmologies or unphysical energy densities are automatically rejected by assigning extremely small likelihood values.

\subsection{Bayesian Sampling}

Posterior sampling is performed using the nested sampling algorithm implemented in the \texttt{dynesty}~\cite{Speagle} package. Nested sampling simultaneously explores the posterior parameter space while providing an estimate of the Bayesian evidence $\ln \mathcal{Z}$.

Equal-weight posterior samples are obtained through importance resampling of the nested sampling chains,
\ben
w_i =
\frac{\exp(\log w_i - \log \mathcal{Z})}
{\sum_j \exp(\log w_j - \log \mathcal{Z})},
\label{63}
\een
The resulting posterior samples are used to derive marginalized parameter constraints and to generate triangle plots using the \texttt{GetDist}~\cite{Lewis} analysis toolkit.

Derived parameters, such as the commonly used clustering amplitude
\ben
S_8 = \sigma_8
\sqrt{\frac{\Omega_{\mathrm{m0}}}{0.3}},
\label{64}
\een
are computed directly from the posterior chains.

The parameter $S_8$ plays a central role in the so-called ``$S_8$-tension''. Cosmic shear i.e weak-lensing surveys typically prefer a lower clustering amplitude than that inferred from Planck 2018 CMB data within $\Lambda$CDM, at the level of a few standard deviations~\cite{Planck2018, Hildebrandt, DESY3}, as also found in the Subaru Hyper Suprime-Cam (HSC) cosmic-shear and joint clustering--lensing analyses~\cite{Hikage2019, Dalal2023, Li2023, Miyatake2023, Sugiyama2023} and in the low galaxy--galaxy lensing amplitude relative to clustering predictions~\cite{More2015, Leauthaud2017}.

\subsection{Model Selection Statistics}

To quantify the goodness of fit and compare models we compute several statistical diagnostics.

The reduced chi-square statistic is
\ben
\chi^2_{\mathrm{red}}= \frac{\chi^2_{\mathrm{min}}}{N-k},
\label{65}
\een
where $N$ denotes the total number of data points and $k$ the number of free parameters.

The Akaike information criterion (AIC) and Bayesian information criterion (BIC) are defined as \cite{Akaike,Schwarz}
\ben
&&\mathrm{AIC} =
\chi^2_{\mathrm{min}} + 2k,\n && \mathrm{BIC} =
\chi^2_{\mathrm{min}} + k \ln N.
\label{66}
\een

We also compute the deviance information criterion (DIC) \cite{Spiegel}
\ben
\mathrm{DIC}= \bar{D} + p_D ,
\label{67}
\een
where $\bar{D}$ is the posterior mean deviance, i.e. $\bar{D}=\langle D(\theta)\rangle$ with $D(\theta)\equiv-2\ln\mathcal{L}(\theta)$, and $p_D$ denotes the effective number of parameters inferred from the posterior distribution.

Finally, the information gain between prior and posterior distributions is quantified through the Kullback--Leibler divergence \cite{Kullback,MacKay}.
\ben
D_{\mathrm{KL}} = \langle \ln \mathcal{L} \rangle - \ln \mathcal{Z} ,
\label{68}
\een
where $\langle \ln \mathcal{L} \rangle$ denotes the posterior expectation value of the log-likelihood and $\mathcal{Z}$ is the Bayesian evidence, consequently $ln \mathcal{Z}$ is log evidence.

\section{Results and Discussion}
\label{S5}
 
We now present the main observational constraints on the tachyonic DBI model, beginning with the posterior distributions of the primary cosmological parameters and their implications for the $S_{8}$ tension, followed by a comparative joint analysis across the three supernova compilations, a physical interpretation of the model parameters, a statistical comparison with $\L$CDM, and finally an investigation of the dark-energy perturbation dynamics.
 
\subsection{Posterior Distributions and Observational 
Consistency of the Tachyonic DBI Model}
\label{SECVA}
 
\begin{figure*}[t]
    \centering
    \includegraphics[width=0.5\textwidth]{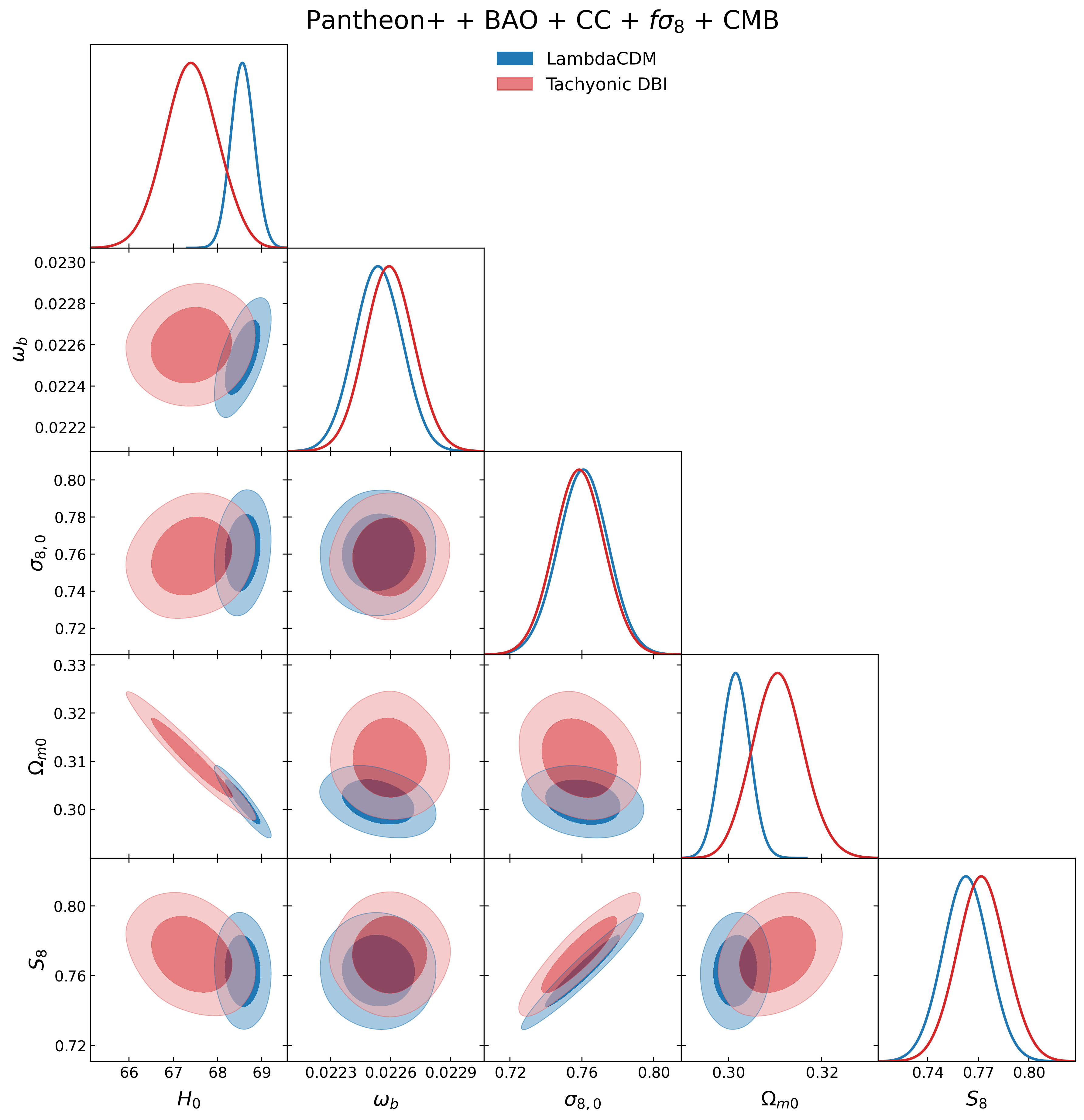}
    \caption{1D marginalized and 2D joint posterios distributions for the primary cosmological parameters comparing the tachyonic DBI DE model (red) and $\Lambda$CDM (blue), utilizing the Pantheon$+$ compilation.}
    \label{Fig1}
\end{figure*}

\begin{figure*}[t]
    \centering
    \includegraphics[width=0.5\textwidth]{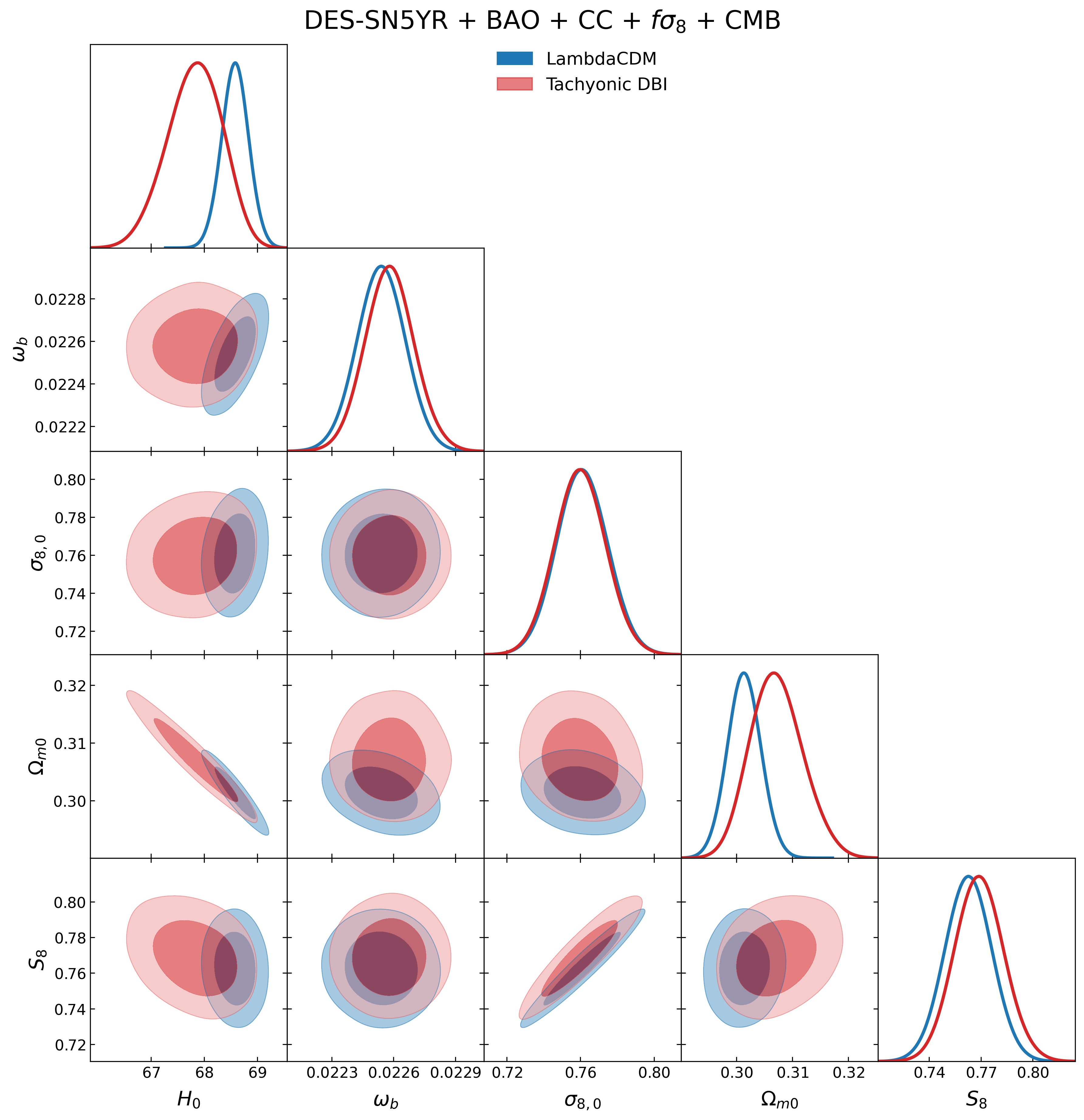}
    \caption{1D marginalized and 2D joint posterior distributions for the primary cosmological parameters comparing the tachyonic DBI DE model (red) and $\Lambda$CDM (blue), utilizing the DES-SN5YR compilation.}
    \label{Fig2}
\end{figure*}

\begin{figure*}[t]
    \centering
    \includegraphics[width=0.5\textwidth]{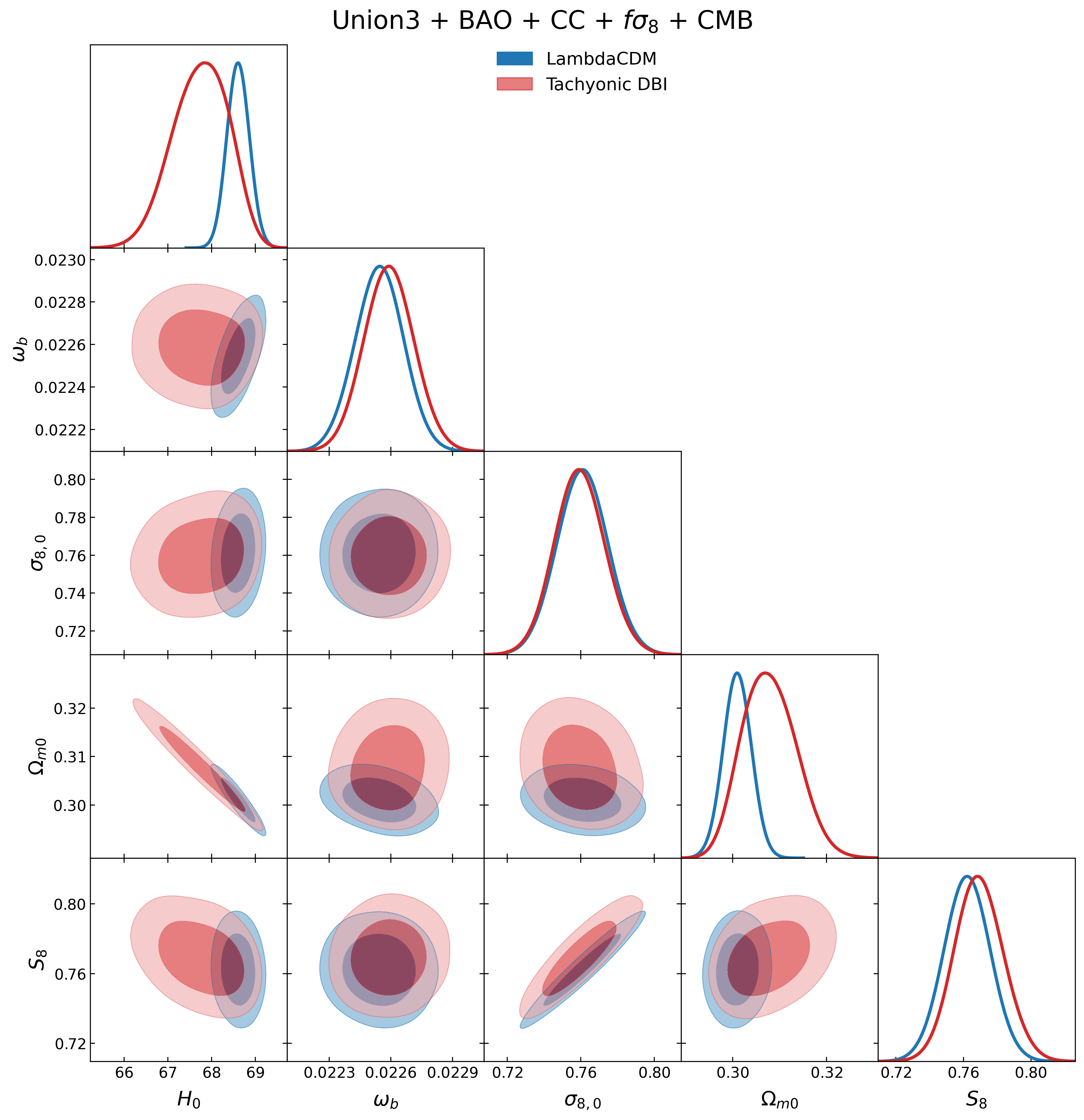}
    \caption{1D marginalized and 2D joint posterior distributions for the primary cosmological parameters comparing the tachyonic DBI DE model (red) and $\Lambda$CDM (blue), utilizing the Union3 compilation.}
    \label{Fig3}
\end{figure*}

To evaluate the observational viability of the tachyonic DBI dark energy model defined in Eq.~\eqref{2}, we perform a comprehensive nested-sampling inference using the full multi-probe joint likelihood described in Sec.~\ref{S4}. Figs.~\ref{Fig1}, \ref{Fig2}, and \ref{Fig3} present the 1D marginalized posterior distributions and 2D joint confidence contours at 68\% and 95\% confidence for the primary cosmological parameters ($H_0$, $\omega_{\rm b}$, $\sigma_{8,0}$, $\Omega_{\rm m0}$, and $S_8$~(derived)). Fig.~\ref{Fig1} compares the tachyonic DBI model against the $\Lambda$CDM baseline using the Pantheon$+$ compilation; Fig.~\ref{Fig2} presents a comparison between $\Lambda$CDM, and the tachyonic DBI model using DES-SN5YR; and Fig.~\ref{Fig3} compares the tachyonic model against $\Lambda$CDM using Union3. In each case, the baseline dataset of compressed CMB distance priors, DESI DR2 BAO, CC, and $f\sigma_8$ is held fixed~\cite{Zhang,Simon,Moresco1,Moresco2,Ratsimbazafy,Stern,Borghi,Moresco3,Scolnic,Brout,Riess1,DESSN5YRCosmo,DESSN5YRDR,DESDovekie,Union3,Chen,Zhai,DESI,DESI2,Nesseris,Kazantzidis}.\\
 
The posterior analysis shows that the tachyonic DBI model and $\Lambda$CDM agree in the growth and early-universe parameters but not in the late-time expansion parameters. The 1D posterior of $\sigma_{8,0}$ is identical in the two models for all three SN compilations, while those of $\omega_{\rm b}$ and $S_8$ are shifted slightly upward in the tachyonic case by less than $1\sigma$, with contours that overlap almost completely with their $\Lambda$CDM counterparts (Figs.~\ref{Fig1}, \ref{Fig2} and \ref{Fig3}). The posteriors of $H_0$ and $\Omega_{m0}$, in contrast, are about twice as broad (e.g.\ $\sigma_{H_0}\simeq0.51$--$0.63$ against $\simeq0.25$) for the tachyonic model in every compilation, and for DES-SN5YR and Union3, their centres also move toward lower $H_0$ and higher $\Omega_{m0}$. For Pantheon$+$ the centers show the same shift, from $H_0=68.57$ to $67.42\,{\rm km\,s^{-1}\,Mpc^{-1}}$ and from $\Omega_{m0}=0.302$ to $0.311$. The $H_0$--$\Omega_{m0}$ panels show the origin of this behavior. Both models lie along the same anticorrelated direction, which preserves $\omega_m=\Omega_{m0}h^2$ and the distance to last scattering. In $\Lambda$CDM, the fixed equation of state $w=-1$ confines the posterior to a short segment of this line, whereas the thawing tachyonic field with $w_\phi>-1$ raises $H(z)$ at intermediate redshift, and therefore the data may compensate by moving along the line toward lower $H_0$ and higher $\Omega_{m0}$. Marginalising over $\lambda_{\rm ini}$, which sets how far the field thaws, may be the reason for this spreading of the posterior along this degeneracy direction. The agreement in $\sigma_{8,0}$, $\omega_{\rm b}$ and $S_8$ follows from the negligible early dark-energy fraction $\Omega_{\rm de,ini}\sim10^{-9}$ at $z=1100$ (Table~\ref{TableI}), which leaves the early universe untouched and produces no detectable change in the growth rate at the current precision $\sigma_{f\sigma_8}/f\sigma_8\sim5$--$20\%$~\cite{Nesseris,Kazantzidis}. The influence of the tachyonic field is therefore confined to the late-time background, which is exactly where the $H_0$ and $\Omega_{m0}$ posteriors differ. 
 
The three SN compilations also differ from one another. Pantheon$+$ (Fig.~\ref{Fig1}) gives $H_0\simeq67.42$ and $68.57\,{\rm km\,s^{-1}\,Mpc^{-1}}$ for the tachyonic and $\Lambda$CDM models respectively, with $\omega_{\rm b}\simeq0.0226$ in both. DES-SN5YR (Fig.~\ref{Fig2}) gives $H_0\simeq67.83\,{\rm km\,s^{-1}\,Mpc^{-1}}$ for the tachyonic model and $\simeq68.58\,{\rm km\,s^{-1}\,Mpc^{-1}}$ for $\Lambda$CDM, with $\omega_{\rm b}\simeq0.0226$ in both. Union3 (Fig.~\ref{Fig3}) follows DES-SN5YR closely but yields the broadest tachyonic $H_0$ and $\Omega_{m0}$ posteriors of the three, consistent with the smaller statistical weight of the binned Union3 Hubble diagram. The baryon density separates the compilations most sharply. Against the marginal uncertainty $1.49\times10^{-4}$ of our compressed CMB likelihood, the DES-SN5YR and Union3 values lie $1.5$--$1.6\sigma$ from the CMB mean, and the Pantheon$+$ value lies $1.6\sigma$ from it, so all three compilations agree in $\omega_{\rm b}$. Because no compilation uses an absolute calibration, $H_0$ is not pinned by a distance-ladder anchor in any of the three analyses, and BAO and the CMB, which constrain the product $H_0 r_d$, are no longer forced to accommodate a high $H_0$ through a smaller sound horizon and a shifted $\omega_{\rm b}$. The consistency of each combination is tested with the posterior predictive check of Sec.~\ref{SecVG}.

Table~\ref{TableI} shows that the tachyonic DBI extension has a negligible impact on $S_8$ relative to $\Lambda$CDM ($S_8\simeq0.763$): we obtain $S_8\simeq0.769\pm0.014$ for DES-SN5YR and Union3, and $S_8\simeq0.772\pm0.014$ for Pantheon$+$. This corresponds to an upward shift of only $\sim0.5\sigma$, with essentially unchanged uncertainty. As demonstrated in Sec.~\ref{S5} and Fig.~\ref{Fig5}, this is because the dark-energy perturbation source term $\Omega_{\rm de}|\delta_{\rm de}|$ is suppressed by four to six orders of magnitude relative to the matter source $\Omega_m|\delta_m|$, so any dependence of matter growth on the tachyonic sound speed is undetectable at the current $f\sigma_8$ precision. The model therefore does not alleviate the $S_8$ tension. Nevertheless, the inferred values are consistent with weak-lensing constraints from KiDS-1000 ($S_8 = 0.766^{+0.020}_{-0.014}$~\cite{Hildebrandt}) and DES-Y3 ($S_8 = 0.759^{+0.024}_{-0.023}$~\cite{DESY3}), both within $1\sigma$.\\
 
Taken together, these posteriors indicate that the tachyonic DBI model fits the cosmological observables at a level comparable to $\Lambda$CDM across three independent SN compilations and five complementary probes. However, the posteriors alone do not prove such a degeneracy: they show where the parameters are allowed to lie, but not whether the corresponding best-fitting models actually reproduce the data. We therefore test this directly using the posterior predictive check in Sec.~\ref{SecVG}. For completeness, we list the individual parameter constraints in Table~\ref{TableI}, and provide the full corner plots for each compilation in \emph{Appendices~\ref{A4}--\ref{A6}.}
 
\subsection{Joint Cosmological Constraints}\label{SecVB}
\begin{figure*}[t]
    \centering
    \includegraphics[width=0.75\textwidth]{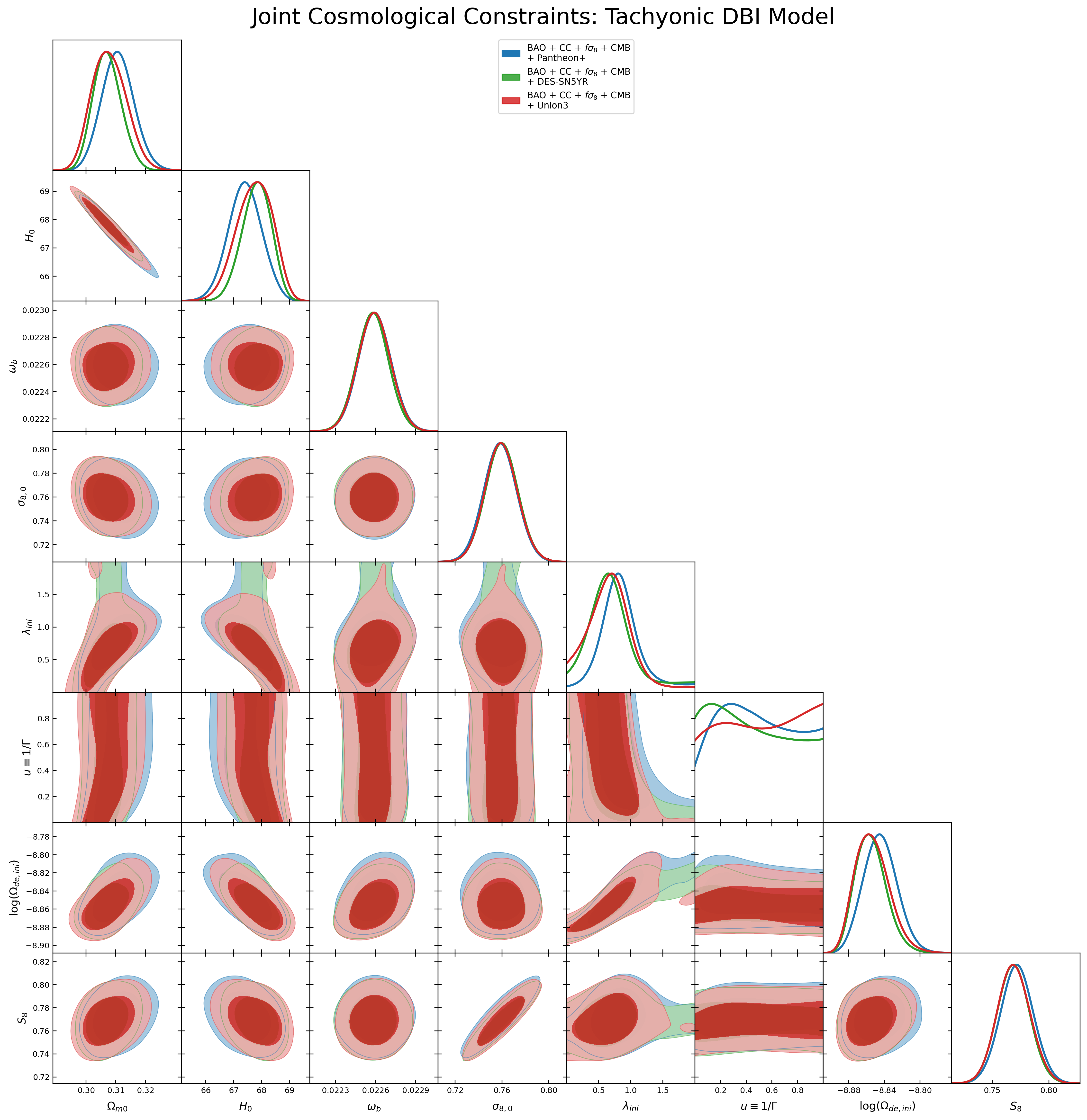}
    \caption{Joint marginalized constraints for the compilation of CMB distance priors $+$ BAO $+$ $f\sigma_8$ $+$ CC combined separately with three SN~Ia compilations: Pantheon$+$ (blue), DES-SN5YR (green), and Union3 (red). The diagonal panels show the 1D marginalized posteriors, while the off-diagonal panels show the corresponding 2D confidence regions with 68\% and 95\% confidence intervals.}
    \label{Fig4}
\end{figure*}
 
\begin{table*}[t]
    \centering
    \renewcommand{\arraystretch}{1.3}
    \caption{Summary of the 68\% marginalized constraints for the three SN compilations, each combined with the same base dataset (compressed CMB distance priors, DESI DR2 BAO, CC, and the $f\sigma_8$ compilation).}
    \label{TableI}
    \begin{tabular}{c c c c c}
        \hline\hline
        Parameter & Range & Pantheon$+$ & DES-SN5YR & Union3 \\
        \hline
        $\Omega_{\mathrm{m0}}$ & $\mathcal{U}(0.20,0.40)$ & $0.3107\pm0.0054$ & $0.3072\pm0.0047$ & $0.3078\pm0.0057$ \\
        $H_0$ & $\mathcal{U}(60,80)$ & $67.4150\pm0.5932$ & $67.8256\pm0.5113$ & $67.7571\pm0.6320$ \\
        $\omega_{\mathrm{b}}\equiv\Omega_b h^2$ & $\mathcal{U}(0.010,0.040)$ & $0.0226\pm0.0001$ & $0.0226\pm0.0001$ & $0.0226\pm0.0001$ \\
        $\sigma_{8,0}$ & $\mathcal{U}(0.5,1.0)$ & $0.7587\pm0.0138$ & $0.7600\pm0.0136$ & $0.7596\pm0.0136$ \\
        $S_8$ & derived & $0.7720\pm0.0144$ & $0.7690\pm0.0141$ & $0.7693\pm0.0143$ \\
        \hline
        $\lambda_{\mathrm{ini}}$ & $\mathcal{U}(0.001,2.0)$ & $0.8552\pm0.3279$ & $0.7141\pm0.3698$ & $0.6946\pm0.3456$ \\
        $u\equiv1/\Gamma$ & $\mathcal{U}(0.01,1.0)$ & $0.4936\pm0.2755$ & $0.4644\pm0.2876$ & $0.5184\pm0.2908$ \\
        $\Gamma$ & derived ($1/u$), median & $2.11^{+3.42}_{-0.90}$ & $2.28^{+5.05}_{-1.05}$ & $1.91^{+3.72}_{-0.75}$ \\
        $\log(\Omega_{de,\mathrm{ini}})$ & derived (flatness shooting) & $-8.8446\pm0.0187$ & $-8.8552\pm0.0172$ & $-8.8536\pm0.0189$ \\
        \hline\hline
    \end{tabular}
\end{table*}
 
Fig.~\ref{Fig4} compares the joint posteriors obtained when the same baseline dataset (compressed CMB distance priors, BAO, $f\sigma_8$, and CC) is combined, in addition to Pantheon$+$, DES-SN5YR, and Union3. The corresponding 68\% marginalized constraints are summarized in Table~\ref{TableI}.
 
DES-SN5YR and Union3 yield statistically consistent results across the full parameter set. Their 1D posteriors and 2D contours are nearly coincident in $\Omega_{\mathrm{m0}}$, $H_0$, $\omega_{\mathrm{b}}$, $\sigma_{8,0}$ and $S_8$, and Union3 has a slightly smaller $\lambda_{\rm ini}$ ($0.69\pm0.35$ against $0.71\pm0.37$). In both compilations $\lambda_{\rm ini}$ is anticorrelated with $H_0$ and positively correlated with $\Omega_{m0}$ and $\log\Omega_{\rm de,ini}$. A steeper initial potential produces a stronger late-time thaw, and the data compensate by lowering $H_0$ and raising $\Omega_{m0}$. This correlation is the origin of the broader $H_0$ and $\Omega_{m0}$ posteriors of the tachyonic model discussed in Sec.~\ref{SECVA}. $\Gamma$, by contrast, shows no correlation with any other parameter.

Since Pantheon$+$ is used without the SH0ES calibration, in all three compilations $H_0$ is set by the CMB distance priors, BAO and CC. Pantheon$+$ now agrees with DES-SN5YR and Union3 within $1\sigma$ in every standard parameter, with $H_0=67.42\pm0.59$, $\Omega_{m0}=0.3107\pm0.0054$, $\omega_{\rm b}=0.0226\pm0.0001$ and $S_8=0.772\pm0.014$, compared with $S_8=0.769\pm0.014$ for DES-SN5YR and Union3 (Fig.~\ref{Fig4} and Table~\ref{TableI}). Without a fixed $H_0$ the $\lambda_{\rm ini}$--$H_0$ degeneracy is no longer broken externally, and Pantheon$+$ gives $\lambda_{\rm ini}\simeq0.86\pm0.33$, of similar width to the other two compilations rather than the ten-fold tighter bound found with the SH0ES calibration.
 
Finally, the early-time dark-energy fraction is tightly localized in every case, with $\log(\Omega_{\rm de,ini})=-8.845\pm0.019$ for Pantheon$+$ and $-8.855\pm0.017$ and $-8.854\pm0.019$ for DES-SN5YR and Union3 (Table~\ref{TableI}). All three SN compilations are now treated as relative distance indicators, so the residual difference between them, below $1\sigma$, reflects the SN samples themselves rather than an external calibration. The tachyonic DBI model remains observationally viable with every compilation.
 
\subsection{Physical Implications of the Tachyonic DBI Model Parameters}\label{SecVC}
 
The extended parameter space of the tachyonic DBI model introduces two free model parameters, $\lambda_{\rm ini}$ and $\Gamma$ (sampled as $u=1/\Gamma$), together with the derived quantity $\log(\Omega_{de,\mathrm{ini}})$ fixed by the flatness shooting. We discuss the physical implications of their statistical bounds.
 
The parameter $\lambda\equiv-V'/V^{3/2}$ characterizes the instantaneous steepness of the tachyonic scalar-field potential $V(\phi)$ and is initialized strictly positive, $\lambda_{\mathrm{ini}}>0$, because we only consider exponential and power-law potentials (see below). Its dynamical evolution is governed by the third equation of the autonomous system in Eq.~\eqref{31}, where $x=\dot\phi>0$ is the field velocity, $y=\sqrt{V}/(\sqrt{3}H)>0$ encodes the potential contribution to the Hubble rate, {\color{blue} and $\Gamma$ is sampled through $u\equiv1/\Gamma\in\mathcal{U}(0.01,1.0)$. Since $\lambda^{2}>0$, $x\geq0$ and $y>0$, the sign of $d\lambda/dN$ is set entirely by $(3/2-\Gamma)$, and $\lambda$ evolves monotonically in one of two ways. For $\Gamma>3/2$ ($u<2/3$), $d\lambda/dN\leq0$ and the potential \emph{flattens} as the universe expands; this branch includes inverse power-law potentials $V\propto\phi^{-n}$ with $n<2$ and admits eternal acceleration. For $\Gamma<3/2$ ($u>2/3$), $d\lambda/dN\geq0$ and the potential \emph{steepens}, this branch includes the exponential string-tachyon potential $V\propto e^{-\phi}$ ($\Gamma=1$), for which $\lambda\propto V^{-1/2}$ grows as the field rolls, and acceleration is transient. The case $\Gamma=3/2$ ($V\propto\phi^{-2}$) keeps $\lambda$ constant. In either branch the field is frozen at early times by strong Hubble friction with $w_\phi\simeq-1$, and $w_\phi$ departs from $-1$ only at late times~\cite{Caldwell,Scherrer,Linder1}.}
 
The posterior constraints on $\lambda_{\mathrm{ini}}$ exhibit clear dataset dependency. {\color{blue}Pantheon$+$ yields $\lambda_{\mathrm{ini}}=0.86\pm0.33$, while DES-SN5YR and Union3 give} initial slopes {\color{blue}$\lambda_{\mathrm{ini}}=0.71\pm0.37$ and $0.69\pm0.35$} respectively, accommodating a more dynamically active dark energy component in the recent universe. {\color{blue}Because $\Gamma$ is not restricted to $\Gamma>3/2$, the posterior contains both flattening and steepening trajectories. The data therefore do not select a branch, and the $\Gamma$ essentially remains unconstrained and prior dependent} (Figs.~\ref{Fig1}--\ref{Fig3} and \emph{Appendices~\ref{A4}--\ref{A6})}.\\
 
The parameter $\Gamma\equiv VV''/(V')^{2}$ governs the curvature of the tachyonic potential relative to its slope, and via Eq.~\eqref{31} directly sets the rate at which $\lambda$ evolves. {\color{blue}The sign of $d\lambda/dN$ is controlled by the factor $(3/2-\Gamma)$: $\Gamma>3/2$ flattens the potential, $\Gamma<3/2$ steepens it, and at $\Gamma=3/2$ the slope freezes at its initial value. Neither sign is required for dynamical consistency, so we do not impose $\Gamma>3/2$. We sample $u\equiv1/\Gamma\in\mathcal{U}(0.01,1.0)$, which spans the cosmological-constant limit ($u\to0$, $\Gamma\to\infty$), the scaling potential ($u=2/3$) and the exponential string tachyon ($u=1$), and avoids the singular point $\Gamma=0$.} The attractor behavior referred to here concerns the insensitivity to initial conditions. For the tachyonic DBI system, the key observation is that at early times, deep in matter domination ($z\approx 1100$), the strong Hubble friction term $3H\dot\phi/\gamma^2$ in Eq.~\eqref{14} drives $\dot\phi \to 0$ regardless of the precise initial field velocity, placing the field near $w_\phi = \dot\phi^2 - 1 \approx -1$ for a wide range of initial conditions. This frozen-field configuration is the early-time attractor of the autonomous system~\eqref{31}. Trajectories originating from different initial values of $x_i = \dot\phi_i$ all converge toward $x \approx 0$ before dark energy becomes dynamically significant. As the universe expands and $H$ decreases, Hubble friction weakens, the field begins to roll, and $w_\phi$ moves away from $-1$ toward less negative values. It is this late-time departure from the frozen attractor that is observationally accessible and encodes how far the field has thawed since matter domination. The physical significance for initial conditions is therefore the following: because the early-time frozen state $\dot\phi \approx 0$ is an attractor, a broad class of initial field velocities all converge to the same near-$\Lambda$CDM trajectory before dark energy turns on, resolving the fine-tuning problem without requiring a precisely tuned $\dot\phi_i$. Simultaneously, the tracking behavior of $\Omega_\phi$ concerns negligibly small during radiation and matter domination and grows to dominate only at late times as encoded by $\Omega_{\rm de,ini} \sim \mathcal{O}(10^{-9})$ thereby addressing the coincidence problem, the near-equality of matter and dark-energy densities today is a dynamical consequence of the phase-space flow rather than an accident. {\color{blue}This early-time attractor operates in both branches, so the thawing behaviour does not depend on whether $\Gamma$ is above or below $3/2$.}
 
Across all three observational compilations, the posteriors for $\Gamma$ are broad and nearly uniform, with values {\color{blue}$\Gamma = 2.11^{+3.42}_{-0.90}$ (Pantheon$+$),} {\color{blue}$2.28^{+5.05}_{-1.05}$ (DES-SN5YR), and $1.91^{+3.72}_{-0.75}$ (Union3) (median and 68\% interval of $\Gamma=1/u$ with $u=0.49\pm0.28$, $0.46\pm0.29$ and $0.52\pm0.29$)}, in each case essentially spanning the full {\color{blue}prior. This insensitivity has a transparent physical interpretation: whether $\lambda$ slowly decreases or slowly increases over the short late-time window in which the field thaws produces negligible differences} in the background expansion history and the matter growth rate at current observational precision (Figs.~\ref{Fig1}--\ref{Fig3}). The model is therefore robust to the precise curvature of the potential and {\color{blue}a broad class of potentials, flattening or steepening,} remain observationally indistinguishable with present multi-probe data. Sharper constraints on $\Gamma$ will likely require next-generation surveys or the full CMB power spectrum to break the residual degeneracies with $\lambda_{\mathrm{ini}}$ and the background expansion.\\
 
{\color{blue}The derived parameter $\log(\Omega_{de,\mathrm{ini}})$} quantifies the fractional energy density of the tachyonic field at the onset of the numerical integration, $z_{\mathrm{ini}}\simeq1100$, deep within the matter-dominated epoch. The constraints are exceptionally tight (except $\Gamma$ and $\lambda_{\rm ini}$ ) and stable across all compilations, which corresponds to initial dark-energy fractions of {\color{blue}$\Omega_{de,\mathrm{ini}}\approx1.4\times10^{-9}$ (Pantheon$+$)} and $\approx1.4\times10^{-9}$ (DES-SN5YR and Union3). {\color{blue}Because it is fixed by the flatness condition rather than sampled, its posterior inherits the constraints on $\Omega_{m0}$, $H_0$ and $\lambda_{\rm ini}$, and is localized within an interval of width $\lesssim0.04$ in every compilation.}
 
The deeply sub-dominant value $\Omega_{de,\mathrm{ini}}\sim\mathcal{O}(10^{-9})$ constitutes a necessary internal consistency validation of the model. With such a negligible initial energy fraction, the tachyonic field contributes minimally to the total energy budget during Big Bang Nucleosynthesis and leaves the photon-baryon plasma dynamics and hence the CMB acoustic peaks~\cite{Dodelson,Weinberg} remain entirely unmodified. The standard thermal history is preserved to well within current observational sensitivity, confirming that the scenario introduces no disruption to the early-universe sector.
 
\subsection{Statistical Analysis}
\label{SecVD}
 
To rigorously evaluate the relative statistical performance of the tachyonic DBI dark energy model against the standard $\Lambda$CDM model, we compute the goodness of fit and a suite of information criteria. This analysis is performed across our three primary joint datasets, which combine a baseline of compressed CMB, BAO, CC, and $f\sigma_8$ with each of the three distinct Type~Ia supernova compilations: {\color{blue}Pantheon$+$}, DES-SN5YR, and Union3.
 
The complete statistical results, including total data points ($N$), degrees of freedom (DoF), best-fit $\chi^2$, and various information criteria, are summarized in Table~\ref{TableII}.\\
 
\begin{table*}[t]
    \centering
    \small
    \renewcommand{\arraystretch}{1.2}
    \setlength{\tabcolsep}{4pt}
    \caption{Comprehensive statistical comparison between the $\Lambda$CDM model and the tachyonic DBI dark energy model evaluated across three distinct data combinations, each pairing the baseline dataset with a different SN~Ia compilation.}
    \label{TableII}
    \resizebox{\textwidth}{!}{%
    \begin{tabular}{l cc cc cc}
        \hline\hline
        & \multicolumn{2}{c|}{Pantheon$+$}
        & \multicolumn{2}{c|}{DES-SN5YR}
        & \multicolumn{2}{c}{Union3} \\
        Statistical Metric
            & $\Lambda$CDM & Tachyonic
            & $\Lambda$CDM & Tachyonic
            & $\Lambda$CDM & Tachyonic \\
        \hline
        Total Data Points ($N$)
            & 1705 & 1705 & 1937 & 1937 & 139 & 139 \\
        Degrees of Freedom
            & 1701 & 1699 & 1933 & 1931 & 135 & 133 \\
        Best-Fit $\chi^2_{\mathrm{min}}$
            & 1510.28 & 1502.89 & 1702.84 & 1698.52 & 96.36 & 95.24 \\
        Reduced $\chi^2$ ($\chi^2_{\mathrm{red}}$)
            & 0.888 & 0.885 & 0.881 & 0.880 & 0.714 & 0.716 \\
        AIC & 1518.28 & 1514.89 & 1710.84 & 1710.52 & 104.36 & 107.24 \\
        BIC & 1540.05 & 1547.54 & 1733.12 & 1743.93 & 116.10 & 124.85 \\
        DIC & 1518.40 & 1514.30 & 1710.80 & 1708.80 & 104.28 & 105.06 \\
        Effective Parameters ($p_D$)
            & 4.07 & 5.47 & 4.01 & 4.87 & 4.00 & 4.65 \\
        Information Gain (KL, nats)
            & 15.42 & 14.79 & 15.41 & 16.21 & 15.39 & 14.67 \\
        Final Log Evidence ($\ln\mathcal{Z}$)
            & $-772.58\pm0.09$ & $-769.21\pm0.09$ & $-868.81\pm0.09$ & $-868.17\pm0.09$ & $-65.54\pm0.09$ & $-64.88\pm0.09$ \\
        \hline\hline
    \end{tabular}%
    }
\end{table*}
 
Table~\ref{TableII} shows that both $\Lambda$CDM and the tachyonic DBI model provide very good fits to the joint data, with reduced chi-squared values that are close to (or below) unity for all three combinations. For DES-SN5YR the tachyonic model achieves a marginally better best-fit with {\color{blue}$\Delta\chi^2_{\mathrm{min}}\simeq-4.32$} (tachyonic preferred at the maximum-likelihood level), and for Union3 with {\color{blue}$\Delta\chi^2_{\mathrm{min}}\simeq-1.12$}; in both cases the improvement is modest and the two models are effectively tied. {\color{blue}For Pantheon$+$ the tachyonic model is also better, with $\Delta\chi^2_{\mathrm{min}}\simeq-7.40$, the largest improvement of the three.}
 
The reduced chi-squared values primarily reflect the effective scatter and error budgeting of each joint likelihood:
\begin{itemize}
    \item {\color{blue}\textit{Pantheon$+$:} $\chi^2_{\mathrm{red}}\simeq0.89$ for both models.}
    \item \textit{DES-SN5YR:} $\chi^2_{\mathrm{red}}\simeq0.88$.
    \item \textit{Union3:} $\chi^2_{\mathrm{red}}\simeq0.72$.
\end{itemize}
{\color{blue} The $\chi^2_{\mathrm{red}}$ value well below unity for DES-SN5YR and Union3 may primarily reflects a conservative error budget across the combined probes rather than any genuine improvement in fit, and the same applies to Pantheon$+$, whose value is now also below unity. Nevertheless, The posterior predictive check of Sec.~\ref{SecVG} may separate these two
possibilities.}\\
 
{\color{blue}After accounting for the number of parameters, the BIC penalizes the tachyonic extension in every dataset, whereas the AIC gives mixed results. The differences (tachyonic minus $\Lambda$CDM) are $\Delta\mathrm{AIC}\simeq-3.40$ (Pantheon$+$; tachyonic preferred), $-0.32$ (DES-SN5YR; essentially tied), and $2.88$ (Union3), while the corresponding BIC differences are $\Delta\mathrm{BIC}\simeq7.49$, $10.81$, and $8.75$. This BIC penalty is driven by the $k\ln N$ term: each additional parameter costs $\ln N\simeq5$--$8$ even if the data do not constrain it; here $u$ is largely prior-dominated, so it pays the penalty without a compensating improvement in fit.}\\
 
A more global Bayesian comparison is provided by the evidence $\mathcal{Z}$, which averages the likelihood over the full prior volume. Using Table~\ref{TableII}, the log-evidence differences {\color{blue}$\Delta\ln\mathcal{Z}\equiv\ln\mathcal{Z}_{\rm DBI}-\ln\mathcal{Z}_{\Lambda{\rm CDM}}$ (statistical uncertainty $\simeq0.13$)} are
\begin{itemize}
    \item {\color{blue}\textit{Pantheon$+$:} $\Delta\ln\mathcal{Z}\simeq+3.38$,}
    \item {\color{blue}\textit{DES-SN5YR:} $\Delta\ln\mathcal{Z}\simeq+0.64$,}
    \item {\color{blue}\textit{Union3:} $\Delta\ln\mathcal{Z}\simeq+0.66$,}
\end{itemize}
{\color{blue}All three are positive, so the Bayesian evidence favours the tachyonic DBI model in every compilation. On the Jeffreys scale the preference is moderate for Pantheon$+$ ($2.5<\Delta\ln\mathcal{Z}<5$, odds of about $29:1$) and inconclusive for DES-SN5YR and Union3 ($\Delta\ln\mathcal{Z}<1$, odds of about $2:1$). Writing $\Delta\ln\mathcal{Z}=\Delta\langle\ln\mathcal{L}\rangle-\Delta D_{\rm KL}$ shows where this comes from: the posterior-averaged log-likelihood changes by $+2.75$, $+1.44$ and $-0.06$ and the information gain by $\Delta D_{\rm KL}=-0.63$, $+0.80$ and $-0.72$ nats for Pantheon$+$, DES-SN5YR and Union3. For Pantheon$+$ and DES-SN5YR the preference may be driven by the better fit, whereas for Union3 the fit is unchanged and the small positive $\Delta\ln\mathcal{Z}$ may comes from the prior-dominated $u(\equiv1/\Gamma)$ direction costing almost no prior volume.}\\
 
The effective number of parameters $p_D$ and the DIC probe how many degrees of freedom are actually being used by the posterior, providing a fairer assessment of model performance than raw evidence ratios. The tachyonic model exhibits a moderate increase in effective complexity, with {\color{blue}$\Delta p_D \simeq 1.40$ (Pantheon$+$),} {\color{blue}$0.87$ (DES-SN5YR), and $0.66$ (Union3)} {\color{blue}smaller than the nominal addition of two new parameters ($\lambda_{\rm ini}$ and $u=1/\Gamma$).} This compression demonstrates that the data actively constrain the tachyonic parameter space rather than leaving it diffuse, the posterior concentrates the effective complexity well below the prior dimensionality, which is precisely the hallmark of a physically meaningful and identifiable extension. The corresponding DIC shifts are {\color{blue}$\Delta\mathrm{DIC}\simeq-4.10$ (Pantheon$+$; tachyonic preferred), {\color{blue}$-2.00$} (DES-SN5YR; tachyonic preferred), and $0.77$ (Union3), implying that DIC prefers the tachyonic model for Pantheon$+$ and DES-SN5YR and assigns an effectively negligible penalty for Union3.}\\
 
The Kullback-Leibler~(KL) \cite{Kullback,MacKay} divergence provides the most direct validation of our tachyonic model—it quantifies how much the data have actually learned about the model parameters, independent of any penalization for prior volume. {\color{blue}The tachyonic model achieves information gains of $14.79$, $16.21$ and $14.67$ nats for Pantheon$+$, DES-SN5YR, and Union3, compared with $15.42$, $15.41$ and $15.39$ nats for $\Lambda$CDM, i.e., $\Delta D_{\rm KL}=-0.63$, $+0.80$ and $-0.72$ nats. The data therefore learn about as much in either model. For the tachyonic model, the constraint on $\lambda_{\rm ini}$ adds information but $u=1/\Gamma$ is prior-dominated and adds none essentially, and the two effects roughly cancel. The additional tachyonic parameters thus do not inflate the prior volume; the data must pay for, which is why the improved fit translates directly into a positive $\Delta\ln\mathcal{Z}$.}
 
{\color{blue} Overall, the statistical picture favours the tachyonic DBI model: it attains a lower $\chi^2_{\rm min}$ in all three compilations, the Bayesian evidence prefers it moderately for Pantheon$+$ and weakly for DES-SN5YR and Union3, and the DIC prefers it for Pantheon$+$ and DES-SN5YR. Only the BIC, whose penalty ignores whether a parameter is constrained, prefers $\Lambda$CDM. The preference may be driven by the thawing background, via $\lambda_{\rm ini}$  the potential curvature $\Gamma$ remains unconstrained.}
 
\subsection{Dark Energy Clustering and Perturbation Dynamics}
\label{SecVE}

\begin{figure*}[t]
    \centering
    \includegraphics[width=0.9\textwidth]{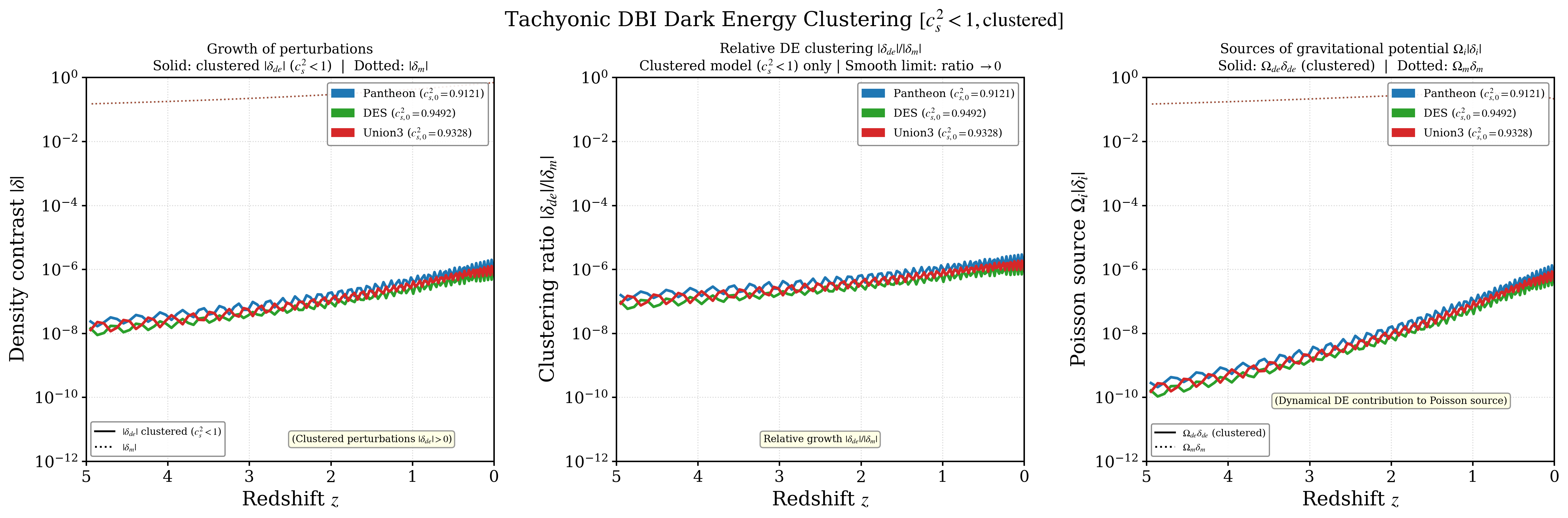}
    \caption{Redshift dependence of linear density perturbations in the tachyonic DBI dark energy scenario, evaluated at the posterior-mean best-fit cosmological parameters for each supernova compilation combined with the baseline probes (compressed CMB, BAO, CC, and $f\sigma_8$). \textit{Left panel:} Absolute density contrasts $|\delta_m|$ (dotted) and $|\delta_{\rm de}|$ (solid) as functions of redshift $z$ for the clustered DBI model ($c_s^2<1$). The three compilations correspond to {\color{blue}Pantheon$+$} ($c_{s,0}^2=0.9925$, blue), DES-SN5YR ($c_{s,0}^2=0.9397$, green), and Union3 ($c_{s,0}^2=0.9360$, red). \textit{Middle panel:} Relative DE clustering ratio $|\delta_{\rm de}|/|\delta_m|$ for the three supernova compilations. \textit{Right panel:} Poisson source terms $\Omega_m\,|\delta_m|$ (dotted) and $\Omega_{\rm de}\,|\delta_{\rm de}|$ (solid) which directly compares the gravitational source contributions of matter and dark energy.}
    \label{Fig5}
\end{figure*}

A distinctive aspect of the tachyonic DBI scenario is that it can, in principle, support spatial fluctuations. For a non-canonical field the relevant control parameter is the effective sound speed, for $c_s^2<1$ the dark-energy component can cluster on sub-horizon scales, whereas in the smooth limit ($c_s^2=1$) one has $\delta_{\rm de}\equiv0$. Figure~\ref{Fig5} summarizes the resulting linear perturbation dynamics for our three joint data combinations, evaluated at their posterior-mean parameter values.\\

The left panel tracks the absolute amplitudes of the matter contrast $|\delta_m|$ (dotted) and the tachyonic dark-energy contrast $|\delta_{\rm de}|$ (solid) from $z=5$ down to $z=0$. Matter grows monotonically, as expected from the standard gravitational instability picture. The dark-energy perturbation tells a richer and more compilation-dependent story. All three compilations produce non-zero $|\delta_{\rm de}|$ throughout, which confirms genuine dark-energy clustering, but the amplitude and oscillation character differ markedly. {\color{blue}DES-SN5YR ($c_{s,0}^2=0.9492$), Union3 ($c_{s,0}^2=0.9328$) and Pantheon$+$ ($c_{s,0}^2=0.9121$) have similar sound speeds and their perturbation curves nearly coincide:} $|\delta_{\rm de}|$ grows monotonically from $\sim\!10^{-8}$ at $z=5$ toward $\sim\!10^{-6}$ at $z=0$, with shallow, slow oscillations superimposed on the growing trend. {\color{blue}Pantheon$+$, with the lowest sound speed, has a slightly larger $|\delta_{\rm de}|$ and slightly deeper oscillations than the other two.} In all cases, the growing secular trend of $|\delta_{\rm de}|$ confirms that the perturbation is not simply decaying away; dark-energy clustering is genuinely active, even if its amplitude remains modest on the absolute scale of matter fluctuations.\\

The ratio $|\delta_{\rm de}|/|\delta_m|$ shown in the middle panel puts the clustering efficiency in sharper relief. The compilation hierarchy established in the left panel is even clearer here - all three compilations reach ratios of $\sim\!10^{-6}$ at $z=0$, {\color{blue}with Pantheon$+$ slightly the highest, reflecting its lower best-fit sound speed $c^2_{s,0}\simeq0.91$ (against $0.95$ and $0.93$ for DES-SN5YR and Union3).} The peak of this ratio occurs at intermediate redshift, during matter domination, when the background Hubble friction is still large enough to drive dark-energy fluctuations but the matter contrast has not yet grown to its present amplitude, it rises thereafter as the secular growth of $|\delta_{\rm de}|$ continues while $|\delta_m|$ levels off. Superimposed on this trend are the characteristic pressure-driven oscillations, which are {\color{blue}slightly faster and deeper for Pantheon$+$} and slow and shallow for DES/Union3, in direct proportion to the proximity of each best-fit sound speed to unity. In no case, however, does the ratio approach unity, and the tachyonic fluid remains a subdominant clustering component throughout cosmic history.\\

The right panel strips away the ratios and compares the raw Poisson source terms $\Omega_m|\delta_m|$ and $\Omega_{\rm de}|\delta_{\rm de}|$ directly. The hierarchy is unambiguous: the matter source (dotted line) sits near $\sim\!0.3$ across the full redshift range, while the dark-energy source peaks at $\sim\!10^{-6}$ for DES/Union3 {\color{blue}and likewise at $\sim\!10^{-6}$ for Pantheon$+$ at $z=0$.} This places the tachyonic contribution eight to ten orders of magnitude below matter in the gravitational potential budget, at every redshift shown. Thus, although the tachyonic fluid does cluster when $c_s^2<1$, so that $\delta_{\rm de}\not\equiv 0$, its contribution to Newtonian potential wells remains negligible on linear scales.\\

{\color{blue}Taken together, these three panels (Fig.~\ref{Fig5}) decisively clarify the implications for the $S_8$ tension. {\color{blue}All of DES-SN5YR, Union3, and Pantheon$+$ compilations prefer sound speeds sufficiently below unity to produce non-negligible DE perturbations up to $\sim\!10^{-6}$ in $|\delta_{\rm de}|$, the Poisson source $\Omega_{\rm de}|\delta_{\rm de}|$ never rises to a level that could perceptibly alter gravitational potential wells or the matter growth rate.} As a result, all three compilations predict essentially the comparable $f\sigma_8$ history and the $S_8$ posterior as $\Lambda$CDM under current observational precision. We do not claim, nor do our results support, any alleviation of the $S_8$ discrepancy. Our analysis establishes a more modest but still meaningful result: the tachyonic DBI model is fully consistent with current large-scale-structure data. The mechanism by which it could, in principle, affect matter growth, a modification of Hubble friction via noncanonical pressure gradients, remains viable, and forthcoming Stage-IV surveys with sub-percent $f\sigma_8$ precision may be able to detect it or rule it out decisively.} \\

\subsection{Consistency checks against DESI DR2 and different supernova compilation}

\begin{figure*}[t]
    \centering
    \includegraphics[width=\textwidth]{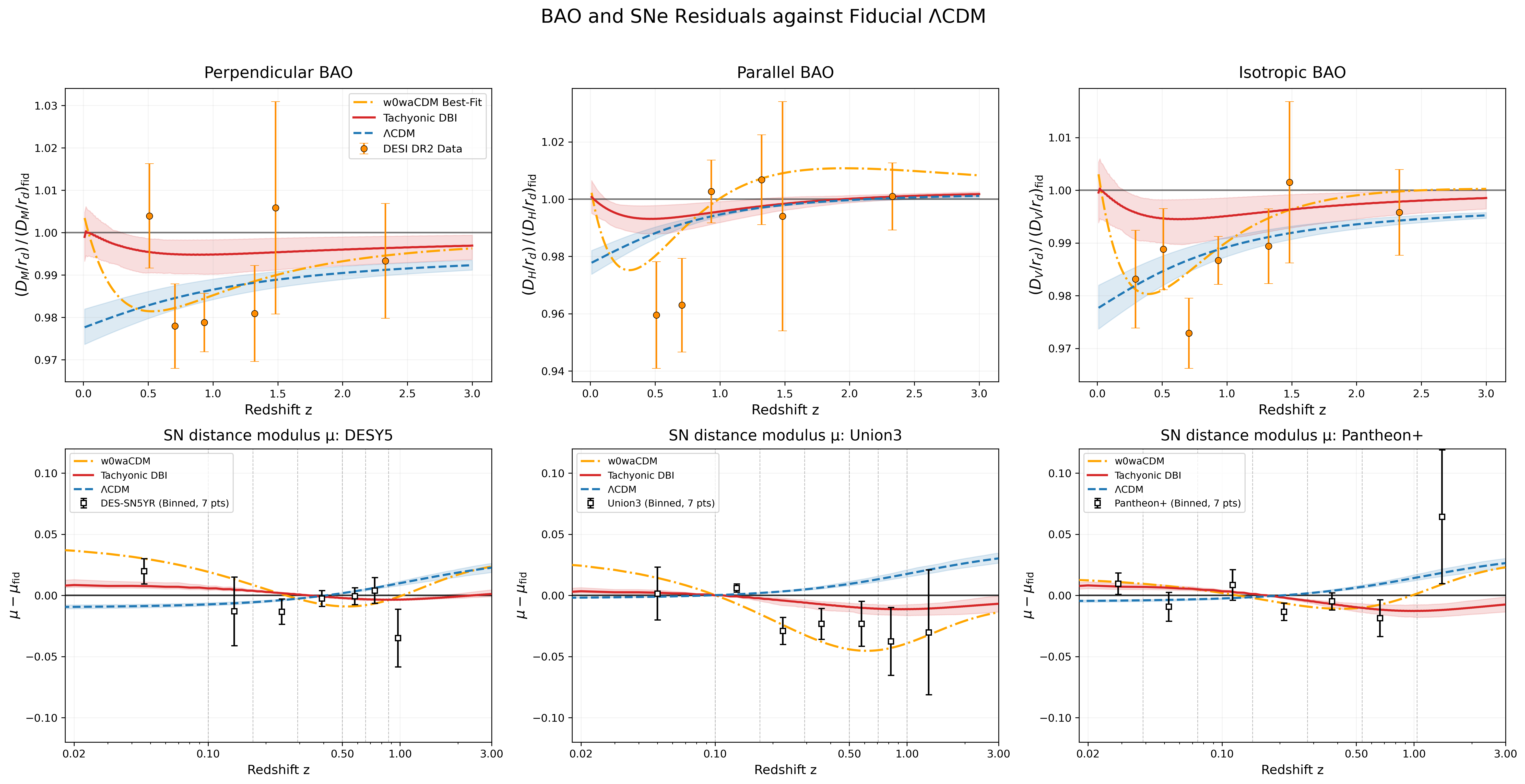}
    \caption{Joint BAO and SNe~Ia residuals against the DESI DR2 flat-$\Lambda$CDM best-fit ($\Omega_{m0}=0.3169$, $H_0=67.14\,{\rm km\,s^{-1}\,Mpc^{-1}}$~\cite{DESI,DESI2}). \textit{Top row:} Ratios of the three BAO distance metrics to the same fiducial, evaluated for the DES-SN5YR joint dataset combination. Red solid with shaded $1\sigma$ band: tachyonic DBI, Blue dashed: $\Lambda$CDM. Orange dash-dot: DES-SN5YR $w_0w_a$CDM best-fit from DESI DR2 ($w_0=-0.752$, $w_a=-0.86$~\cite{DESI,DESI2}). Orange circles: DESI DR2 data~\cite{DESI,DESI2}. \textit{Bottom row:} Distance-modulus residuals $\mu-\mu_{\rm fid}$ for {\color{blue}Pantheon$+$} (right), Union3 (centre), and DES-SN5YR (left). In each panel $w_0w_a$CDM best-fit from \cite{DESI, DESI2} is shown. Gray points: covariance-compressed binned SN data~\cite{Scolnic,DESSN5YRCosmo,Union3}.}
    \label{Fig5a}
\end{figure*}

\begin{figure*}[t]
    \centering
    \includegraphics[width=\textwidth]{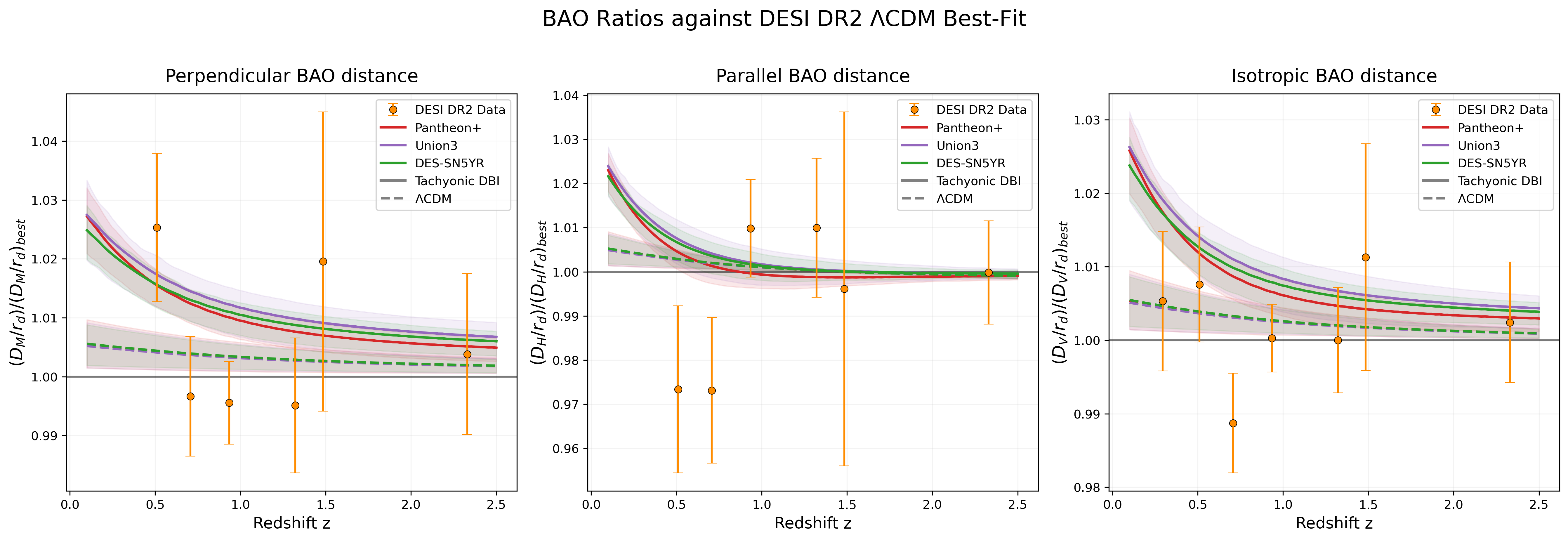}
    \caption{BAO distance ratios for all three SN compilations, normalized to the DESI DR2 $\Lambda$CDM best-fit obtained from a BAO-only nested-sampling inference (mean $\Omega_{m0}\approx0.295$~\cite{DESI,DESI2}). \textit{Left:} $D_M/r_d$. \textit{Centre:} $D_H/r_d$. \textit{Right:} $D_V/r_d$. Solid curves: tachyonic DBI, dashed curves: $\Lambda$CDM. Colors denote compilations: {\color{blue}Pantheon$+$} (red), Union3 (purple), DES-SN5YR (green) along with base compilation: CC + BAO + f$\sigma_8$ + CMB. Orange circles with error bars: DESI DR2 data~\cite{DESI,DESI2}.}
    \label{Fig5b}
\end{figure*}

It is worth emphasizing that the recent DESI DR2~\cite{DESI, DESI2} results indicate a preference for phantom-like behavior in the dark energy sector. This motivates a direct comparison of our model with these latest observations, in order to assess its viability as an alternative, field-theoretic description of cosmic acceleration.  To this end, Figures~\ref{Fig5a} and~\ref{Fig5b} provide two complementary consistency checks of our inferred expansion histories, going beyond the compressed likelihoods used in the main inference. Figure~\ref{Fig5a} is a $2\times3$ grid whose top row shows ratios of each BAO distance metric ($D_M/r_d$, $D_H/r_d$, and $D_V/r_d$) to the same DESI DR2 flat-$\Lambda$CDM best-fit ($\Omega_{m0}=0.3169$, $H_0=67.14\,{\rm km\,s^{-1}\,Mpc^{-1}}$~\cite{DESI,DESI2}), while the bottom row shows the distance-modulus residuals $\mu-\mu_{\rm fid}$ for each SN compilation separately. In every panel, the tachyonic DBI (red solid with shaded $1\sigma$ posterior band) and $\Lambda$CDM (blue dashed) predictions are evaluated at the posterior-mean parameters of Table~\ref{TableI} for the corresponding joint dataset combination used throughout the paper, and the orange dash-dot curve is the DESI DR2 compilation-specific $w_0w_a$CDM best-fit~\cite{DESI,DESI2} ($w_0=-0.752$, $w_a=-0.86$ for DES-SN5YR in both the top and bottom row, $w_0=-0.838$, $w_a=-0.62$ for {\color{blue}Pantheon$+$}, $w_0=-0.667$, $w_a=-1.09$ for Union3 in the bottom panels). The DESI DR2 data points (orange circles with error bars, Table~4 of \cite{DESI,DESI2}) are overplotted on the top row, and the SN data are shown as weighted and covariance-compressed binned residuals~\cite{Scolnic,DESSN5YRCosmo,Union3} in the bottom row. Details of our optimal linear compression of the supernova data are given in Appendix~\ref{A7}.

The top row of Fig.~\ref{Fig5a} reveals a clear and physically transparent picture of how the tachyonic and $\Lambda$CDM models compare against both the fiducial and the DESI-preferred CPL cosmology. {\color{blue}In all three BAO metrics the tachyonic DBI curve starts at unity at $z\sim0$, dips to $\approx0.993$ around $z\sim0.5$--$1$ and returns to $\approx0.997$--$1.001$ by $z=3$, placing it within $\simeq0.7\%$ of the fiducial across the full DESI redshift baseline.} The $\Lambda$CDM model for the same DES-SN5YR combination sits $\sim1$--$2\%$ below the fiducial at low redshift, converging toward unity only at $z>2$. The CPL curve displays the most distinctive behaviour. It starts just above unity near $z\sim0$, dives to $\approx0.97$ in the parallel metric $D_H/r_d$ around $z\sim0.30$ where the DESI tension is most pronounced, then crosses unity and rises to $\approx1.01$--$1.02$ at $z=2.5$. This non-monotonicity is the geometric imprint of the dynamical equation of state preferred by DESI, with $w_0>-1$ and $w_a<0$~\cite{DESI,DESI2}. Over the same redshift range, the tachyonic DBI prediction stays closer to unity and follows the data more smoothly, whereas the CPL best-fit exhibits a sharp pivot. Taken at face value, this suggests that our tachyonic scalar-field trajectory does not reproduce the detailed DESI DR2 behaviour. In the bottom row the SN residuals further reinforce this picture. For DES-SN5YR and Union3 the tachyonic band more or less encloses the binned data at essentially all redshifts, with the tachyonic model sitting slightly above $\Lambda$CDM at low-redshift in a manner consistent with its mildly higher inferred $\mu$ from the thawing equation of state. {\color{blue}For Pantheon$+$ the tachyonic band follows the binned data within their errors at all redshifts, while $\Lambda$CDM rises above the data at $z\gtrsim0.3$, the same pattern as for Union3.}

Figure~\ref{Fig5b} brings all three SN compilations together in a single comparison, normalizing by the mean of the DESI DR2 BAO-only $\Lambda$CDM posterior values provided in ~\cite{DESI,DESI2}. This choice of fiducial isolates how much of the residual variation in the predicted distance ratios is driven by the SN prior versus intrinsic model differences. The result is consistent across all three metrics. DES-SN5YR (green) lies $\sim1$-$2.5\%$ above the fiducial at low redshift in all three panels,  converging toward unity by $z\approx2$--$2.5$. Union3 (purple) traces an almost identical path to DES-SN5YR, consistent with their near-identical posterior parameters (Table~\ref{TableI}), and likewise traces the data throughout. {\color{blue}Pantheon$+$ (red) follows the same path as DES-SN5YR and Union3, with its tachyonic curve slightly below theirs at $0.5\lesssim z\lesssim2$ and its $\Lambda$CDM curve indistinguishable from theirs, so no compilation shows a systematic geometric offset from the DESI data.} Crucially, in every compilation and every metric the tachyonic DBI curve (solid) sits \emph{consistently above} the corresponding $\Lambda$CDM curve (dashed) by a $\sim0.5$--$1\%$ margin, a pattern that is uniform across the full $z\in[0.1,\,2.5]$ baseline. This systematic upward shift has a origin that comes from the thawing DBI equation of state $w_\phi>-1$ with $\dot{w}_\phi>0$ produces slightly larger comoving distances at any given redshift compared to a cosmological constant with identical background parameters, and this accumulated difference is precisely what separates the tachyonic and $\Lambda$CDM predictions in Fig.~\ref{Fig5b}. 

Taken together, Figs.~\ref{Fig5a} and~\ref{Fig5b} establish that: (i) although the wCDM model with the CPL parametrization provides the best fit to the DESI DR2 data, {\color{blue}our tachyonic DBI model tracks the DESI DR2 points more closely than $\Lambda$CDM at $z\lesssim1$,} (ii) across all three SN compilations, the tachyonic DBI model does not match the DESI DR2 points uniformly well, which may indicate that the restriction imposed theoretically to our dark-energy EoS to remain in the quintessence regime may indeed be too limiting, {\color{blue}and (iii) with Pantheon$+$ used without the SH0ES calibration, all three SN compilations give mutually consistent BAO predictions.}

By contrast, theoretically motivated models, ranging from dark-sector interactions in coupled scalar-field scenarios~\cite{Chakraborty2025, Chakraborty2026, Antusch2026, Ganguly2026} to scalar-tensor modifications of gravity in the Horndeski/beyond-Horndeski class~\cite{Horndeski, Zumalacárregui, Gleyzes, Tsujikawa2013DE}, can realize phantom-divide crossing or mimic an observationally phantom-like effective equation of state, thereby producing a $w(z)$ evolution (derived from an action principle) that closely resembles the phenomenological $w_0w_a$CDM parametrization. In the case of our tachyonic DBI model, however, the intrinsic constraint $w_{\phi}\geq -1$ restricts the available parameter space and limits its flexibility in reproducing the phantom-like evolution suggested by the DESI DR2 data. Consequently, the model does not provide sufficient freedom to achieve a comparably good fit to the DESI DR2 observations. Also, we note that within the broad class of single-field, minimally coupled scalar models described by a regular low-energy Lagrangian, crossing the phantom divide $w=-1$ is generically forbidden (the single-field ``no-go'' theorem)~\cite{Vikman2}.

\subsection{Posterior Predictive Checks}
\label{SecVG}

For consistency between the data and the best-fit posterior prediction we have done a posterior predictive check for the full posterior through both of our model to generate replicated datasets under each cosmology, then compares the resulting distribution of $\chi^2$ discrepancy statistics to the value realized by the actual data~\cite{Gelman}. The procedure avoids any particular condition on a single parameter point estimate, because the replicates average over posterior uncertainty, therefore the diagnostic is sensitive to whether the model family itself is misspecified, not merely to the quality of the MAP fit.

We draw $N_{\rm rep}=10000$ equal-weight samples from the \texttt{dynesty} posterior~\cite{dynesty}, evaluate each through the identical ODE integration used at inference, and compute predictions for all seven observables entering the joint likelihood namely the Hubble rate $H(z)$, supernova distance moduli $\mu$, the growth rate combination $f\sigma_8(z)$, compressed CMB distance priors, and the BAO ratios $D_M/r_d$, $D_H/r_d$, and $D_V/r_d$. Thereafter, correlated Gaussian noise sampled from the relevant covariance matrix is then added to each prediction, which yield a synthetic replicated dataset $\mathbf{d}^{\rm rep}$. The Bayesian predictive $p$-value is
\ben
p_B =  P\!\left(\chi^2_{\rm rep} > \chi^2_{\rm obs}\big|\mathbf{d}\right),
\label{68a}
\een
estimated from the $10000$ replicates. The check goes as follows: values near $0.5$ reflect a well-calibrated fit, $p_B \lesssim 0.05$ indicates genuine tension between the posterior and the data and value close to $1$, indicates that the observed $\chi^2$ is unusually small compared to the posterior-predictive $\chi^2$ distribution.

Table~\ref{TableIIa} presents $\chi^2_{\rm med}$, $\chi^2_{\rm med}/N$, and $p_B$ for every run-observable pair. The $H(z)$ and $f\sigma_8$ compilations return $p_B \approx 1$ with $\chi^2_{\rm med}/N\approx 0.46$--$0.53$ across all six runs without exception. Conservative systematic budgets in cosmic-chronometer and RSD datasets are the established source of this behaviour~\cite{Verde}, the low $\chi^2_{\rm med}/N$ signals over-estimated errors, not a poorly specified model. {\color{blue}Supernova distance moduli~($\mu$) are reproduced well by both models for all three compilations as follows: the tachyonic DBI $\chi^2_{\rm med}/N$ ranges from $0.896$ for DES-SN5YR to $1.260$ for Union3}, while the $\Lambda$CDM counterparts span a nearly identical $0.898$--$1.391$. The Union3 compilation shows the mildest predictive tension of the three SN datasets, with $p_B = 0.177$ and $p_B = 0.105$ for DBI and $\Lambda$CDM respectively, lying well above the $0.05$ threshold in both cases.

For DES-SN5YR and Union3, both models achieve CMB {\color{blue}$\chi^2_{\rm med}/N$ values of $1.65$--$2.29$ over three data points, with $p_B$ in the range $0.109$--$0.212$.} These values are not catastrophic, though the mild compression relative to the $p_B=0.5$ is consistent with the same marginal tension with phantom-crossing behaviour that DESI DR2 has signalled~\cite{DESI,DESI2}. All three BAO observables are reproduced well for these two compilations, {\color{blue}with $\chi^2_{\rm med}/N \lesssim 1.07$ and $p_B \in [0.385,\,0.615]$ throughout.}

{\color{blue}For Pantheon$+$, the CMB $\chi^2_{\rm med}/N$ is $2.40$ ($p_B=0.101$) for the tachyonic DBI model and $1.63$ ($p_B=0.218$) for $\Lambda$CDM, and the BAO ratios $D_M/r_d$, $D_H/r_d$ and $D_V/r_d$ give $\chi^2_{\rm med}/N=0.82$--$1.07$ with $p_B\in[0.377,\,0.591]$. Pantheon$+$ therefore behaves similar to DES-SN5YR and Union3.}

\begin{table*}[t]
\centering
\caption{Posterior predictive summary for all six joint analyses}
\label{TableIIa}
\setlength{\tabcolsep}{6pt}
\begin{tabular}{llrrrr}
\hline\hline
Model & Observable & $N$ & $\chi^2_{\rm med}$ & $\chi^2_{\rm med}/N$ & $p_B$ \\
\hline
\multirow{7}{*}{Tachyonic DBI (DES-SN5YR)}
  & $H(z)$ & 32 & {\color{blue}14.72} & {\color{blue}0.460} & {\color{blue}0.996} \\
  & $\mu$ & 1820 & {\color{blue}1631.23} & {\color{blue}0.896} & {\color{blue}0.999} \\
  & $f\sigma_8$ & 63 & {\color{blue}33.13} & {\color{blue}0.526} & {\color{blue}0.999} \\
  & CMB & 3 & {\color{blue}6.58} & {\color{blue}2.192} & {\color{blue}0.116} \\
  & $D_M/r_d$ & 6 & {\color{blue}5.03} & {\color{blue}0.838} & {\color{blue}0.525} \\
  & $D_H/r_d$ & 6 & {\color{blue}6.11} & {\color{blue}1.019} & {\color{blue}0.412} \\
  & $D_V/r_d$ & 7 & {\color{blue}5.06} & {\color{blue}0.722} & {\color{blue}0.610} \\
\hline
\multirow{7}{*}{Tachyonic DBI ({\color{blue}Pantheon$+$})}
  & $H(z)$ & 32 & {\color{blue}14.74} & {\color{blue}0.461} & {\color{blue}0.996} \\
  & $\mu$ & 1588 & {\color{blue}1435.02} & {\color{blue}0.904} & {\color{blue}0.997} \\
  & $f\sigma_8$ & 63 & {\color{blue}33.27} & {\color{blue}0.528} & {\color{blue}0.998} \\
  & CMB & 3 & {\color{blue}7.21} & {\color{blue}2.404} & {\color{blue}0.101} \\
  & $D_M/r_d$ & 6 & {\color{blue}4.90} & {\color{blue}0.817} & {\color{blue}0.535} \\
  & $D_H/r_d$ & 6 & {\color{blue}6.21} & {\color{blue}1.036} & {\color{blue}0.395} \\
  & $D_V/r_d$ & 7 & {\color{blue}5.18} & {\color{blue}0.739} & {\color{blue}0.591} \\
\hline
\multirow{7}{*}{Tachyonic DBI (Union3)}
  & $H(z)$ & 32 & {\color{blue}14.73} & {\color{blue}0.460} & {\color{blue}0.996} \\
  & $\mu$ & 22 & {\color{blue}27.72} & {\color{blue}1.260} & {\color{blue}0.177} \\
  & $f\sigma_8$ & 63 & {\color{blue}33.16} & {\color{blue}0.526} & {\color{blue}0.999} \\
  & CMB & 3 & {\color{blue}6.88} & {\color{blue}2.293} & {\color{blue}0.109} \\
  & $D_M/r_d$ & 6 & {\color{blue}5.01} & {\color{blue}0.834} & {\color{blue}0.537} \\
  & $D_H/r_d$ & 6 & {\color{blue}6.11} & {\color{blue}1.019} & {\color{blue}0.408} \\
  & $D_V/r_d$ & 7 & {\color{blue}5.03} & {\color{blue}0.718} & {\color{blue}0.615} \\
\hline
\multirow{7}{*}{$\Lambda$CDM (DES-SN5YR)}
  & $H(z)$ & 32 & {\color{blue}14.70} & {\color{blue}0.459} & {\color{blue}0.996} \\
  & $\mu$ & 1820 & {\color{blue}1635.14} & {\color{blue}0.898} & {\color{blue}0.999} \\
  & $f\sigma_8$ & 63 & {\color{blue}32.91} & {\color{blue}0.522} & {\color{blue}0.999} \\
  & CMB & 3 & {\color{blue}4.94} & {\color{blue}1.648} & {\color{blue}0.212} \\
  & $D_M/r_d$ & 6 & {\color{blue}5.41} & {\color{blue}0.902} & {\color{blue}0.467} \\
  & $D_H/r_d$ & 6 & {\color{blue}6.38} & {\color{blue}1.064} & {\color{blue}0.385} \\
  & $D_V/r_d$ & 7 & {\color{blue}5.58} & {\color{blue}0.797} & {\color{blue}0.544} \\
\hline
\multirow{7}{*}{$\Lambda$CDM ({\color{blue}Pantheon$+$})}
  & $H(z)$ & 32 & {\color{blue}14.70} & {\color{blue}0.459} & {\color{blue}0.996} \\
  & $\mu$ & 1588 & {\color{blue}1442.52} & {\color{blue}0.908} & {\color{blue}0.995} \\
  & $f\sigma_8$ & 63 & {\color{blue}32.91} & {\color{blue}0.522} & {\color{blue}0.999} \\
  & CMB & 3 & {\color{blue}4.90} & {\color{blue}1.632} & {\color{blue}0.218} \\
  & $D_M/r_d$ & 6 & {\color{blue}5.44} & {\color{blue}0.906} & {\color{blue}0.466} \\
  & $D_H/r_d$ & 6 & {\color{blue}6.41} & {\color{blue}1.069} & {\color{blue}0.377} \\
  & $D_V/r_d$ & 7 & {\color{blue}5.66} & {\color{blue}0.808} & {\color{blue}0.533} \\
\hline
\multirow{7}{*}{$\Lambda$CDM (Union3)}
  & $H(z)$ & 32 & {\color{blue}14.71} & {\color{blue}0.460} & {\color{blue}0.996} \\
  & $\mu$ & 22 & {\color{blue}30.59} & {\color{blue}1.391} & {\color{blue}0.105} \\
  & $f\sigma_8$ & 63 & {\color{blue}32.90} & {\color{blue}0.522} & {\color{blue}0.999} \\
  & CMB & 3 & {\color{blue}5.12} & {\color{blue}1.705} & {\color{blue}0.198} \\
  & $D_M/r_d$ & 6 & {\color{blue}5.39} & {\color{blue}0.898} & {\color{blue}0.474} \\
  & $D_H/r_d$ & 6 & {\color{blue}6.32} & {\color{blue}1.053} & {\color{blue}0.389} \\
  & $D_V/r_d$ & 7 & {\color{blue}5.43} & {\color{blue}0.776} & {\color{blue}0.557} \\
\hline\hline
\end{tabular}
\end{table*}

Three conclusions we can draw from Table~\ref{TableIIa}. {\color{blue}For all three SN compilations, both models pass all seven observable tests without exception;} there is no evidence of systematic misspecification anywhere in the fitted cosmological model. {\color{blue}The mild CMB compression in these analyses aligns} with the same phantom-crossing preference visible in the DESI DR2 comparison of Section~\ref{SecVE}. {\color{blue}The values of $\chi^2_{\rm red}$ below unity for all three compilations can be} partly attributed to the relatively conservative uncertainties in the CC ($H(z)$) and $f\sigma_8$ measurements. As can be observed from Table.\ref{TableIIa} that across all model-dataset combinations, the posterior predictive $p$-values for both $H(z)$ and $f\sigma_8$ remain close to unity, which indicates that these datasets are comparatively weakly constraining and contribute little to distinguishing between the parameter combinations preferred by the different models and combination of datasets used here.

{\color{red}Therefore, although both $\Lambda$CDM and the tachyonic DBI model remain statistically viable for the joint DESI DR2 and CMB analysis, their comparison with the DESI DR2 BAO~\cite{DESI, DESI2} distance ratios indicates that the tachyonic DBI model tracks the observations more closely than $\Lambda$CDM at $z\lesssim 1$. This improvement is not uniform across the full redshift range or all supernova compilations. Nevertheless, the results suggest that dynamical dark energy can improve the description of the low-redshift BAO observations without requiring phantom crossing.}

\subsection{Verification of the dynamical nature of our Tachyonic Model}
\label{SecVF}
 
Although the tachyonic DBI model was shown in the preceding subsections to exhibit a near-complete observational degeneracy with $\Lambda$CDM for standard cosmological parameters, this statistical consistency does not imply that the underlying dark-energy dynamics are trivial or $\Lambda$-like. In this subsection, we provide a set of independent, analytically exact diagnostics computed at the posterior-mean best-fit parameters of Table~\ref{TableI}, that directly reveal the dynamical nature of the tachyonic scalar field and distinguish it from a cosmological constant. Specifically, we examine: (i) the CPL best-fit projection of the tachyonic equation-of-state trajectory $w_\phi(z)$, (ii) the $\text{Om}(z)$ null diagnostic, (iii) the Statefinder hierarchy in the $(r,s)$ plane, and (iv) the matter growth index $\gamma(z)$. Together, these probes will confirm that the model is dynamical in nature, residing strictly in the thawing quintessence region, while remaining consistent with all current observational data.

Here, we use the term \emph{thawing} in the classification sense of Caldwell \& Linder~\cite{Caldwell}: a dark-energy field frozen near $w \approx -1$ at early times that evolves toward less negative $w$ at late times, regardless of whether the field is canonical or noncanonical. For the tachyonic DBI field, $w_\phi = \dot\phi^2 - 1 \in (-1,0)$ by construction, since the DBI speed limit enforces $0 < \dot\phi^2 < 1$, so the field necessarily resides in the $w > -1$ half-plane, conventionally called the quintessence region of EoS space. We emphasize, however, that this label carries a different physical meaning than in canonical quintessence. For a canonical field, $w_\phi > -1$ implies potential energy dominates over the kinetic energy, since $w_\phi = (\frac{1}{2}\dot\phi^2 - V)/(\frac{1}{2}\dot\phi^2 + V)$. For the tachyonic DBI field in our case, $w_\phi = \dot\phi^2 - 1$ depends solely on the field velocity, independently of $V(\phi)$. $w_\phi \to -1$ whenever $\dot\phi^2 \to 0$, irrespective of the potential magnitude. The quintessence region and thawing classification are therefore purely statements about the \emph{value} of $w_\phi$, not about the kinetic structure or the kinetic-to-potential energy ratio of the DBI field.\\
 
\begin{figure}[htbp]
    \centering
    \includegraphics[width=\columnwidth]{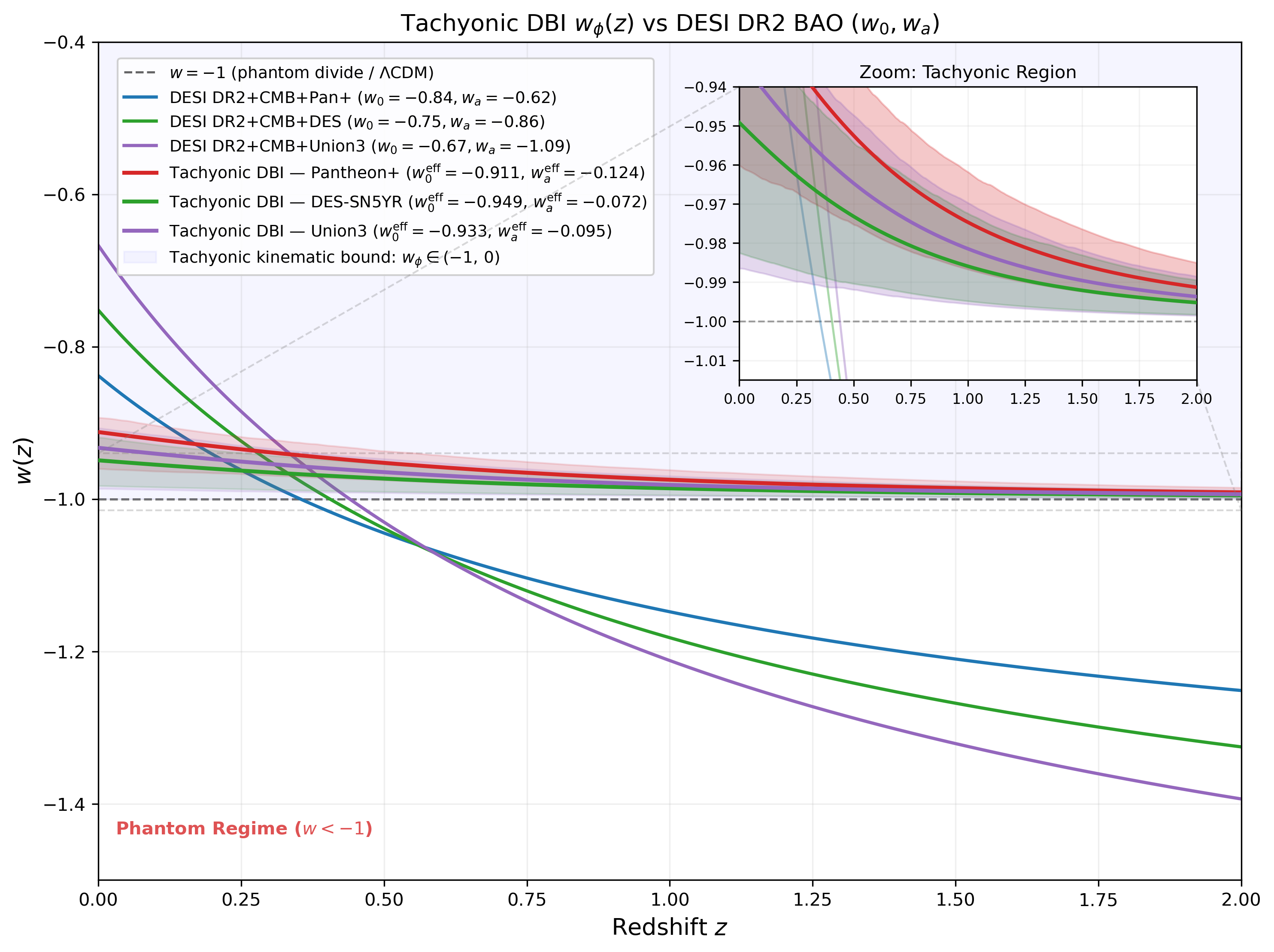}
    \caption{CPL best-fit projection of the tachyonic DBI equation of state $w_\phi(z) = \dot{\phi}^2 - 1$ evaluated at the posterior-mean parameters for each supernova compilation combined with the baseline probes.}
    \label{Fig6a}
\end{figure}

\textit{CPL best-fit:}~The tachyonic DBI equation of state, given by Eq.~\eqref{10}, is bounded by construction to the kinematic interval $w_\phi \in (-1, 0)$, with $w_\phi \to -1$ as $\dot{\phi}^2 \to 0$. To quantify how far the best-fit evolution departs from a cosmological constant and to provide a model-agnostic representation of the dark-energy dynamics, we project $w_\phi(z)$ onto the Chevallier--Polarski--Linder (CPL) parametrization~\cite{Caldwell, Linder1}:
\ben
w_{\rm CPL}(z) = w_0^{\rm eff} + w_a^{\rm eff}\,\frac{z}{1+z},
\label{69}
\een
by fitting $w_{\rm CPL}(z)$ to the numerically computed $w_\phi(z)$ over the redshift range $z \in [0,2]$. The effective CPL parameters extracted from the three joint datasets are:
\begin{itemize}
    \item {\color{blue}Pantheon$+$: $w_0^{\rm eff} = -0.911$,\; $w_a^{\rm eff} = -0.124$,}
    \item {\color{blue}DES-SN5YR: $w_0^{\rm eff} = -0.949$,\; $w_a^{\rm eff} = -0.072$,}
    \item {\color{blue}Union3: $w_0^{\rm eff} = -0.933$,\; $w_a^{\rm eff} = -0.095$.}
\end{itemize}
Figure~\ref{Fig6a} presents the resulting $w_\phi(z)$ curves together with the blue-shaded region marking the kinematic bound $w_\phi \in (-1, 0)$. Three key features are apparent. First, $w_\phi(z)$ always stays above $-1$, so the tachyonic field remains in the quintessence regime. This is guaranteed by the DBI form: reality of the action enforces $0<\dot\phi^2<1$, implying $w_\phi=\dot\phi^2-1\in(-1,0)$ and ruling out phantom crossing, in line with the single-field \emph{no-go theorem}~\cite{Vikman2}. Second, the negative sign of $w_a^{\rm eff}$ encodes the \emph{thawing} character of the dynamics, in which the field is frozen near $w_\phi \approx -1$ at high redshift by strong Hubble friction, and only begins to depart from the de Sitter limit at late times as the Hubble rate decreases~\cite{Caldwell,Scherrer,Linder1}. Third, the magnitude of the dynamical departure is dataset-dependent. {\color{blue}Pantheon$+$ gives the most active dynamics (largest $|w_a^{\rm eff}|$ and $w_0^{\rm eff}$ furthest from $-1$), consistent with its largest $\lambda_{\rm ini}$ in Table~\ref{TableI}, followed by Union3 and DES-SN5YR.} The DBI kinematic constraint $\dot{\phi}^2 < 1$ of Eq.~\eqref{4} ensures that these trajectories remain physically bounded and self-consistent throughout cosmic history.\\
 
\begin{figure*}[htbp]
    \centering
    \includegraphics[width=\textwidth]{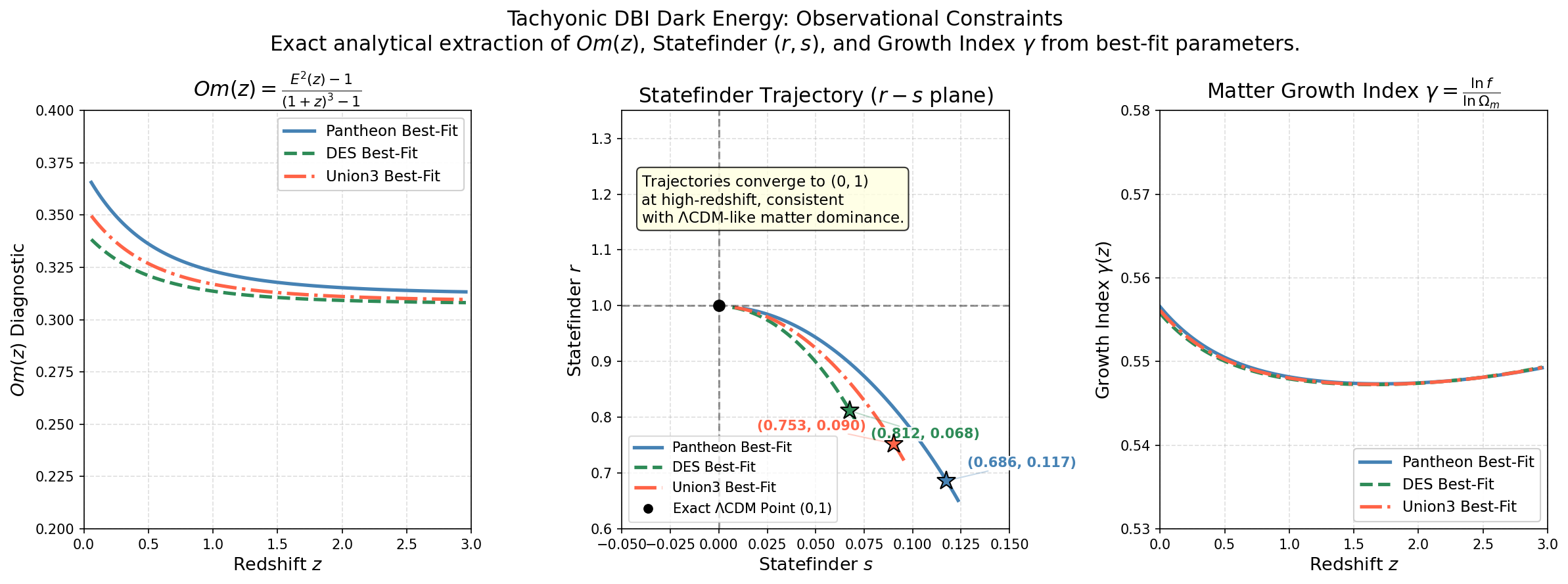}
    \caption{Exact analytical diagnostics of the dynamical nature of the tachyonic DBI dark energy model, evaluated at the posterior-mean best-fit parameters for each supernova compilation combined with the baseline probes. \textit{Left:} $\text{Om}(z)$ diagnostic (Eq.~\eqref{70}).  \textit{Centre:} Statefinder trajectory in the $(s, r)$ plane (Eq.~\eqref{71}). \textit{Right:} Matter growth index $\gamma(z)$ (Eq.~\eqref{72}).}
    \label{Fig6b}
\end{figure*}

\textit{Om$(z)$ diagnostic, Statefinder hierarchy, and growth index:} The three panels of Fig.~\ref{Fig6b} provide complementary diagnostics based on exact analytical expressions evaluated at the posterior-mean parameters.\\
 
\underline{\textit{Om$(z)$ diagnostic:}} The $\text{Om}(z)$ statistic~\cite{Sahni1} is defined as
\ben
\text{Om}(z) = \frac{E^2(z) - 1}{(1+z)^3 - 1},
\label{70}
\een
where $E(z) \equiv H(z)/H_0$. For a flat $\Lambda$CDM universe, $\text{Om}(z) = \Omega_{m0}$ exactly, a positive slope indicates phantom-like behavior ($w < -1$), while a negative slope signals quintessence ($w > -1$). {\color{blue}The left panel of Fig.~\ref{Fig6b} shows that all three compilations give $\text{Om}(z)$ above their $\Omega_{m0}$ at low redshift (Table~\ref{TableI}).} {\color{blue}All three fits exhibit a negative slope, with $\text{Om}(z)$ decreasing from $\approx0.366$ (Pantheon$+$), $0.350$ (Union3) and $0.338$ (DES-SN5YR) at $z\simeq0$} toward $\approx 0.30$--$0.31$ at $z = 3$, directly confirming quintessence-like dynamics ($w_\phi > -1$) in a model-independent manner.\\
 
\underline{\textit{Statefinder hierarchy:}} 
The Statefinder pair $(r, s)$~\cite{Sahni2} is constructed from the third time derivative of the scale factor:
\ben
r \equiv \frac{\dddot{a}}{aH^3}, \qquad
s \equiv \frac{r - 1}{3\bm{\big(}q - \tfrac{1}{2}\bm{\big)}},
\label{71}
\een
where $q \equiv -\ddot{a}/(aH^2)$ is the deceleration parameter. For $\Lambda$CDM, $(r, s) = (1, 0)$ exactly at all epochs. The central panel of Fig.~\ref{Fig6b} displays the Statefinder trajectories in the $(s, r)$ plane, with the $\Lambda$CDM fixed point marked by a filled circle. All three trajectories originate near $(1, 0)$ at early times reflecting the model's $\Lambda$CDM-like attractor behavior deep in matter domination and evolve into the quintessence region as dark energy becomes dynamically significant. The present-epoch coordinates $(r_0, s_0)$ are:
\begin{itemize}
    \item {\color{blue}Pantheon$+$: $(r_0,\,s_0) = (0.686,\; 0.118)$,}
    \item {\color{blue}DES-SN5YR: $(r_0,\,s_0) = (0.812,\; 0.068)$,}
    \item {\color{blue}Union3: $(r_0,\,s_0) = (0.753,\; 0.090)$,}
\end{itemize}
all lying strictly below the $\Lambda$CDM point in the $(r, s)$ plane therefore unambiguously placing the model in the thawing quintessence region. The spread among the three compilations in the $s$-direction is proportional to the inferred initial potential slope $\lambda_{\rm ini}$, {\color{blue}with Pantheon$+$ exhibiting the greatest Statefinder deviation,} consistent with its broader posterior for dynamical dark energy.\\
 
\underline{\textit{Growth index:}} The matter growth index $\gamma(z)$, defined through
\ben
f \equiv \frac{d\ln\delta_m}{d\ln a} = \Omega_m(z)^{\gamma(z)},
\label{72}
\een
provides a direct measure of the growth rate normalized by the instantaneous matter fraction. For standard general relativity with $\Lambda$CDM, $\gamma_{\rm GR} \approx 6/11 \approx 0.545$~\cite{Linder}. The right panel of Fig.~\ref{Fig6b} shows that across all three compilations $\gamma(z)$ is nearly compilation-independent, {\color{blue}beginning at $\gamma_0 \approx 0.555$--$0.556$ at $z = 0$, passing through a shallow minimum $\gamma\approx0.547$ near $z \sim 1.7$, and recovering toward $\gamma \approx 0.549$ at $z = 3$.} This mild but systematic upward shift relative to the pure $\Lambda$CDM expectation is a direct imprint of the tachyonic sound speed modifying the effective Hubble friction in the matter perturbation equation. Since $c_{s,\phi}^2 < 1$ alters the feedback between dark-energy pressure gradients and matter growth (Sec.~\ref{S5}), the growth index provides an independent observational handle on dark-energy microphysics that will become accessible to next-generation surveys achieving sub-percent precision in growth-rate measurements.
 
Taken together, Figs.~\ref{Fig6a} and \ref{Fig6b} constitute a rigorous verification that the tachyonic DBI model is not a cosmological constant in disguise. Although the standard cosmological parameters exhibit near-perfect posterior degeneracy with $\Lambda$CDM at current precision (Sec.~\ref{SECVA}), the four independent diagnostics presented here namely the CPL EoS trajectory, the $\text{Om}(z)$ slope, the Statefinder displacement, and the growth-index offset, all point consistently toward a thawing quintessence interpretation, fully in accord with the theoretical prediction of the autonomous dynamical system of Eq.~\eqref{31}. As next-generation surveys such as Euclid and the final DESI releases achieve percent-level precision in these quantities, these signatures will become directly testable and may provide the first definitive observational constraints on the tachyonic DBI dark energy scenario beyond the background-level degeneracy with $\Lambda$CDM.

\section{Conclusion}
\label{S6}
 
In this work, we have investigated the observational viability of a clustering tachyonic DBI dark-energy scenario through a comprehensive multi-probe Bayesian analysis. The model is motivated by the need to address the fine-tuning and coincidence problems of $\Lambda$CDM through a theoretically grounded, ghost-free noncanonical scalar field with a subluminal sound speed $c_{s,\phi}^2 < 1$, which enables dark-energy clustering on subhorizon scales, a physical mechanism absent in both the cosmological constant and canonical quintessence. Our analysis yields the following key results:
 
\begin{enumerate}
    \item {\color{blue}The tachyonic DBI model agrees with $\Lambda$CDM in $\omega_b$ and $\sigma_{8,0}$ and within $\sim2\sigma$ in $H_0$ and $\Omega_{m0}$ for all three SN compilations, preferring $H_0\simeq67.4$--$67.8$ against $\simeq68.6\,{\rm km\,s^{-1}\,Mpc^{-1}}$ and a correspondingly higher $\Omega_{m0}$. The noncanonical sound speed and dark-energy clustering remain undetectable at the present $f\sigma_8$ precision of $\sim 5$--$20\%$, and both models pass every posterior predictive test.}
 
    \item {\color{blue}The Bayesian evidence favours the tachyonic DBI model in all three compilations, moderately for Pantheon$+$ ($\Delta\ln\mathcal{Z}\simeq+3.38$) and inconclusively for DES-SN5YR ($+0.64$) and Union3 ($+0.66$). The tachyonic model also has the lower $\chi^2_{\rm min}$ everywhere, is preferred by DIC for Pantheon$+$ ($-4.10$) and DES-SN5YR ($-2.00$) and by AIC for Pantheon$+$ ($-3.40$); only the BIC prefers $\Lambda$CDM. The Kullback--Leibler information gain is nearly the same in both models ($\Delta D_{\rm KL}$ between $-0.72$ and $+0.80$ nats), so the extra parameters cost little prior volume: $\lambda_{\rm ini}$ is constrained, while $\Gamma$ remains prior-dominated.}
 
    \item The dark-energy perturbation source $\Omega_{\rm de}|\delta_{\rm de}|$ remains four to six orders of magnitude below the matter source across all redshifts, so the tachyonic fluid does not directly deepen gravitational-potential wells. Rather, pressure gradients associated with $c_{s,\phi}^2 < 1$ subtly modify the Hubble friction in the matter growth equation, {\color{blue}shifting $S_8$ upward by only $\approx0.5\sigma$ with an essentially unchanged width.} The model is fully consistent with KiDS-1000 and DES-Y3 weak-lensing constraints within $1\sigma$.
 
    \item Despite the background-level degeneracy with $\Lambda$CDM, four independent model-agnostic diagnostics confirm the genuinely dynamical nature of the field: the CPL projection yields $w_a^{\rm eff} < 0$ in every dataset (thawing behavior); the $\text{Om}(z)$ slope is negative (quintessence); the statefinder coordinates $(r_0, s_0)$ lie strictly off the $\Lambda$CDM fixed point $(1,0)$; and {\color{blue}the growth index $\gamma_0 \approx 0.555$--$0.556$ exceeds} the general-relativistic prediction $\gamma_{\rm GR} \approx 6/11$.
 
    \item The consistently inferred $\Omega_{\rm de,ini} \approx \mathcal{O}(10^{-9})$ at $z \approx 1100$ confirms that attractor solutions drive a wide range of initial conditions toward the same late-time trajectory, with the near equality of matter and dark-energy densities today emerging dynamically from the noncanonical DBI kinetic structure rather than by coincidence.

    \item {\color{red}Both $\Lambda$CDM and the tachyonic DBI model provide statistically acceptable descriptions of the combined observational dataset, which demonstrates the overall viability of our constrained cosmological scenarios. Nevertheless, the phenomenological $w_0w_a$CDM parametrization provides the best fit to the DESI DR2~\cite{DESI, DESI2} observations, owing to its greater freedom in describing the time evolution of the dark-energy EoS and its ability to accommodate phantom-divide crossing. Such flexibility is absent in both $\Lambda$CDM, with its constant $w=-1$, and our tachyonic DBI model, whose theoretical constraint $w_\phi\geq -1$ prevents phantom crossing. Consequently, although both models remain consistent with the combined dataset, their restricted dark-energy dynamics may limit their ability to reproduce the full phenomenological preference indicated by DESI DR2.}
 
\end{enumerate}
 
Although we have not make a consistent statistical analysis with respect to $w$CDM or CPL parametrization but we do want to make sure that our analysis is robust, in that sense we make a comparison for fitting DESI DR2 BAO data against $w_0w_a$CDM model but what we got is that CPL parameterization is consistently preferred. Nevertheless we have a reason to not consider the $w_0w_a$CDM model and that reason is - the $w$CDM and CPL parametrizations are background level constructs. They specify an equation of state without a Lagrangian, a physical sound speed, or perturbation equations. By convention, these fluids are assigned $c_s^2=1$, forcing dark energy to be perfectly smooth on subhorizon scales, an arbitrary prescription with no field-theoretic justification for $w\neq-1$~\cite{Garriga, Picon1}. Whereas, the Tachyonic DBI model is a complete theory with a well-defined action, a time-dependent sound speed $c_{s,\phi}^2<1$, and a self-consistent perturbation sector. The perturbation observables that form the physical core of this work - the $f\sigma_8$ growth history, the dark-energy density contrast $|\delta_{\rm de}|$, and the growth index $\gamma$ are all sourced by $c_{s,\phi}^2<1$ and are identically absent in any smooth $w$CDM fluid. A model-selection comparison at the background level alone would therefore be potentially misleading. The natural null hypothesis is $\Lambda$CDM, which has zero dynamical freedom, $c_s^2=1$, and no dark-energy perturbations. At the background level, however, the CPL comparison is not deferred- Fig.~\ref{Fig6a} projects the tachyonic $w_\phi(z)$ trajectory onto the CPL plane, yielding $(w_0^{\rm eff},\,w_a^{\rm eff})$ values directionally consistent with the DESI DR2 preference for $w_0>-1$, $w_a<0$~\cite{DESI, DESI2} which is precisely the thawing-quintessence regime that this class of models naturally occupies. 
 
The current implementation has several clearly identified limitations. The tracking parameter $\Gamma$ remains largely unconstrained which reflects insensitivity to the higher-order curvature of the tachyonic potential at current observational precision. This unconstrained prior volume inflates the Bayes-factor penalty, although it does not signal a genuine prediction failure. The use of CMB distance priors, while computationally efficient, discards acoustic peak and damping-tail shape information that could probe early-universe tachyonic dynamics if any were present. The perturbation analysis is restricted to the linear regime and a single representative Fourier mode $k = 0.05\,h\,\mathrm{Mpc}^{-1}$, {\color{blue}and the inferred $\Omega_{\rm de,ini}$ agrees across all three SN compilations once they are treated as relative distance indicators.} These limitations define a concrete roadmap. Integrating the tachyonic fluid into a Boltzmann solver (\textsc{class} or \textsc{camb}) would enable direct confrontation with the full Planck CMB likelihood including lensing. Including weak-lensing data at the full likelihood level would provide the sensitivity needed to test the $S_8$-broadening mechanism identified here. Other than that adapting $N$-body codes to evolve a noncanonical dark-energy fluid alongside cold dark matter would extend the analysis beyond the linear regime. Most ambitiously, incorporating an Early Dark Energy component active near matter-radiation equality alongside the late-time tachyonic dynamics~\cite{Poulin, Niedermann, Guendelman} offers a theoretically well-motivated route toward a unified framework capable of simultaneously addressing both the $H_0$ and $S_8$ tensions, which we identify as the primary scientific priority for the next phase of this research program.

Another key limitation is that recent observations have been argued to favor an epoch of phantom-like evolution, a behavior that is absent by construction in the tachyonic DBI scenario. Thus, despite its theoretical appeal, our model remains largely consistent with $\Lambda$CDM. Consequently, while we have pursued the best possible fit to the DESI DR2 measurements within this framework, the level of agreement achievable is ultimately limited by the model's restriction to $w_\phi\geq -1$. In any case, having a phantom-crossing field that simultaneously avoid gradient instabilities and maintain a subluminal sound speed within a single scalar field $k$-essence model are difficult to realize in a consistent theory. These findings satisfy the no-go theorem~\cite{Vikman2}, which establishes that a single, minimally coupled  $k$-essence field with a well-defined Lagrangian cannot cross the phantom divide without encountering gradient instabilities, and the tachyonic DBI model is no exception. Nevertheless, $k$-essence represents only a subclass of the broader Horndeski family~\cite{Horndeski}, where genuinely higher-derivative interactions can, in principle, permit stable phantom crossing while preserving a subluminal sound speed. Exploring this direction would move beyond the $k$-essence sector considered here and may provide additional flexibility to improve the DESI DR2 fit. A systematic investigation of such higher-order Horndeski interactions, however, lies outside the scope of the present work and is deferred to future studies.

\begin{acknowledgments}
G.M. (an IUCAA Associate) and S.G. acknowledge the Inter-University Centre for Astronomy and Astrophysics (IUCAA), Pune, India, for facilitating part of this work during their visit. G.M. and E.G. thank the COST Association (CA21136 CosmoVerse), European Union, for the opportunity to participate in this international association as group members. E.G. also thanks COST  Association  (CA23130 BridgeQG), the European Union, for the opportunity to participate in this international association as a group member.  K.B. appreciates partial support by JSPS KAKENHI Grant Nos.\ 24KF0100 and 25KF0176 and by the Competitive Research Funds for Fukushima University Faculty (25RK011). A.B. acknowledges partial support from the RUSA project at the University of Calcutta. G.M. extends his gratitude to all undergraduate, postgraduate, and doctoral students, along with his teachers, collaborators, well-wishers, and his son (Aditya), whose support has significantly enhanced his academic journey. S.G. thanks his mother and elder sister for their unwavering support and love.\\

\textbf{Conflicts of interest:} The authors declare no conflicts of interest.\\

\textbf{Funding information:} Not available.\\

\textbf{Data availability:} The data used in this study are readily accessible from public sources for validation of our model; however, we did not generate any new data sets for this research.\\

\end{acknowledgments}

\appendix
\section{Hubble cosmic chronometer dataset}\label{A1}

In our observational analysis, we include measurements of the Hubble expansion rate $H(z)$ obtained with the cosmic chronometer (CC) technique, i.e., the differential-age method based on passively evolving galaxies. The method exploits
\begin{equation}
H(z) = -\frac{1}{1+z}\,\frac{dz}{dt},
\end{equation}
so that $H(z)$ can be inferred from estimates of $dt/dz$ without assuming a specific background cosmological model beyond the FLRW relation between time and redshift.

Table~\ref{TableIII} lists the CC compilation used in this work, covering the redshift range $0.07\le z\le 1.965$. The quoted uncertainties $\sigma_H$ are treated as Gaussian $1\sigma$ errors, and are commonly done for heterogeneous CC compilations. 

\begin{table}[ht]
    \centering
    \caption{Cosmic-chronometer measurements of the Hubble expansion rate used in this work. Units: $H(z)$ in $\mathrm{km\,s^{-1}\,Mpc^{-1}}$.}
    \label{TableIII}
    \begin{tabular}{c c c c}
        \hline
        $z$ & $H(z)$ & $\sigma_H$ & Ref. \\
        \hline
        0.07   & 69.0  & $\pm 19.68$ & \cite{Zhang} \\
        0.09   & 69.0  & $\pm 12.0$  & \cite{Simon} \\
        0.12   & 68.6  & $\pm 26.2$  & \cite{Zhang} \\
        0.17   & 83.0  & $\pm 8.0$   & \cite{Simon} \\
        0.179  & 75.0  & $\pm 4.0$   & \cite{Moresco1} \\
        0.199  & 75.0  & $\pm 5.0$   & \cite{Moresco1} \\
        0.2    & 72.9  & $\pm 29.6$  & \cite{Zhang} \\
        0.27   & 77.0  & $\pm 14.0$  & \cite{Simon} \\
        0.28   & 88.8  & $\pm 36.6$  & \cite{Zhang} \\
        0.352  & 83.0  & $\pm 14.0$  & \cite{Moresco1} \\
        0.38   & 83.0  & $\pm 13.5$  & \cite{Moresco2} \\
        0.4    & 95.0  & $\pm 17.0$  & \cite{Simon} \\
        0.4004 & 77.0  & $\pm  10.2$ & \cite{Moresco2} \\
        0.425  & 87.1  & $\pm 11.2$  & \cite{Moresco2} \\
        0.445  & 92.8  & $\pm 12.9$  & \cite{Moresco2} \\
        0.47   & 89.0  & $\pm 50.0$  & \cite{Ratsimbazafy} \\
        0.4783 & 80.9  & $\pm  9.0$  & \cite{Moresco2} \\
        0.48   & 97.0  & $\pm 62.0$  & \cite{Stern} \\
        0.593  & 104.0 & $\pm 13.0$  & \cite{Moresco1} \\
        0.68   & 92.0  & $\pm  8.0$  & \cite{Moresco1} \\
        0.75   & 98.8  & $\pm 33.6$  & \cite{Borghi} \\
        0.781  & 105.0 & $\pm 12.0$  & \cite{Moresco1} \\
        0.875  & 125.0 & $\pm 17.0$  & \cite{Moresco1} \\
        0.88   & 90.0  & $\pm 40.0$  & \cite{Stern} \\
        0.9    & 117.0 & $\pm 23.0$  & \cite{Simon} \\
        1.037  & 154.0 & $\pm 20.0$  & \cite{Moresco1} \\
        1.3    & 168.0 & $\pm 17.0$  & \cite{Simon} \\
        1.363  & 160.0 & $\pm 33.6$  & \cite{Moresco3} \\
        1.43   & 177.0 & $\pm 18.0$  & \cite{Simon} \\
        1.53   & 140.0 & $\pm 14.0$  & \cite{Simon} \\
        1.75   & 202.0 & $\pm 40.0$  & \cite{Simon} \\
        1.965  & 186.5 & $\pm 50.4$  & \cite{Moresco3} \\
        \hline
    \end{tabular}
\end{table}

\section{DESI DR2 BAO dataset}
\label{A2}

Baryon acoustic oscillations (BAO) provide a standard ruler set by the sound horizon at the drag epoch, $r_d$, and therefore constrain the late-time geometry through distances measured at multiple redshifts. In this work we use the BAO measurements reported by the DESI Collaboration in Data Release~2 (DR2)~\cite{DESI,DESI2}.

DESI reports BAO constraints in terms of the dimensionless ratios $D_V(\tilde z)/r_d$ (isotropic BAO) and/or the anisotropic combinations $D_M(\tilde z)/r_d$ and $D_H(\tilde z)/r_d$, where
\ben
D_M(z)=(1+z)D_A(z),\qquad D_H(z)=\frac{c}{H(z)},
\label{B1}
\een
and the volume-averaged distance is
\ben
D_V(z)=\bm{\Big[}z\,D_M^2(z)\,D_H(z)\bm{\Big]}^{1/3}
\label{B2}
\een.
Here $\tilde z$ denotes the effective redshift of each DESI tracer bin.

The sound horizon at the drag epoch, $r_d$, is computed from
\ben
r_d\equiv r_s(z_d)=\int_{z_d}^{\infty}\frac{c_s(z)}{H(z)}\,dz,
\label{B3}
\een
with the baryon-photon sound speed
\ben
&& c_s(z)=\frac{c}{\sqrt{3\bm{\big[}1+R_b(z)\bm{\big]}}} \n && R_b(z)=\frac{3\rho_b}{4\rho_\gamma}=\frac{3\Omega_b}{4\Omega_\gamma}\,\frac{1}{1+z}
\label{B4}
\een.
For the drag redshift $z_d$ we adopt the fitting formula of Eisenstein \& Hu~\cite{EisensteinHu1998},
\ben
z_d=1291\,\frac{(\Omega_{\mathrm{m0}}h^2)^{0.251}}{1+0.659(\Omega_{\mathrm{m0}}h^2)^{0.828}}\bm{\big[}1+b_1\,(\omega_{\mathrm{b}})^{b_2}\bm{\big]},
\label{B5}
\een
where $h\equiv H_0/(100\,\mathrm{km\,s^{-1}\,Mpc^{-1}})$, $\omega_{\mathrm{b}}\equiv\Omega_bh^2$, and
\ben
&&b_1=0.313\,(\Omega_{\mathrm{m0}}h^2)^{-0.419}\bm{\big[}1+0.607(\Omega_{\mathrm{m0}}h^2)^{0.674}\bm{\big]}
\n &&
b_2=0.238\,(\Omega_{\mathrm{m0}}h^2)^{0.223}
\label{B6}
\een.

For each effective redshift the DESI DR2 measurements is provided in Table~\ref{TableIV}.

\begin{table*}[ht]
    \centering
    \renewcommand{\arraystretch}{1.4}
    \caption{DESI BAO measurements from DR2 used in this work. The $z=0.295$ entry is an isotropic BAO constraint reported as $D_V/r_d$, while the remaining redshifts include anisotropic BAO information through $(D_M/r_d,\, D_H/r_d)$.}
    \label{TableIV}
    \resizebox{\textwidth}{!}{%
    \begin{tabular}{cccccccc}
        \toprule
        $\tilde z$ & \textbf{0.295} & \textbf{0.510} & \textbf{0.706} & \textbf{0.934} & \textbf{1.321} & \textbf{1.484} & \textbf{2.330} \\ \midrule
        $D_V(\tilde z)/r_d$ & $7.942 \pm 0.075$ & $12.720 \pm 0.099$ & $16.050 \pm 0.110$ & $19.721 \pm 0.091$ & $24.252 \pm 0.174$ & $26.055 \pm 0.398$ & $31.267 \pm 0.256$ \\ 
        $D_M(\tilde z)/r_d$ & --- & $13.588 \pm 0.167$ & $17.351 \pm 0.330$ & $21.576 \pm 0.152$ & $27.601 \pm 0.318$ & $30.512 \pm 0.760$ & $38.988 \pm 0.531$ \\ 
        $D_H(\tilde z)/r_d$ & --- & $21.863 \pm 0.425$ & $19.455 \pm 0.330$ & $17.641 \pm 0.193$ & $14.176 \pm 0.221$ & $12.817 \pm 0.516$ & $8.632 \pm 0.101$ \\ 
        \bottomrule
    \end{tabular}}
\end{table*}

\section{2D Spatial Correlation Function and Intrinsic Clustering}
\label{A3}

The primary motivation for visualizing the 2D spatial two-point correlation function, $\xi(r_p, \pi)$, is to explicitly demonstrate how the modified perturbation dynamics of the tachyonic DBI model manifest in large-scale structure observables. While the 1D marginalized posteriors and 2D confidence contours provide a statistical measure of model viability, the uncorrected $\xi(r_p, \pi)$ plots offer a direct physical visualization of RSD.

Peculiar velocities of galaxies falling into gravitational potential wells squash the clustering signal along the line-of-sight at large scales (the Kaiser effect). The magnitude of this effect is governed by the linear growth rate, $f$. Because the tachyonic fluid's clustering ($c_s^2 < 1$) modifies the Hubble friction and the growth of matter perturbations, it leaves a distinct, intrinsic signature on the RSD profile prior to any geometric transformations.\\

\textit{Decoupling the Alcock-Paczynski (AP) Effect}:

It is important to note that translating observed redshifts and angles into comoving physical separations requires assuming a cosmological background. Differences in the Hubble parameter $H(z)$ and angular diameter distance $D_A(z)$ between models induce apparent geometric distortions (the AP effect). 

\begin{figure*}[t]
    \centering
    \begin{subfigure}[t]{0.48\textwidth}
        \centering
        \includegraphics[width=\linewidth]{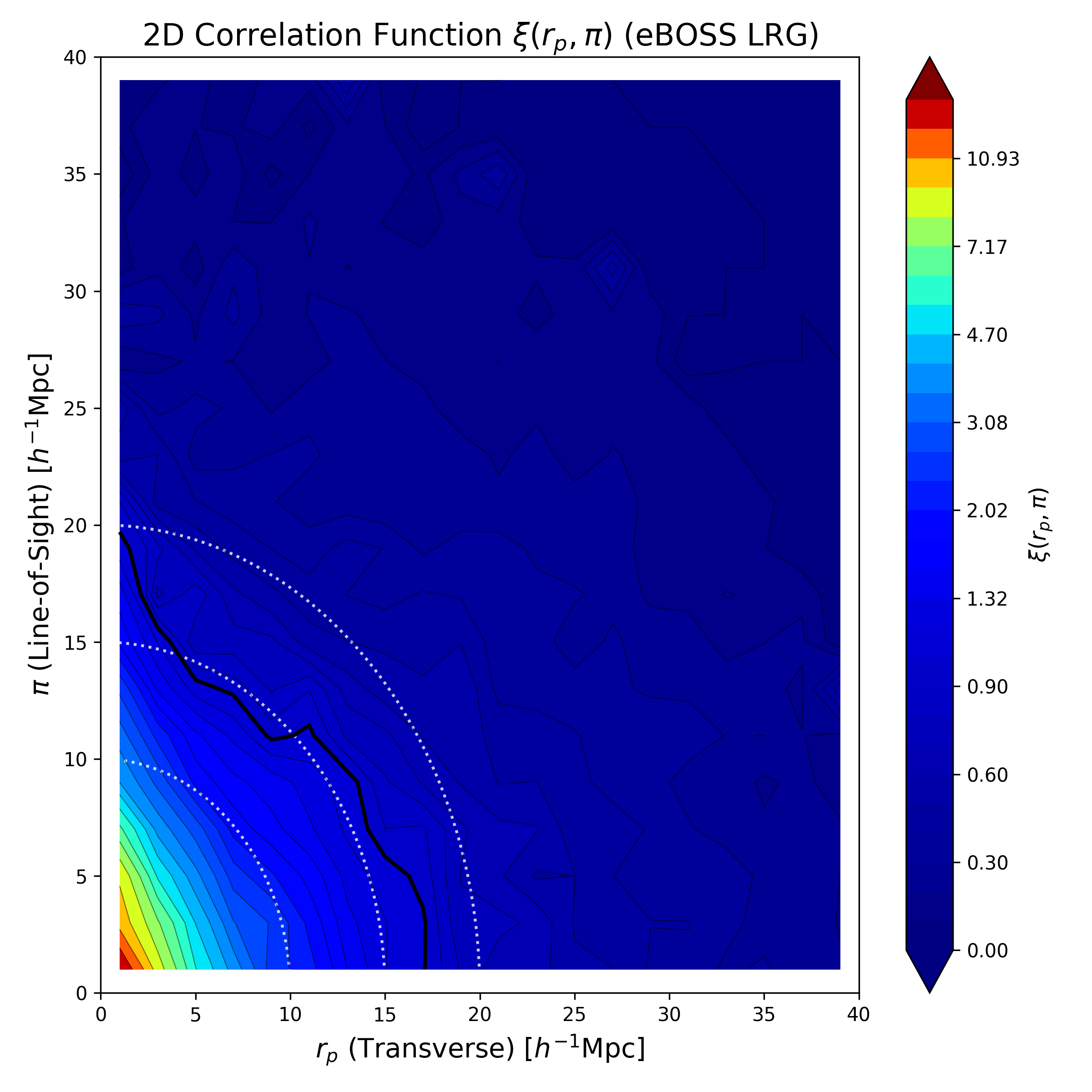}
        \caption{Tachyonic DBI model}
        \label{FigA4T}
    \end{subfigure}\hfill
    \begin{subfigure}[t]{0.48\textwidth}
        \centering
        \includegraphics[width=\linewidth]{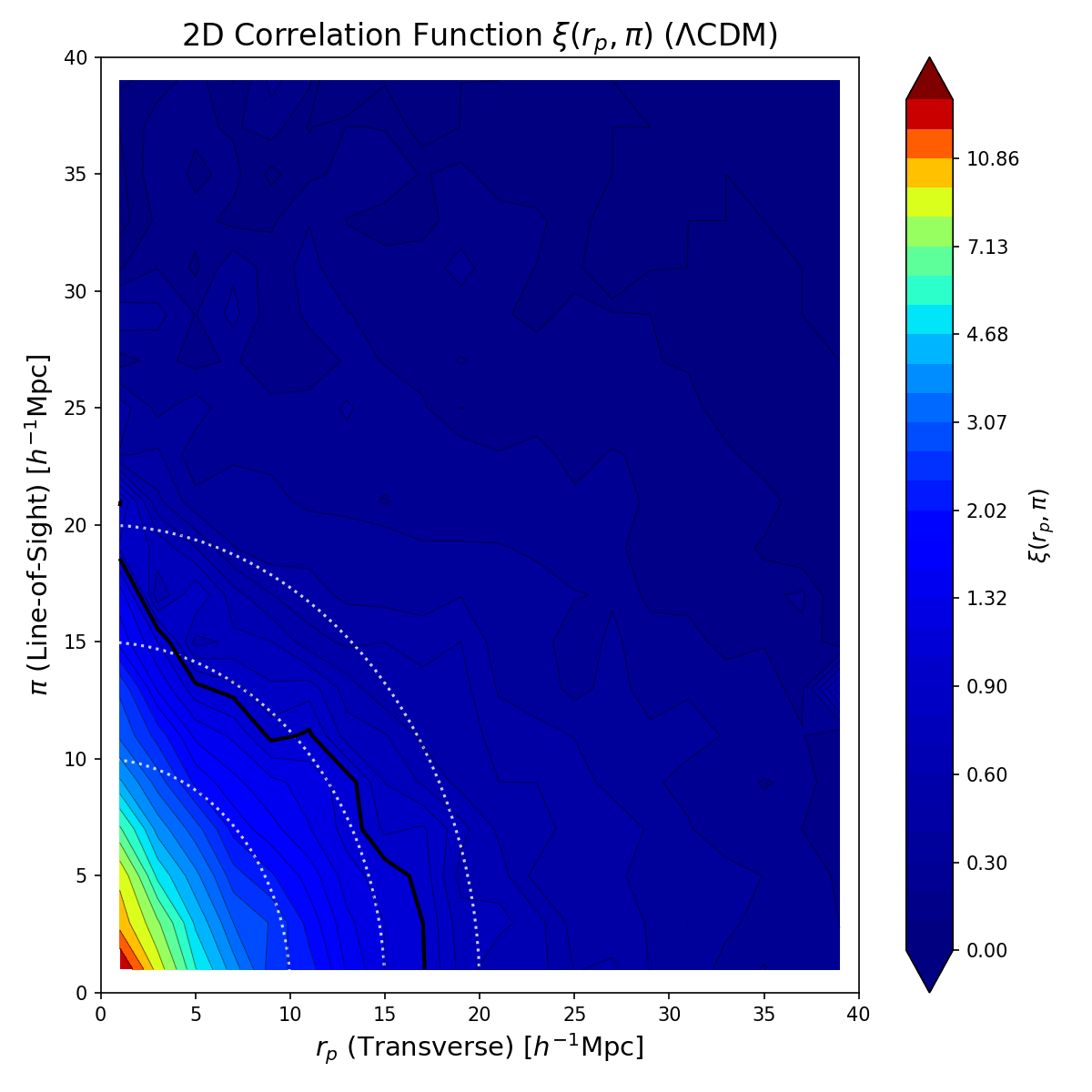}
        \caption{$\L$CDM model}
        \label{FigA4L}
    \end{subfigure}
    \caption{The 2D spatial correlation function
$\xi(r_p, \pi)$ evaluated at the best-fit parameters for the
DES-SN5YR joint dataset. The color mapping
indicates the amplitude of clustering at intermediate
to non-linear scales.}
    \label{FigA4LT}
\end{figure*}

To visually isolate the dynamical physics from geometric artifacts, the correlation functions plotted in Figs.~\ref{FigA4LT} and \ref{Fig1A} in the main text specifically map the intrinsic RSD signal \textit{without} applying the AP geometric distortions. The necessary AP adjustments are handled purely analytically within our likelihood pipeline [Eq.~\eqref{57}], allowing us to maintain a clean visual comparison of how structures physically assemble in each scenario.\\

\textit{Validation of the AP Rescaling:}

To verify that the first-order AP rescaling adopted in Eq.~\eqref{57} is adequate for the parameter ranges inferred in this work, we compute the fractional deviation
\ben
\mathcal{R}(z) \equiv 
\frac{H_{\mathrm{th}}(z)\,D_{A,\mathrm{th}}(z)}
{H_{\mathrm{fid}}(z)\,D_{A,\mathrm{fid}}(z)} - 1,
\label{D1}
\een
evaluated at the posterior mean parameters reported in Table~\ref{TableI} for the tachyonic DBI model and at the corresponding posterior mean parameters for the $\L$CDM baseline, for all three SN compilations. The fiducial cosmology is $\Lambda$CDM with $\Omega_{m0}^{\mathrm{fid}}$ value is consistent with the fiducial model adopted in the $f\sigma_8$ compilation of~\cite{Nesseris,Kazantzidis}.

\begin{figure*}[t]
    \centering
    \includegraphics[width=\textwidth]{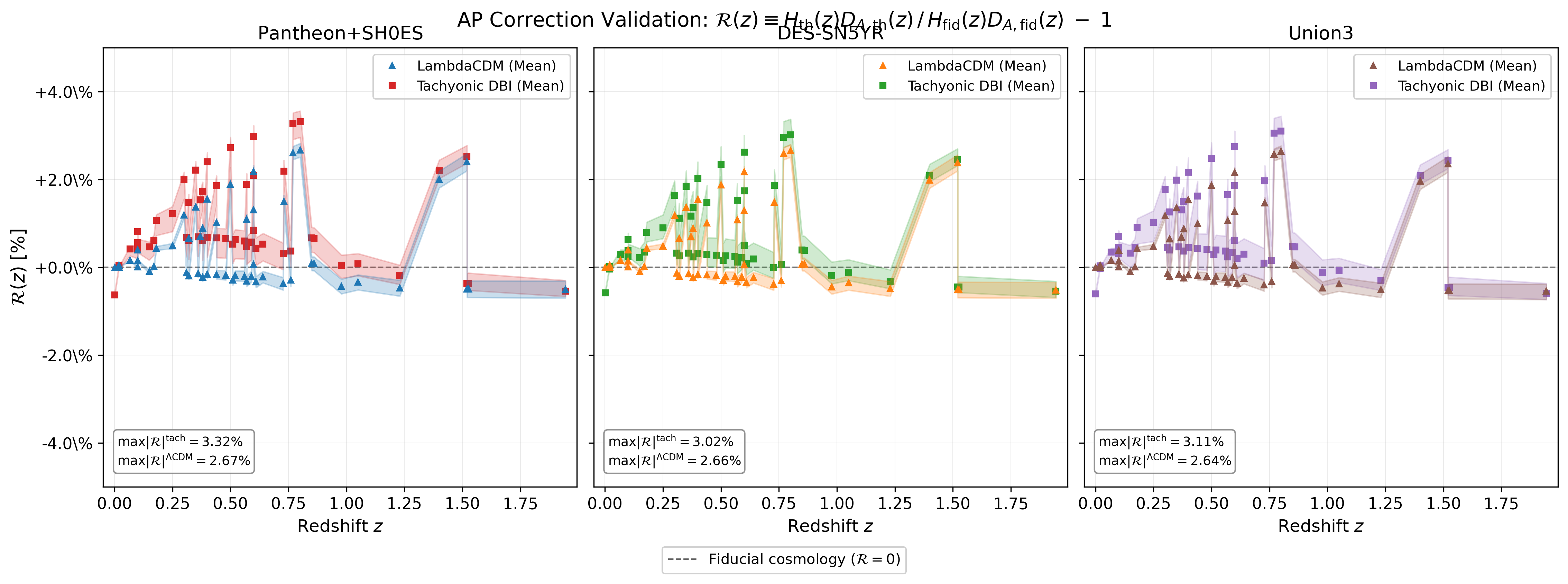}
    \caption{A posteriori validation of the first-order AP rescaling of Eq.~\eqref{57}, evaluated using the posterior mean parameters of Table~\ref{TableI}. Each panel shows the fractional deviation $\mathcal{R}(z)$ defined in Eq.~\eqref{D1} for the tachyonic DBI model (red) and the $\L$CDM model (blue), with $1\sigma$ posterior bands (shaded), for each of the three SN compilations. The tachyonic model satisfies $|\mathcal{R}(z)|\leq 3.32\%$ across all datasets and the full redshift range shown. The higher-order AP corrections~\cite{Ballinger} enter at $\mathcal{O}(\mathcal{R}^2)\lesssim 0.10\%$, negligible relative to the observational uncertainties~\cite{Nesseris,Kazantzidis}.}
    \label{Fig15}
\end{figure*}

Fig.~\ref{Fig15} shows $\mathcal{R}(z)$ together with $1\sigma$ posterior bands across the redshift range $z\in[0.05,\,1.5]$ relevant to the $f\sigma_8$ data. For the tachyonic DBI model, we find $|\mathcal{R}(z)|\leq 3.32\%$ across all three compilations. The higher-order corrections to the full AP transformation — specifically the volume factor $\alpha_\perp^2\alpha_\parallel$ and the spectral-index term~\cite{Ballinger} — enter at $\mathcal{O}(\mathcal{R}^2)$, which for $|\mathcal{R}|\lesssim 3.5\%$ amounts to at most $\sim 0.10\%$ in the predicted $f\sigma_8$. Since the typical fractional uncertainty in the compiled measurements is $\sigma_{f\sigma_8}/f\sigma_8 \sim 5$--$20\%$~\cite{Nesseris,Kazantzidis}, the residual systematic from the first-order approximation is negligible and does not affect our conclusions.

\section{Pantheon$+$ joint constraints}
\label{A4}

In this appendix, we present the marginalized constraints obtained from the full joint likelihood
\ben
&&\mathcal{L}_{\mathrm{tot}}\propto \exp\!\Big[-\frac12\big(\chi^2_{\mathrm{CMB}}+\chi^2_{\mathrm{BAO}}+\chi^2_{f\sigma_8}+\chi^2_{\mathrm{CC}}\n &&+\chi^2_{\mathrm{SN}}\big)\Big],
\label{C1}
\een
{\color{blue}specialized to the case in which the SN component is given by the Pantheon$+$ Hubble diagram without the SH0ES calibration.}

\subsection*{Marginalized posterior constraints}

\begin{figure*}[t]
    \centering
    \begin{subfigure}[t]{0.48\textwidth}
        \centering
        \includegraphics[width=\textwidth]{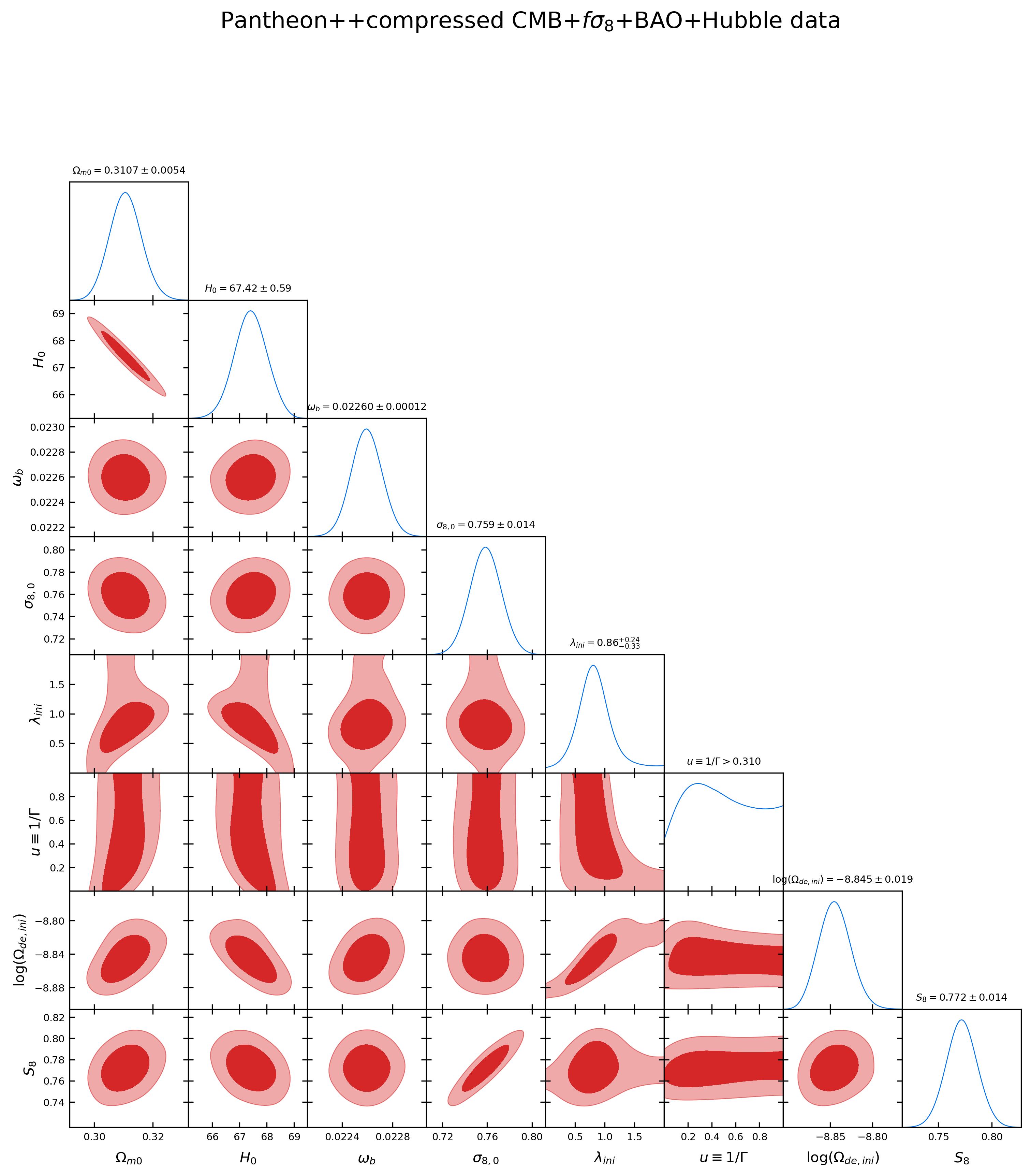}
        \caption{Tachyonic DBI model.}
        \label{FigA1T}
    \end{subfigure}\hfill
    \begin{subfigure}[t]{0.48\textwidth}
        \centering
        \includegraphics[width=\textwidth]{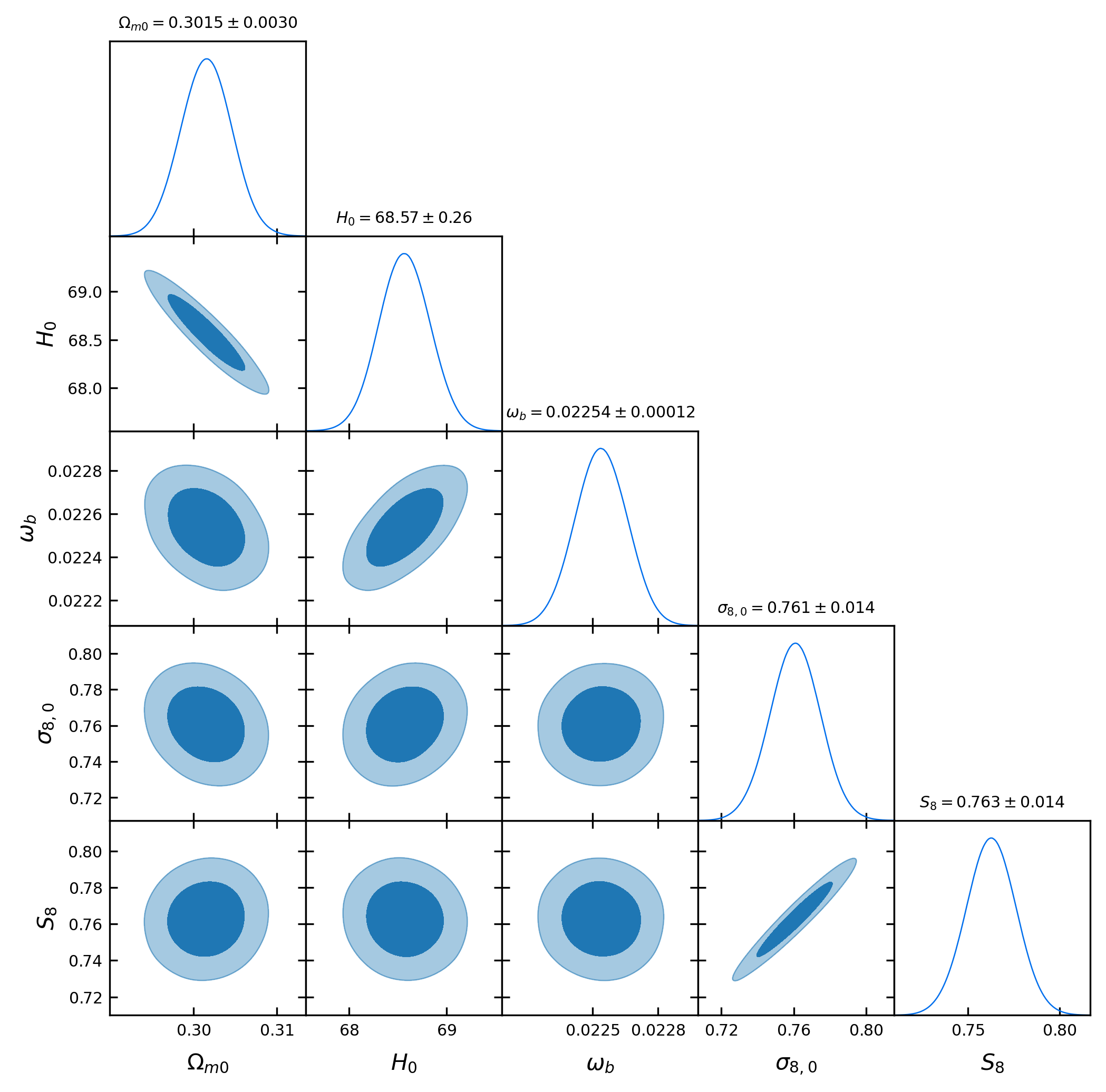}
        \caption{$\Lambda$CDM model.}
        \label{FigA1L}
    \end{subfigure}
    \caption{Marginalized constraints for {\color{blue}Pantheon$+$} combined with CMB distance priors, DESI DR2 BAO, CC, and $f\sigma_8$. Contours show the 68\% and 95\% confidence regions.}
\end{figure*}

Figures~\ref{FigA1T} and \ref{FigA1L} show the 1D and 2D marginalized posteriors for the same {\color{blue}Pantheon$+$}$+$CC$+$BAO$+$CMB distance-prior$+$ $f\sigma_8$ dataset, comparing the Tachyonic DBI model (Fig.~\ref{FigA1T}) against the baseline $\Lambda$CDM scenario (Fig.~\ref{FigA1L}).

For convenience, Table~\ref{TableV} summarizes the 68\% marginalized constraints for the {\color{blue}Pantheon$+$} joint dataset, shown side-by-side for the Tachyonic DBI model (Fig.~\ref{FigA1T}) and the baseline $\Lambda$CDM model (Fig.~\ref{FigA1L}).

\begin{table*}[t]
    \centering
    \renewcommand{\arraystretch}{1.25}
    \caption{Mean posterior along with 68\% marginalized constraints for the {\color{blue}Pantheon$+$} joint dataset.}
    \label{TableV}
    \begin{tabular}{lcc}
        \hline\hline
        Parameter & Tachyonic DBI & $\Lambda$CDM \\
        \hline
        $\Omega_{\mathrm{m0}}$ & {\color{blue}$0.3107\pm0.0054$} & {\color{blue}$0.3015\pm0.0030$} \\
        $H_0$ & {\color{blue}$67.4150\pm0.5932$} & {\color{blue}$68.5718\pm0.2594$} \\
        $\omega_{\mathrm{b}}$ & {\color{blue}$0.0226\pm0.0001$} & {\color{blue}$0.0225\pm0.0001$} \\
        $\sigma_{8,0}$ & {\color{blue}$0.7587\pm0.0138$} & {\color{blue}$0.7608\pm0.0138$} \\
        $S_8$ & {\color{blue}$0.7720\pm0.0144$} & {\color{blue}$0.7627\pm0.0135$} \\
        \hline
        $\lambda_{\mathrm{ini}}$ & {\color{blue}$0.8552\pm0.3279$} & --- \\
        {\color{blue}$u\equiv1/\Gamma$} & {\color{blue}$0.4936\pm0.2755$} & --- \\
        {\color{blue}$\Gamma$ (derived, median)} & {\color{blue}$2.11^{+3.42}_{-0.90}$} & --- \\
        $\log(\Omega_{de,\mathrm{ini}})$ & {\color{blue}$-8.8446\pm0.0187$} & --- \\
        \hline\hline
    \end{tabular}
\end{table*}

\section{DES-SN5YR joint constraints}
\label{A5}

In this appendix, we present the marginalized constraints obtained from the full joint likelihood in Eq.~\eqref{C1}, specialized to the case in which the SN component is given by the DES-SN5YR data set.

\subsection*{Marginalized posterior constraints}

\begin{figure*}[t]
    \centering
    \begin{subfigure}[t]{0.48\textwidth}
        \centering
\includegraphics[width=\textwidth]{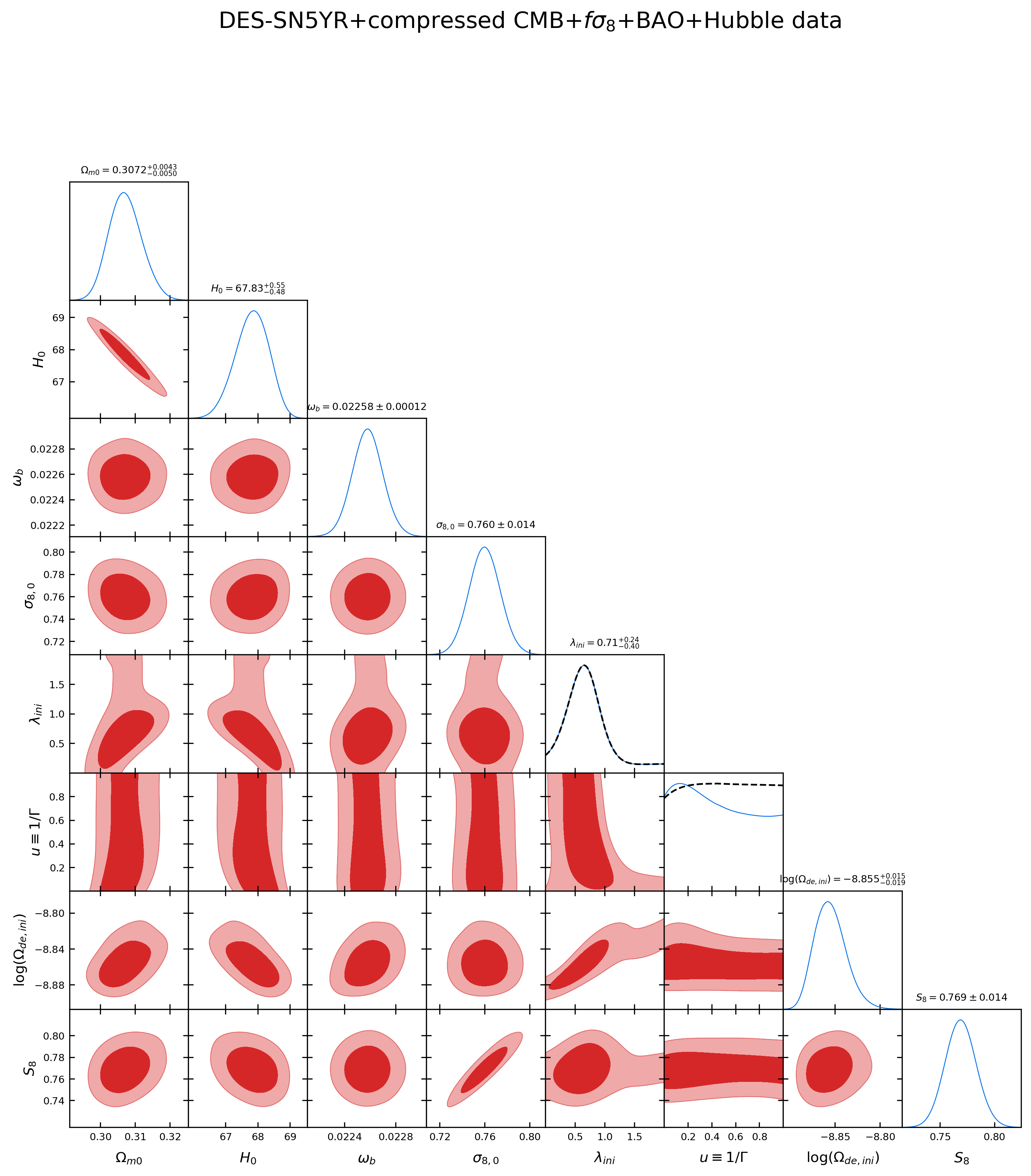}
        \caption{Tachyonic DBI model.}
        \label{FigA2T}
    \end{subfigure}\hfill
    \begin{subfigure}[t]{0.48\textwidth}
        \centering
        \includegraphics[width=\textwidth]{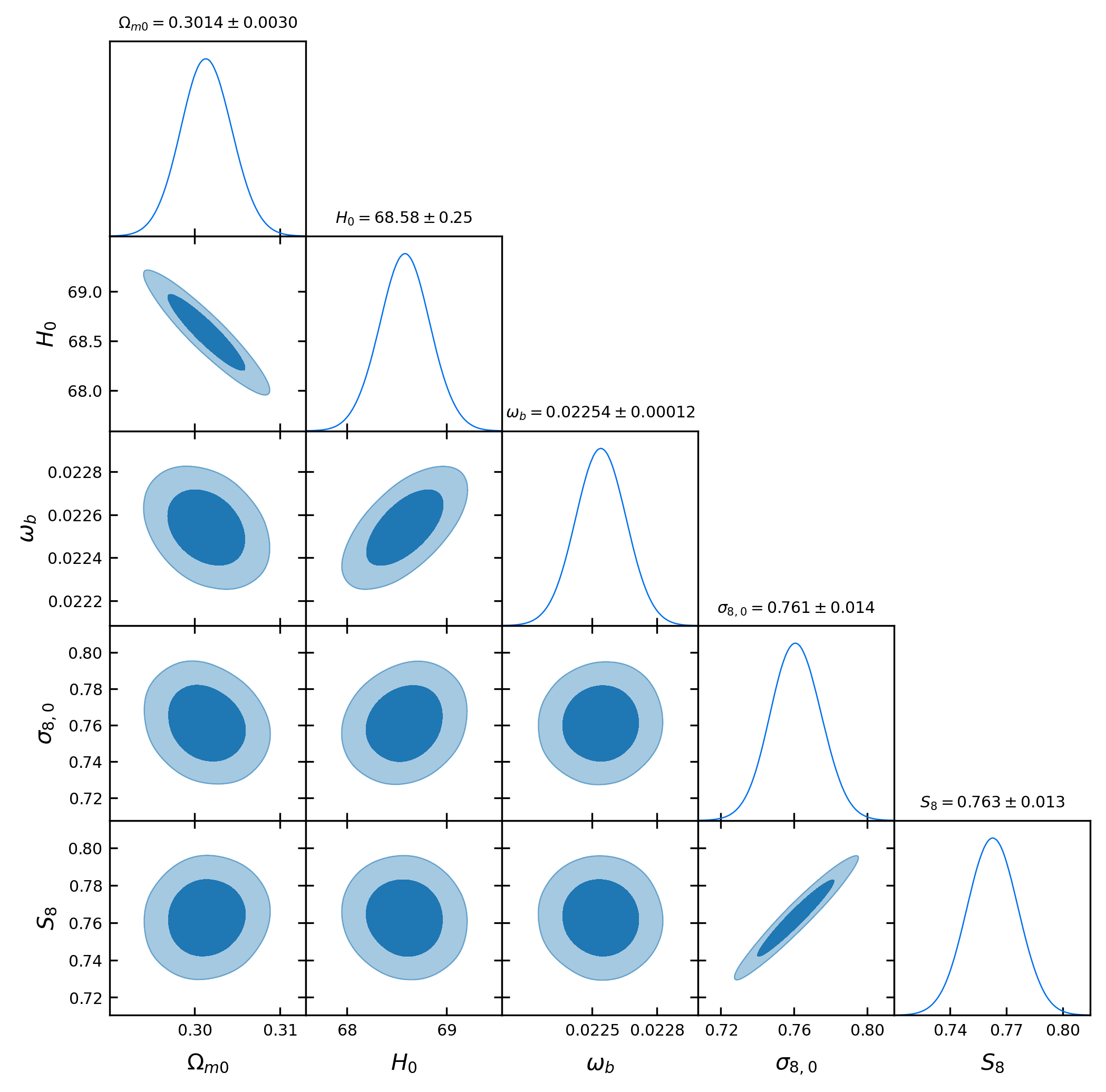}
        \caption{$\Lambda$CDM model.}
        \label{FigA2L}
    \end{subfigure}
    \caption{Marginalized constraints for DES-SN5YR combined with CMB distance priors, DESI DR2 BAO, CC, and $f\sigma_8$. Contours show the 68\% and 95\% confidence regions.}
\end{figure*}

Figures~\ref{FigA2T} and \ref{FigA2L} show the 1D and 2D marginalized posteriors for the same DES-SN5YR$+$CC$+$BAO$+$CMB distance-prior$+$ $f\sigma_8$ dataset, comparing the Tachyonic DBI model (Fig.~\ref{FigA2T}) against the baseline $\Lambda$CDM scenario (Fig.~\ref{FigA2L}).

For convenience, Table~\ref{TableVI} summarizes the 68\% marginalized constraints for the DES-SN5YR joint dataset, shown side-by-side for the Tachyonic DBI model (Fig.~\ref{FigA2T}) and the baseline $\Lambda$CDM model (Fig.~\ref{FigA2L}).

\begin{table*}[t]
    \centering
    \renewcommand{\arraystretch}{1.25}
    \caption{Mean posterior along with 68\% marginalized constraints for the DES-SN5YR joint dataset. }
    \label{TableVI}
    \begin{tabular}{lcc}
        \hline\hline
        Parameter & Tachyonic DBI & $\Lambda$CDM \\
        \hline
        $\Omega_{\mathrm{m0}}$ & {\color{blue}$0.3072\pm0.0047$} & {\color{blue}$0.3014\pm0.0030$} \\
        $H_0$ & {\color{blue}$67.8256\pm0.5113$} & {\color{blue}$68.5829\pm0.2533$} \\
        $\omega_{\mathrm{b}}$ & {\color{blue}$0.0226\pm0.0001$} & {\color{blue}$0.0225\pm0.0001$} \\
        $\sigma_{8,0}$ & {\color{blue}$0.7600\pm0.0136$} & {\color{blue}$0.7610\pm0.0136$} \\
        $S_8$ & {\color{blue}$0.7690\pm0.0141$} & {\color{blue}$0.7627\pm0.0134$} \\
        \hline
        $\lambda_{\mathrm{ini}}$ & {\color{blue}$0.7141\pm0.3698$} & --- \\
        {\color{blue}$u\equiv1/\Gamma$} & {\color{blue}$0.4644\pm0.2876$} & --- \\
        {\color{blue}$\Gamma$ (derived, median)} & {\color{blue}$2.28^{+5.05}_{-1.05}$} & --- \\
        $\log(\Omega_{de,\mathrm{ini}})$ & {\color{blue}$-8.8552\pm0.0172$} & --- \\
        \hline\hline
    \end{tabular}
\end{table*}

\section{Union3 joint constraints}
\label{A6}

In this appendix we present the marginalized constraints obtained from the full joint likelihood in Eq.~\eqref{C1}, specialized to the case in which the SN component is given by the Union3 data set.

\subsection*{Marginalized posterior constraints}

\begin{figure*}[t]
    \centering
    \begin{subfigure}[t]{0.48\textwidth}
        \centering
        \includegraphics[width=\textwidth]{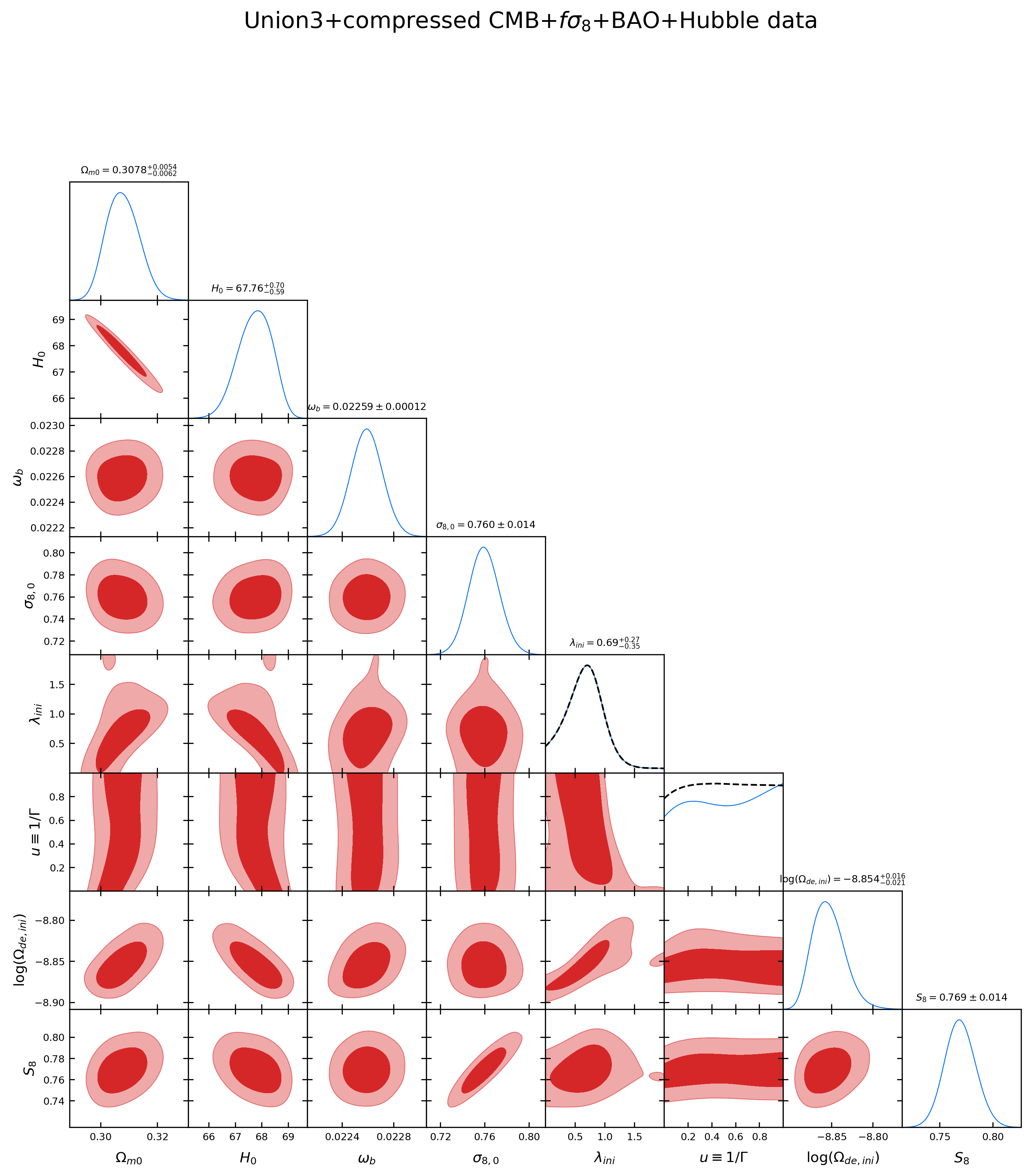}
        \caption{Tachyonic DBI model.}
        \label{FigA3T}
    \end{subfigure}\hfill
    \begin{subfigure}[t]{0.48\textwidth}
        \centering
        \includegraphics[width=\textwidth]{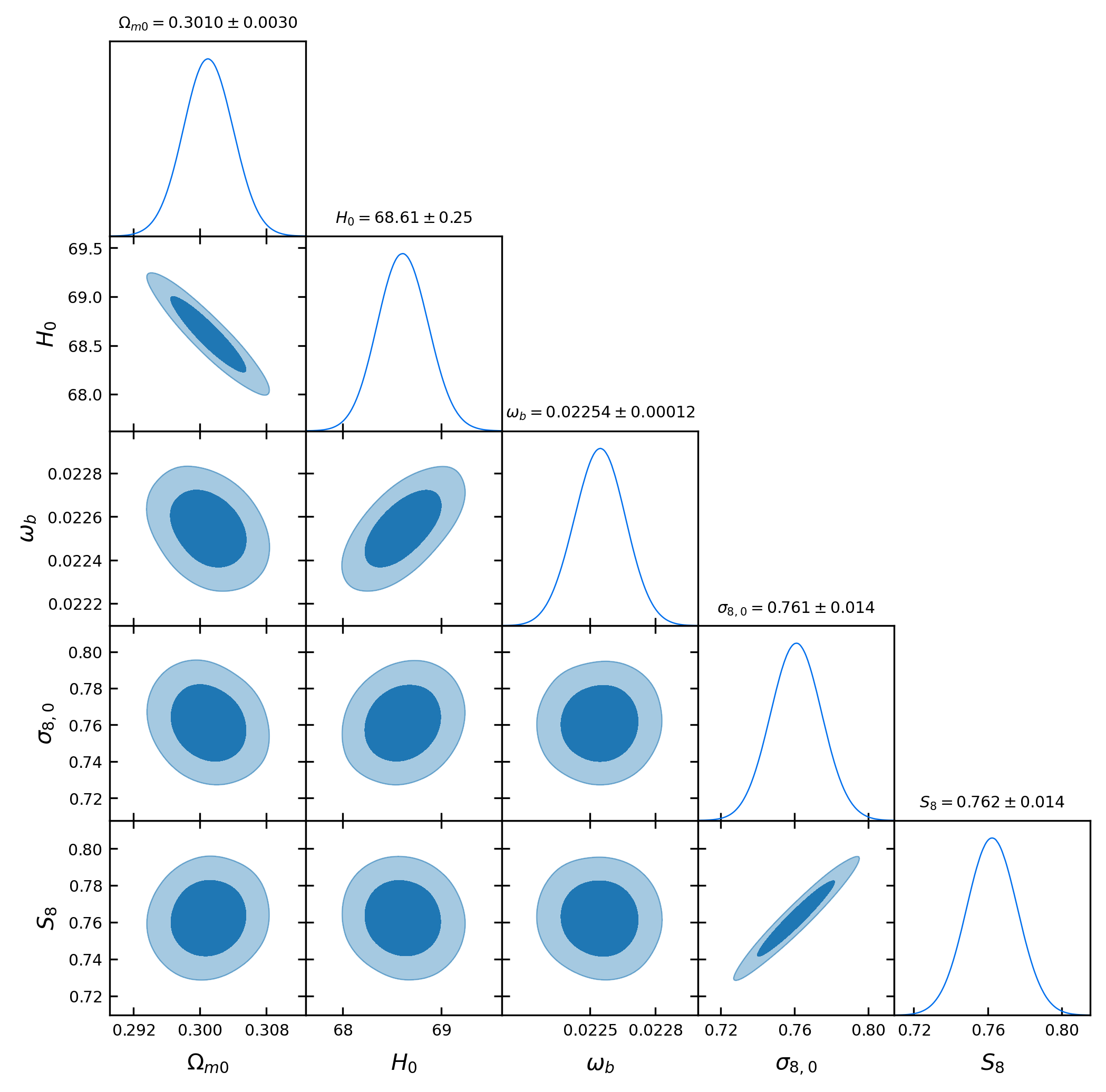}
        \caption{$\Lambda$CDM model.}
        \label{FigA3L}
    \end{subfigure}
    \caption{Marginalized constraints for Union3 combined with CMB distance priors, DESI DR2 BAO, CC, and $f\sigma_8$. Contours show the 68\% and 95\% confidence regions.}
\end{figure*}

Figures~\ref{FigA3T} and \ref{FigA3L} show the 1D and 2D marginalized posteriors for the same Union3$+$CC$+$BAO$+$CMB distance-prior$+$ $f\sigma_8$ dataset, comparing the Tachyonic DBI model (Fig.~\ref{FigA3T}) against the baseline $\Lambda$CDM scenario (Fig.~\ref{FigA3L}).

For convenience, Table~\ref{TableVII} summarizes the 68\% marginalized constraints for the Union3 joint dataset, shown side-by-side for the Tachyonic DBI model (Fig.~\ref{FigA3T}) and the baseline $\Lambda$CDM model (Fig.~\ref{FigA3L}).

\begin{table*}[t]
    \centering
    \renewcommand{\arraystretch}{1.25}
    \caption{Mean posterior along with 68\% marginalized constraints for the Union3 joint dataset.}
    \label{TableVII}
    \begin{tabular}{lcc}
        \hline\hline
        Parameter & Tachyonic DBI & $\Lambda$CDM \\
        \hline
        $\Omega_{\mathrm{m0}}$ & {\color{blue}$0.3078\pm0.0057$} & {\color{blue}$0.3010\pm0.0030$} \\
        $H_0$ & {\color{blue}$67.7571\pm0.6320$} & {\color{blue}$68.6132\pm0.2538$} \\
        $\omega_{\mathrm{b}}$ & {\color{blue}$0.0226\pm0.0001$} & {\color{blue}$0.0225\pm0.0001$} \\
        $\sigma_{8,0}$ & {\color{blue}$0.7596\pm0.0136$} & {\color{blue}$0.7610\pm0.0137$} \\
        $S_8$ & {\color{blue}$0.7693\pm0.0143$} & {\color{blue}$0.7623\pm0.0136$} \\
        \hline
        $\lambda_{\mathrm{ini}}$ & {\color{blue}$0.6946\pm0.3456$} & --- \\
        {\color{blue}$u\equiv1/\Gamma$} & {\color{blue}$0.5184\pm0.2908$} & --- \\
        {\color{blue}$\Gamma$ (derived, median)} & {\color{blue}$1.91^{+3.72}_{-0.75}$} & --- \\
        $\log(\Omega_{de,\mathrm{ini}})$ & {\color{blue}$-8.8536\pm0.0189$} & --- \\
        \hline\hline
    \end{tabular}
\end{table*}

\section{Optimal Linear Compression of Supernova Data}
\label{A7}

Raw supernova catalogs contain $N \sim 10^2$–$10^3$ highly correlated data points. We compress them into $M \ll N$ redshift bins via optimal linear compression, which preserves the full Fisher information of the original dataset. The procedure follows five sequential steps.

Let $\boldsymbol{\mu}_\mathrm{data}$ be the $N$-vector of observed distance moduli and $\boldsymbol{\mu}_\mathrm{fid}(z)$ the prediction of a reference flat $\Lambda$CDM cosmology, the fiducial model. The raw residuals are
\ben
\boldsymbol{\delta}_\mathrm{raw} = \boldsymbol{\mu}_\mathrm{data} - \boldsymbol{\mu}_\mathrm{fid}(\mathbf{z}).
\label{G1}
\een
Let $\mathbf{C}$ denote the $N\times N$ covariance matrix of the dataset (statistical + systematic), and $\mathbf{C}^{-1}$ its precision matrix.

We know that, the SN distance modulus $\mu = 5\log_{10}d_L + 25$ depends on the SN absolute magnitude $M_B$, which is a global additive unknown common to all data points. It manifests as an overall vertical shift in residual plots and therefore must be removed before compression to avoid biasing the binned values.

The inverse-covariance weighted mean of the raw residuals gives the optimal estimate of this offset:
\ben
\delta_0
= \frac{\mathbf{1}^\top \mathbf{C}^{-1}\,\boldsymbol{\delta}_\mathrm{raw}}{\mathbf{1}^\top \mathbf{C}^{-1}\,\mathbf{1}} \equiv \frac{\mathbf{1}^\top \mathbf{C}^{-1}\,\boldsymbol{\delta}_\mathrm{raw}}{W_\mathrm{raw}},
\label{G2}
\een
where $\mathbf{1}$ is an $N$-vector of ones and $W_\mathrm{raw} \equiv \mathbf{1}^\top \mathbf{C}^{-1} \mathbf{1}$ is the total raw precision. In all the subsequent steps the centered residuals are treated as
\ben
\boldsymbol{\delta}_\mathrm{c}
= \boldsymbol{\delta}_\mathrm{raw} - \delta_0\,\mathbf{1}
\label{G3}
\een

Now, we partition the $N$ supernovae into $M$ redshift bins. In order to do so, we define the $N \times M$ indicator matrix $\mathbf{A}$ by
\ben
    A_{ij} =
    \begin{cases}
        1 & \text{if supernova } i \text{ falls in bin } j, \\
        0 & \text{otherwise.}
    \end{cases}
\label{G4}
\een
Then, the $M\times M$ Fisher information matrix projected into the binned space is
\ben
\mathbf{F} = \mathbf{A}^\top \mathbf{C}^{-1} \mathbf{A},
\label{G5}
\een
and the binned covariance matrix is its inverse,
\ben
\mathbf{C}_\mathrm{bin} = \mathbf{F}^{-1}
= \left(\mathbf{A}^\top \mathbf{C}^{-1} \mathbf{A}\right)^{-1}.
\label{G6}
\een
With respect to this binned covariance defined in Eq.(\ref{G6}), the optimally compressed residuals are
\ben
\boldsymbol{\delta}_\mathrm{bin}
= \mathbf{C}_\mathrm{bin}\,\mathbf{A}^\top \mathbf{C}^{-1}\,
\boldsymbol{\delta}_\mathrm{c}.
\label{G7}
\een
This estimator is completely unbiased.

Now to binned the redshift bins we represents redshift of bin $j$ is the diagonal-precision weighted mean of the member supernovae:
\ben
\bar{z}_j
= \frac{\displaystyle\sum_{i \in j}\,\left(\mathbf{C}^{-1}\right)_{ii}\, z_i}{\displaystyle\sum_{i \in j}\,\left(\mathbf{C}^{-1}\right)_{ii}},
\label{G8}
\een
where the sum runs over all supernovae assigned to bin $j$.

Now, the diagonal of $\mathbf{C}_\mathrm{bin}$ defined in Eq.(\ref{G6}) gives the standard error in each bin. However, since the absolute zero-point has been removed, the residuals retain a residual common-mode variance of $1/W_\mathrm{bin}$, where
\ben
W_\mathrm{bin}
\equiv \mathbf{1}_M^\top\,\mathbf{C}_\mathrm{bin}^{-1}\,\mathbf{1}_M
= \mathbf{1}_M^\top \mathbf{F}\,\mathbf{1}_M,
\label{G9}
\een
and $\mathbf{1}_M$ is an $M$-vector of ones. Subtracting this shared variance yields the minimised per-bin uncertainty
\ben
\sigma_{\mathrm{min},j}^2
= \left(\mathbf{C}_\mathrm{bin}\right)_{jj} - \frac{1}{W_\mathrm{bin}},
\label{G10}
\een
which serves as the plotted error bar. Note that $W_\mathrm{raw}$ defined in Eq.(\ref{G2}) and $W_\mathrm{bin}$ in Eq.(\ref{G9}) are distinct quantities defined in different spaces and are not interchangeable.

\textit{Model Curve :} Due to the $M_B$ degeneracy, the vertical zero-point of any supernova residual plot is arbitrary. A theoretical model predicts only the shape of the distance-redshift relation, not its absolute calibration. Any constant vertical shift of $\mu_\mathrm{model}(z)$ is equally valid from the model's perspective, since it is indistinguishable from a change in $M_B$.

The data points $\boldsymbol{\delta}_\mathrm{bin}$ have already had their own $M_B$ removed via the offset $\delta_0$ in Eq.~\eqref{G2}, which anchors them to a precisely defined weighted zero. If the theoretical curve is not subjected to the identical anchoring, the comparison becomes meaningless. A model could appear perfectly consistent or strongly discrepant depending entirely on an unphysical calibration constant rather than genuine cosmological content. Pinning enforces a common reference frame for both data and theory.

Concretely, any uniform shift $\Delta\mu_\mathrm{model}(z)
= \mathrm{const}$ is physically unobservable as it is degenerate with $M_B$, so it must be projected out before visual comparison. Only \emph{redshift-dependent} deviations survive this operation and carry genuine information about the expansion history.

For a model predicting $\boldsymbol{\mu}_\mathrm{model}(z)$,
the theoretical residual at the raw supernova redshifts is defined in Eq.(\ref{G1}). The optimal vertical offset is determined using the same full precision matrix $\mathbf{C}^{-1}$ as in Eq.~\ref{G2}:
\ben
\Delta_\mathrm{pin}
= \frac{\mathbf{1}^\top \mathbf{C}^{-1}\,
\Delta\boldsymbol{\mu}_\mathrm{model}}
{W_\mathrm{raw}}.
\label{G11}
\een
Subtracting this scalar enforces the constraint $\mathbf{1}^\top \mathbf{C}^{-1}\,\Delta\mu_\mathrm{pin}(\mathbf{z}) = 0$, identical to the constraint already imposed on the data. The pinned model residual curve, evaluated over a continuous redshift grid, is then
\ben
\Delta\mu_\mathrm{pin}(z)
= \mu_\mathrm{model}(z) - \mu_\mathrm{fid}(z) - \Delta_\mathrm{pin}.
\label{G12}
\een
By construction, any discrepancy between this curve and the binned data at a specific redshift reflects a genuine difference in expansion dynamics rather than a calibration offset.

\end{document}